%% file: main.tex
\def\NAMEDBUILD{}
\ifdefined\NAMEDBUILD
\documentclass[a4paper,UKenglish,cleveref,autoref,nosubfigcap]{lipics-v2021}
\else
\documentclass[a4paper,UKenglish,cleveref,autoref,nosubfigcap,notab,anonymous]{socg-lipics-v2021}
\fi
\usepackage{booktabs,longtable,float,array}
\usepackage{tikz}
\usetikzlibrary{decorations.pathreplacing}
\let\paragraph\subparagraph
\input{gt/numbers.tex}

\input{gt/numbers_v2.tex}

\input{gt/numbers_2.tex}
\renewcommand{\PcattLazyN}{90{,}000}
\renewcommand{\PcattSteps}{35{,}544}
\renewcommand{\PcattTOne}{1{,}543{,}232}
\renewcommand{\JBestTieWords}{88{,}000}
\newcommand{\OPT}{\mathrm{OPT}}
\newcommand{\GT}{\textsc{GreedTok}}
\newcommand{\GTv}{\mathrm{GT}}
\newcommand{\Sc}{\mathrm{Sc}}
\newcommand{\cov}{\mathrm{cov}}
\newcommand{\JLIM}{\tfrac12-\tfrac{3^{1/4}}{3e}}
\newcommand{\ByteRatio}{0.340124}
\newcommand{\ByteRatioBig}{0.339466}

\newcommand{\ourname}{Baliga, Xie and Byju}
\ifdefined\NAMEDBUILD\else\fi

\title{GreedTok Is a 0.3287-Approximation, and Greedy Tokenizers Lose Order $1/\log k$ on Fixed Alphabets}
\titlerunning{Worst-Case Savings of Greedy Tokenizers}
\ifdefined\NAMEDBUILD
\author{Kundan Baliga}{University of Illinois Urbana-Champaign, USA}{kbaliga2@illinois.edu}{}{}
\author{Aiden Xie}{Georgia Institute of Technology, USA}{axie82@gatech.edu}{}{}
\author{Aditya Byju}{Georgia Institute of Technology, USA}{abyju6@gatech.edu}{}{}
\authorrunning{K. Baliga, A. Xie, and A. Byju}
\Copyright{Kundan Baliga, Aiden Xie, and Aditya Byju}
\newcommand{\pdfauthors}{Kundan Baliga, Aiden Xie, Aditya Byju}
\else
\author{Anonymous author(s)}{Anonymous affiliation(s)}{}{}{}
\authorrunning{Anonymous author(s)}
\Copyright{Anonymous author(s)}
\newcommand{\pdfauthors}{}
\fi
\pdfinfoomitdate=1 \pdftrailerid{}
\hypersetup{pdftitle={GreedTok Is a 0.3287-Approximation, and Greedy Tokenizers Lose Order 1/log k on Fixed Alphabets},
  pdfauthor={\pdfauthors},pdfcreator={},pdfproducer={},pdfkeywords={}}
\hideLIPIcs

\begin{document}
\ifdefined\NAMEDBUILD\nolinenumbers\fi
\DeclareFontShape{T1}{lmr}{bx}{sc}{<->ssub*lmr/bx/n}{}
\DeclareFontShape{T1}{lmr}{m}{scit}{<->ssub*lmr/m/scsl}{}
\ifdefined\NAMEDBUILD
\supplement{Code and checkers for the computed statements (Appendix~\ref{app:comp}).}
\supplementdetails[subcategory={GreedTok}]{Software}{https://doi.org/10.5281/zenodo.24451478}
\else
\supplement{Code and checkers for the computed statements (Appendix~\ref{app:comp}): the anonymous supplementary archive accompanying this submission.}
\fi
\fontfamily{lmss}\selectfont\DeclareFontShape{T1}{lmss}{bx}{sc}{<->ssub*lmss/bx/n}{}\normalfont
\maketitle
\hypersetup{pdftitle={GreedTok Is a 0.3287-Approximation, and Greedy Tokenizers Lose Order 1/log k on Fixed Alphabets}}

\begin{abstract}
A tokenizer's vocabulary is built once and used on every input, so it matters how much compression a greedy
construction can lose. We study two greedy constructions: \GT{} (Lim et al., 2025), which adds the string that
saves the most tokens, and byte-pair encoding (BPE), which merges a most frequent adjacent pair. We prove that \GT{} saves at least
$0.3287$ of the optimal number of tokens saved on every corpus, while no constant above $0.33862$ is possible. The key idea
is that, given the order of greedy's choices, the damage inside an optimal token is fixed by two numbers. On a fixed
alphabet both algorithms are asymptotically optimal and their worst deficit with budget $k$ has order $1/\log k$: one
counting argument gives the upper bound for both, and one binary corpus, whose first greedy commitment erases a boundary
between optimal tokens, gives the matching lower bound for both (for \GT{}, on its committed segmentation), even if greedy's continuation may use
almost twice the budget. For BPE on one
binary string, a forced first merge can cost $\Omega(k)$ final tokens, yet $2k+3$ rules always repair it.
\end{abstract}

\input{sections/intro}
\input{sections/model}
\input{sections/overview}

\input{sections/warmup}
\input{sections/uniform}
\input{sections/fixed}
\input{sections/discussion}

\bibliographystyle{plainurl}
\ifdefined\NAMEDBUILD\bibliography{refs}\else\bibliography{refs-anon}\fi

\clearpage
\appendix
\crefalias{section}{appendix}\crefalias{subsection}{appendix}
\input{app/notation}
\input{app/model}

\input{app/windows}
\input{app/main}

\input{app/upper}

\input{app/alphabet}

\input{app/fixed}
\input{app/string}

\input{app/hard}

\input{app/comp}
\end{document}

%% file: gt/numbers.tex
\newcommand{\CovKmax}{16}
\newcommand{\CovKmin}{6}
\newcommand{\CovN}{10}

\newcommand{\CsKmax}{12}

\newcommand{\EXSTATES}{72,205,022}

\newcommand{\HardCtrl}{16}

\newcommand{\HardInst}{37}
\newcommand{\LZBIGchecks}{1,780,818,426}

\newcommand{\LZinstances}{4,965}

\newcommand{\PcattLazyN}{90,000}

\newcommand{\PcattSteps}{35,544}
\newcommand{\PcattTOne}{1,543,232}
\newcommand{\PcattThree}{3/5}

\newcommand{\VSINST}{103,713}

%% file: gt/numbers_v2.tex
\newcommand{\JBestTie}{0.350316}
\newcommand{\JBestTieFrac}{14348933742199/40960000000000}
\newcommand{\JBestTieK}{400}
\newcommand{\JBestTieWords}{88,000}

\newcommand{\JFourLimSeven}{0.3386145}

\newcommand{\JTwoLimSix}{0.359416}
\newcommand{\KfourUni}{0.437769}
\newcommand{\KthreeUni}{0.500371}
\newcommand{\KtwoTie}{7/12}

\newcommand{\KtwoUni}{0.587669}
\newcommand{\KtwoWords}{17}
\newcommand{\SrhoFour}{101/256}
\newcommand{\SrhoThree}{65/162}
\newcommand{\SrhoTwo}{5/12}

\newcommand{\SyOneSixteen}{0.3386144758582598}
\newcommand{\TauStar}{0.5623}

\newcommand{\JFourHundred}{0.367140}
\newcommand{\JFourHundredFrac}{1468559/4000000}

\newcommand{\MAINconst}{0.3043}

\newcommand{\OcertJfour}{8, 9, 10, 12}
\newcommand{\OcertJtwo}{7, 8, 9, 12, 14, 20}

%% file: gt/numbers_2.tex
\renewcommand{\MAINconst}{0.3287}
\newcommand{\MAINyOneFull}{0.3287413432932835}
\newcommand{\ByteThresh}{104{,}910}
\newcommand{\ByteTheta}{0.4292}
\newcommand{\ByteMargin}{8\cdot10^{-8}}
\newcommand{\ByteGa}{0.32874}
\newcommand{\ByteGb}{0.32874}
\newcommand{\ByteGc}{0.32874}
\newcommand{\ByteGd}{0.35400}
\newcommand{\ByteGe}{0.39200}
\newcommand{\ByteGf}{0.41346}
\newcommand{\ByteGg}{0.42495}
\newcommand{\ByteGh}{0.43116}
\newcommand{\ByteGi}{0.43456}

%% file: sections/intro.tex
\section{Introduction}\label{sec:intro}
A tokenizer fixes the units in which a language model reads text. Its vocabulary is chosen once, on a training corpus,
and then used for every training and inference token; fewer tokens mean shorter sequences and less computation.
Choosing the vocabulary is a combinatorial optimization problem, and the methods used in practice solve it greedily.
Byte-pair encoding (BPE) \cite{sennrich2016} repeatedly merges a most frequent adjacent pair of symbols. \GT{}
\cite{lim2025} adds, one at a time, the string whose commitment saves the most tokens. An empirical ratio measured on a
few corpora says little about other languages, scripts or domains. We ask the worst-case question: \emph{how much
compression can a greedy vocabulary lose, on any corpus at all?} We measure compression by the number of tokens saved
relative to one token per character, and compare greedy with the best vocabulary of the same budget $k$.

\begin{figure}[t]
\centering
\begin{tikzpicture}[x=0.95cm,y=0.95cm,font=\normalsize]
  \draw[thick,fill=gray!30,rounded corners=3pt] (0.58,-0.4) rectangle (2.42,0.4);
  \foreach \c [count=\i from 0] in {x,a,b,y} \node[minimum size=0.5cm,inner sep=0] at (\i,0) {\texttt{\c}};
  \draw[thick] (-0.42,0.75) -- (-0.42,0.95) -- (1.42,0.95) -- (1.42,0.75);
  \draw[thick] (1.58,0.75) -- (1.58,0.95) -- (3.42,0.95) -- (3.42,0.75);
  \node at (0.5,1.35) {\texttt{xa}};
  \node at (2.5,1.35) {\texttt{by}};
  \node[font=\small] at (1.5,-0.85) {committed \texttt{ab}: gain $1$};
  \node[font=\small,align=left,anchor=west] at (4.2,0.6) {optimal: \texttt{xa|by} saves $2$};
  \node[font=\small,align=left,anchor=west] at (4.2,-0.1) {after greedy commits \texttt{ab}:};
  \node[font=\small,align=left,anchor=west] at (4.2,-0.6) {\texttt{xa} and \texttt{by} would cut through \texttt{ab}: both blocked};
\end{tikzpicture}
\caption{A boundary merge. The pair \texttt{ab} straddles the junction between the optimal tokens \texttt{xa} and
\texttt{by}; committing it gains one pair and blocks both. Every bad instance in this paper is built from this
pattern: the junction corpora of \GT{} (Appendix~\ref{sec:upper}) and the binary corpus of
Theorem~\ref{thm:binary}, where the merged pair is \texttt{aa}.}
\label{fig:xaby}
\end{figure}
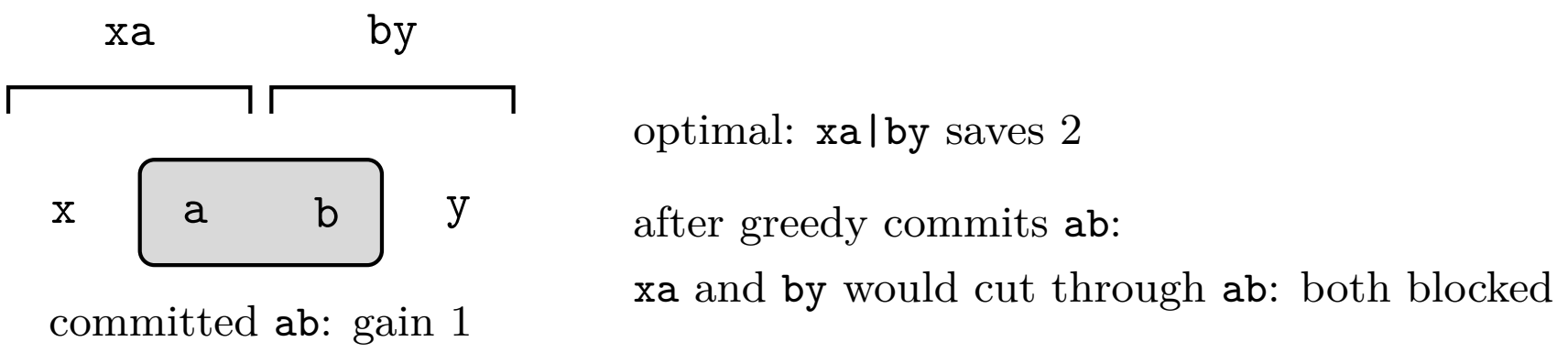

\paragraph{What goes wrong: boundary merges.}
Take the word \texttt{xaby} (Figure~\ref{fig:xaby}). The vocabulary $\{\texttt{xa},\texttt{by}\}$ segments it as
\texttt{xa|by} and saves two tokens. If \texttt{ab} is frequent elsewhere, greedy commits it first. This freezes
\texttt{a} and \texttt{b} together in \texttt{xaby}, and neither \texttt{xa} nor \texttt{by} can be used there, because each
would cut through \texttt{ab}. Greedy gained one pair and blocked two. This is why the textbook $1-1/e$ analysis fails:
the objective is not submodular, not even weakly (the submodularity ratio is $0$ on this word). Lim et al.\ observed good
empirical ratios for \GT{} and left a guarantee open \cite[\S6]{lim2025}. Kozma and Voderholzer asked whether BPE
guarantees ``can be strengthened in the small-alphabet case'' \cite{kozma2024arxiv}. We answer the first question and, for
large budgets, the second.

\paragraph{Results.}
Table~\ref{tab:results} lists the results. Write $\GTv_k$ and $\OPT_k$ for the savings of \GT{} and of the best
vocabulary of $k$ tokens. The three headline statements are these.
\begin{enumerate}
\item \textbf{\GT{} is a $0.3287$-approximation} (Theorem~\ref{thm:main}). On every corpus and for every $k$,
$\GTv_k\ge0.3287\,\OPT_k$, for any tie-breaking and for the lazy loop of their Algorithm~1. Junction corpora show that no
constant above $\JLIM\approx0.33861$ is possible, and for optimal tokens of at most three characters this constant is exact.
\item \textbf{On a fixed alphabet both greedy algorithms lose exactly order $1/\log k$} (Theorems~\ref{thm:fixed}
and~\ref{thm:binary}). Over $\sigma$ letters \GT{} and greedy BPE keep at least $1-(2+o(1))/\log_\sigma k$ of the optimum.
One binary corpus forces a deficit of order $1/\log k$ on both (for \GT{}, on its committed segmentation), and the deficit
survives even if a continuation that keeps greedy's first step may use almost $2k$ tokens or rules.
\item \textbf{One greedy merge can be expensive, but is repairable on one string} (Theorems~\ref{thm:string}
and~\ref{thm:repair}). On a binary string that $k$ merges reduce to one token, every continuation of greedy BPE's first
merge with $k$ rules leaves $\Omega(k)$ tokens; yet a continuation with $2k+3$ partial rules always recovers the comparator's final length.
\end{enumerate}
Further results: the exact worst ratio $1-1/M$ ($k\ge2$) of exhausted \GT{} when tokens are capped at length $M$
against the capped optimum (Theorem~\ref{thm:capped-exhaustion}); an EPTAS for merge-based savings on fixed alphabets (Corollary~\ref{cor:eptas});
byte-alphabet bounds for \GT{}; and APX-hardness of the \GT{} objective over unbounded alphabets (Theorem~\ref{thm:hard}).

\begin{table}[t]
\caption{Results. ``Deficit'' is $1-\text{greedy}/\text{optimum}$ in the worst case; $\sigma$ is the alphabet size and
$k$ the budget. Greedy BPE on unbounded alphabets is analysed in \cite{bpetight}.}\label{tab:results}
\centering\small
\begin{tabular}{@{}p{0.21\textwidth}p{0.38\textwidth}p{0.35\textwidth}@{}}\toprule
Regime & \GT{} & greedy BPE (both count rules)\\\midrule
unbounded $\sigma$ & $\ge0.3287$ (Thm~\ref{thm:main}); $\limsup\le0.33862$ (Thms~\ref{thm:j4}, \ref{thm:unique}), exact for tokens of $\le3$ characters (Thm~\ref{thm:short}) & see \cite{bpetight}\\
fixed $\sigma$: guarantee & deficit $\le2/\log_\sigma(k/4\ln k)+1/k$ (Thm~\ref{thm:fixed}) & deficit $\le(2+o(1))/\log_\sigma k$ (Thm~\ref{thm:fixed})\\
binary corpus: bad case & \multicolumn{2}{p{0.75\textwidth}@{}}{one corpus for both: deficit $\Omega(1/\log k)$ after the first greedy step, even with $(2-\epsilon)k$ entries or rules (Thm~\ref{thm:binary})}\\
one binary string & --- & first merge: deficit $\Omega(1/\log k)$ with $(\frac43-\epsilon)k$ rules, and $\Omega(k)$ final tokens against $1$; repair with $2k+3$ rules (Thms~\ref{thm:string}, \ref{thm:repair})\\
cap $M$, exhausted & exactly $1-1/M$ for $k\ge2$ (Thm~\ref{thm:capped-exhaustion}) & ---\\
256 symbols & $>0.33861$ for $k\ge\ByteThresh$ (Cor.~\ref{cor:byte}) & ---\\
complexity & objective APX-hard for unbounded $\sigma$ (Thm~\ref{thm:hard}) & EPTAS for fixed $\sigma$ (Cor.~\ref{cor:eptas})\\\bottomrule
\end{tabular}
\end{table}

\paragraph{Key ideas.}
A commit destroys optimal value only locally: a covered pair borders at most two optimal occurrences, and inside an
occurrence it blocks only the substrings (\emph{windows}) that start or end at it. So we credit greedy with random windows
of the optimal tokens and obtain a recursion like the one behind $1-1/e$. The difficulty is long tokens, where an
adversary covering arbitrary pairs would win. Our key observation is that greedy's own history pins the covered pairs
down: given the order of its choices, the pattern inside an optimal occurrence is determined by its first and last
uncovered pair. On fixed alphabets a second, simpler mechanism takes over: there are few short strings, so every
uncovered pair between two short units is cheap for greedy to collect, and every other uncovered pair is paid for by
greedy's own savings (Lemma~\ref{lem:dom}). Both guarantees charge an uncovered or covered pair to at most two neighbouring pieces, the units or the optimal
occurrences it borders, and every bad instance uses Figure~\ref{fig:xaby} on a large scale.
Section~\ref{sec:overview} explains all three.

\paragraph{Prior and concurrent work.}
Zouhar et al.~\cite{zouhar2023formal} bound greedy BPE through a curvature parameter. Kozma and Voderholzer
\cite{kozma2026esa,kozma2024arxiv} prove that optimal pair encoding is APX-complete and bound BPE's worst-case ratio
below by $1/3$ and, in the extended version, above by $(1-e^{-2})/2$; \ourname{} \cite{bpetight} show that the second constant is exact for the
replacement-count rule. For \GT{}, an earlier unrefereed preprint by the same group, superseded by this paper, proved $\GTv_k\ge0.3043\,\OPT_k$ with computer-verified certificates, using the
shift lemma (Lemma~\ref{lem:shift}), and gave the junction corpora with their limit $\JLIM$, exact for tokens of at most three characters. For direct tokenization the decision problem is NP-complete \cite{whittington2025}, and Kastreva et
al.~\cite{kastreva2025} prove APX-hardness for the token-count objective, even over a binary alphabet. Three concurrent
papers study the complexity of the problems themselves, not \GT{} or BPE. Ion, Mateiu, Miclaus and Popa
\cite{ion2026} give, for direct savings (our objective $f$), a deterministic EPTAS on every fixed alphabet
\cite[Theorem~5.6]{ion2026} and guarantees $(1-1/e)/2$ and $(1-e^{-3})/3$ for two greedy algorithms that differ from
\GT{} \cite[Theorems~5.2, 5.4]{ion2026}; for the direct token count they rule out $n^{1-\epsilon}$-approximations on binary
inputs of total length $n$. \ourname{} \cite{complexity} prove that for full or partial pair merges on one binary string
of length $n$ the final token count cannot be approximated within $cn/\log n$ unless P$=$NP \cite[Theorem~10]{complexity}. Bastan, He, Pavan and Vinodchandran \cite{bastan2026} rule out
$n^{1-\epsilon}$-approximations for binary direct tokenization with the token-count objective and give
bicriteria algorithms that exceed both the budget and the optimal token count by a factor $\sqrt{M'}\log n$, where $M'$ is
the maximum document length. Section~\ref{sec:related} compares each with our results.

\paragraph{What is new.}
Beyond that earlier preprint: the history family and the peeling induction, which prove $0.3287$ with no computer certificates;
stronger small-alphabet bounds for \GT{}; the shared fixed-alphabet analysis of both greedy algorithms with its single binary
bad corpus; the length-capped exhaustion ratio; the forced-merge cost and its repair on one string; and the
EPTAS for full merge rules. None of the concurrent
papers analyses \GT{} or greedy BPE, and their EPTAS concerns direct vocabularies, not merge sequences.

\paragraph{Roadmap.}
Section~\ref{sec:model} defines the model, Section~\ref{sec:overview} gives a technical overview, and
Section~\ref{sec:warmup} proves the exact bound for short tokens by hand. Section~\ref{sec:uniform} states the uniform bound,
Section~\ref{sec:fixed} treats fixed alphabets for both algorithms, and Section~\ref{sec:discussion} discusses related
work and limitations. The appendices contain all remaining proofs; Table~\ref{tab:notation} collects the notation.

%% file: sections/model.tex
\section{Model}\label{sec:model}
\paragraph{Corpus, pairs, savings.}
A corpus $\mathcal C$ is a finite set of words $w=w_0\cdots w_{n-1}$ over an alphabet $\Sigma$ of size $\sigma$, with counts
$c_w\ge1$. For $1\le i\le n-1$, \emph{pair $i$} of $w$ is the adjacent pair $(w_{i-1},w_i)$; $P(\mathcal C)=\sum_wc_w(|w|-1)$ is
the total number of pairs. Every tokenizer below ends with each word cut into consecutive pieces, its \emph{units}; it
\emph{saves} $|w|$ minus the number of units, which is the number of pairs inside units, and savings are weighted by the
counts. So $P(\mathcal C)$ bounds every saving, and a pair between two consecutive units is \emph{uncovered}. For
$M\ge2$ such a pair between units $U,V$ is \emph{$M$-short} if $|U|+|V|\le M$, and $N_M$ is the number of distinct
substrings of corpus words with length in $[2,M]$, so $N_M<2\sigma^M$.

\paragraph{Direct savings and \GT{}.}
An \emph{occurrence} of a string $x$ ($|x|\ge2$) is an interval $[s,e)$, $e-s=|x|$, with $w_s\cdots w_{e-1}=x$; its
\emph{interior pairs} are $s+1,\dots,e-1$, and its \emph{boundary pairs} are $s$ (if $s\ge1$) and $e$ (if $e\le n-1$). For a
token set $S$, $f(S)=\sum_wc_w\,\mathrm{cover}(w,S)$, where $\mathrm{cover}(w,S)$ is the maximum number of interior pairs of a
family of pairwise disjoint occurrences of tokens of $S$ in $w$; this is the saving of an optimal segmentation of every
word with $S$ and the single characters, Problem~1 of \cite{lim2025}. The candidates are all substrings of length $\ge2$,
and $\OPT_k=\max_{|S|\le k}f(S)$. The state of \GT{} is, for every word, a family of \emph{committed occurrences}; a pair is
\emph{covered} if it is an interior pair of some committed occurrence. An occurrence is \emph{valid} if neither of its
boundary pairs is covered, and its \emph{gain} is its number of uncovered interior pairs. The \emph{leftmost scan} of $x$
visits, word by word, the occurrences of $x$ in increasing order of start and \emph{places} an occurrence iff it is valid
and does not overlap the previously placed occurrence of $x$ in that word. The \emph{score} $\Sc_t(x)$ is the
count-weighted total gain of the placed occurrences. At step $t+1$, \GT{} picks
$x_{t+1}\in\arg\max_{x\notin S_t}\Sc_t(x)$, with ties broken arbitrarily, and commits every placed occurrence of $x_{t+1}$;
$S_t=\{x_1,\dots,x_t\}$, $g_{t+1}=\Sc_t(x_{t+1})$, $G_t=g_1+\dots+g_t$, and $\GTv_k=G_k$. We measure the committed
segmentation; re-encoding optimally with $S_k$ gives $f(S_k)\ge G_k$. The \emph{lazy loop} of Algorithm~1 of
\cite{lim2025} rescores a candidate only when it reaches the top of a priority queue; it builds the same vocabulary up to
ties (Corollary~\ref{cor:lazy}). Appendix~\ref{app:impl} compares the model with the released implementation.

\paragraph{Merge savings and greedy BPE.}
A \emph{rule} replaces disjoint occurrences of one ordered pair of token types by a fresh type, with no merge across a word
boundary. A \emph{full} rule replaces occurrences in every word in a left-to-right scan; a \emph{partial} rule may replace
any disjoint subset. Starting from the letters, a run saves one token per replacement. For a budget $k$ let
$\OPT^{\mathrm{full}}_k$ and $\OPT^{\mathrm{part}}_k$ be the maximum savings of at most $k$ full and partial rules. Greedy BPE chooses a pair
that maximizes either its overlapping occurrence count (\emph{raw greedy}) or its number of disjoint replacements
(\emph{saving greedy}), then applies the full rule, while an adjacent pair and an unused rule remain; $G_k$ denotes its
saving after $k$ rules. In both models the units of a word are its current tokens, so the notions of covered,
uncovered and $M$-short pairs apply to both algorithms at every step.

%% file: sections/overview.tex
\section{Technical overview}\label{sec:overview}
\paragraph{1. Windows and a recursion.}
Fix an optimal vocabulary and its segmentation. At step $t$, take an optimal token $y$ and a random window $x$ of $y$ (a
substring, drawn from a distribution $D$). The next greedy gain $g_{t+1}$ is at least the total gain of the copies of $x$
inside the optimal occurrences of $y$; a \emph{shift lemma} shows that \GT{}'s left-to-right scan loses nothing when such
copies overlap. If $D$ keeps an $\alpha$ fraction of the pairs of $y$, and each covered pair costs at most $\beta$, then,
summing over the $k$ optimal tokens and using that greedy has covered $G_t$ pairs,
\[
k\,g_{t+1}\ \ge\ \alpha\,\OPT_k-\beta\,G_t .
\]
Several such lines $(\alpha,\beta)$ can be used, the best one at each step, and solving the recursion gives a constant.
Section~\ref{sec:warmup} carries this out by hand for tokens of at most three characters, where it is tight.

\paragraph{2. Greedy's history.}
For long tokens the adversary could cover any set of pairs inside an occurrence, and no distribution $D$ would give a good
line. But a covered stretch strictly between two uncovered pairs must be a string greedy has chosen, and greedy commits
one chosen string at a time by a deterministic scan. So, \emph{given the order of greedy's choices}, the covered pairs of an
occurrence are determined by two numbers, its first and its last uncovered pair (Figure~\ref{fig:idea}b). Against this
two-parameter \emph{history family} we build $D$ by \emph{peeling}: the first chosen window falls into one of three cases,
each reducing to a shorter token (Figure~\ref{fig:idea}c). This proves the lines $(\frac\beta{\beta+1},\beta)$ for
$\beta=1,\dots,6$ and the constant $0.3287$. The order is essential: without it, two chosen windows whose covered runs kill
each other break the line $\beta=1$ at $20$ characters (Figure~\ref{fig:idea}a, Proposition~\ref{prop:cxL20}). For the upper
bound, junction corpora realize Figure~\ref{fig:xaby} at scale: every optimal token is bordered by two junctions whose
cut strings greedy prefers, and in the limit the run follows the extremal trajectory of the short-token recursion step by step.

\paragraph{3. Fixed alphabets: few short strings.}
On $\sigma$ letters there are fewer than $2\sigma^M$ strings of length at most $M$. Every uncovered pair is $M$-short or
not. The two units around a non-short pair contain at least $M-1$ covered pairs, so there are at most $\frac2{M-1}$
non-short pairs per covered pair (Lemma~\ref{lem:dom}). The short pairs are spelled by at most $N_M$ strings. For \GT{},
each such string is a candidate that collects at least half of its short pairs at once, so greedy's next gain is large
while many short pairs remain. For BPE, each spelling belongs to a single typed pair, whose count is at most the final
maximum pair frequency $T$, and each of greedy's $k$ merges has saved at least $T/2$. Both give a deficit of
about $2/M$ once $k\gg N_M$, and $M\approx\log_\sigma k$ gives $(2+o(1))/\log_\sigma k$ (Theorem~\ref{thm:fixed}).

\paragraph{4. One lost boundary.}
The matching bad instance encodes $2m\approx k$ binary codewords $L_i$ and $R_j$ (Theorem~\ref{thm:binary}). The corpus has
many copies of each codeword, one copy of each concatenation $L_iR_j$, and a heavy word \texttt{aa} that forces greedy's
first choice. In $L_iR_j$ the cut lies inside a run of four \texttt{a}'s, and greedy's left-to-right commitment of
\texttt{aa} straddles it, so no codeword token fits in any grid word afterwards. The optimum spends one token per
codeword and segments $L_iR_j$ into two tokens; after the first greedy step, any vocabulary has to pay for codewords and
for $m^2$ grid words separately, and a counting argument shows that even almost $2k$ entries cannot do both. The deficit
is $\Omega(m^2)$ tokens out of $O(m^2\log k)$ savings. The same corpus, the same first step and the same count serve \GT{}
and both BPE rules; only the comparator differs (a direct vocabulary of $2m+1$ strings, or $2m+O(\sqrt m)$ full merges).
On a single string the cells of the construction are arranged along an Euler circuit with unique cut signatures, which
gives Theorem~\ref{thm:string}.

\input{gt/fig_idea}

%% file: gt/fig_idea.tex
\begin{figure}[t]
\centering
\resizebox{\linewidth}{!}{\begin{tikzpicture}[x=0.45cm,y=0.5cm,font=\small]
  \begin{scope}
    \node[anchor=west] at (-1,4.6) {(a) Without history: a mutual kill ($L=20$, $\beta=1$, chosen windows $(6,8)$ and $(7,14)$)};
    \draw[thick] (6,3.3) -- (6,3.6) -- (8,3.6) -- (8,3.3); \node[font=\scriptsize] at (5.6,4.0) {$(6,8)$};
    \draw[thick] (7,2.5) -- (7,2.8) -- (14,2.8) -- (14,2.5); \node[font=\scriptsize] at (10.5,3.2) {$(7,14)$};
    \node[anchor=east,font=\scriptsize] at (-0.3,1.5) {run $\{7\}$};
    \foreach \e in {0,...,20} \draw (\e,1.5) circle (2.4pt);
    \fill (7,1.5) circle (2.4pt);
    \node[anchor=west,font=\scriptsize] at (20.6,1.5) {left end of $(7,14)$ covered: both dead};
    \node[anchor=east,font=\scriptsize] at (-0.3,0.5) {run $\{8..13\}$};
    \foreach \e in {0,...,20} \draw (\e,0.5) circle (2.4pt);
    \foreach \e in {8,...,13} \fill (\e,0.5) circle (2.4pt);
    \node[anchor=west,font=\scriptsize] at (20.6,0.5) {right end of $(6,8)$ covered: both dead};
    \foreach \e in {0,5,10,15,20} \node[font=\scriptsize,gray] at (\e,-0.2) {\e};
  \end{scope}
  \begin{scope}[yshift=-3.6cm]
    \node[anchor=west] at (-1,4.0) {(b) With history: two parameters $p,q$ determine the pattern};
    \foreach \e in {0,...,20} \draw (\e,2.4) circle (2.4pt);
    \foreach \e in {0,1,2,3} \fill (\e,2.4) circle (2.4pt);
    \foreach \e in {18,19,20} \fill (\e,2.4) circle (2.4pt);
    \foreach \e in {7,10,11,12} \fill (\e,2.4) circle (2.4pt);
    \node[font=\scriptsize] at (4,1.7) {$p$}; \node[font=\scriptsize] at (17,1.7) {$q$};
    \draw[decorate,decoration={brace,amplitude=3pt,mirror}] (-0.2,1.9) -- (3.2,1.9) node[midway,below=3pt,font=\scriptsize] {prefix};
    \draw[decorate,decoration={brace,amplitude=3pt,mirror}] (17.8,1.9) -- (20.2,1.9) node[midway,below=3pt,font=\scriptsize] {suffix};
    \draw[decorate,decoration={brace,amplitude=3pt,mirror}] (5.5,1.9) -- (15.5,1.9) node[midway,below=3pt,font=\scriptsize,align=center] {$M(p,q)$: greedy's scan of the chosen\\windows inside $[p,q]$, in time order};
  \end{scope}
  \begin{scope}[yshift=-7.0cm]
    \node[anchor=west] at (-1,4.5) {(c) Peeling the first chosen window $w^*=(a,b)$: three cases, three smaller tokens};
    \draw[thick] (6,3.0) -- (6,3.3) -- (11,3.3) -- (11,3.0); \node[font=\scriptsize] at (8.5,3.65) {$w^*$};
    \foreach \e in {0,...,20} \draw (\e,2.4) circle (2.4pt);
    \node[font=\scriptsize,anchor=east] at (-0.3,2.4) {offsets};
    \node[font=\scriptsize] at (6,1.85) {$a$}; \node[font=\scriptsize] at (11,1.85) {$b$};
    \node[font=\scriptsize,align=left,anchor=north west] at (-1,1.3) {(i) prefix covers $a$\\$\Rightarrow$ sub-token $[a+1,L]$};
    \node[font=\scriptsize,align=left,anchor=north west] at (6.2,1.3) {(ii) a run contains $w^*$\\$\Rightarrow$ contract $a+1..b-1$};
    \node[font=\scriptsize,align=left,anchor=north west] at (13.6,1.3) {(iii) suffix covers $b$\\$\Rightarrow$ sub-token $[0,b-1]$};
  \end{scope}
\end{tikzpicture}}
\caption{The key idea.  (a) Without the order of greedy's choices, the adversary can mix patterns in which either chosen
window's covered run kills the other; at $L=20$ this pushes the best $\alpha$ for the line $\beta=1$ below $\frac12$ (Proposition~\ref{prop:cxL20}).  (b) Greedy
chooses strings at different steps, and whichever comes first is committed first: given the time order, an occurrence's
pattern is fixed by its first and last uncovered offsets $p$ and $q$ (Lemma~\ref{lem:2P}); the example shows a covered
prefix, a covered suffix and two committed windows between them.  (c) The window
$w^*$ chosen first is committed whenever it lies between $p$ and $q$, so every pattern falls into one of three cases, each
of which embeds into a smaller instance (Lemmas~\ref{lem:embed}--\ref{lem:restrict}).  Mixing, or for $\beta=1$ sometimes adding, window distributions
for the smaller instances proves the bound by induction (Lemma~\ref{lem:peel}, Theorem~\ref{thm:PH}).  In (a) and (b), filled dots are covered offsets.}
\label{fig:idea}
\end{figure}
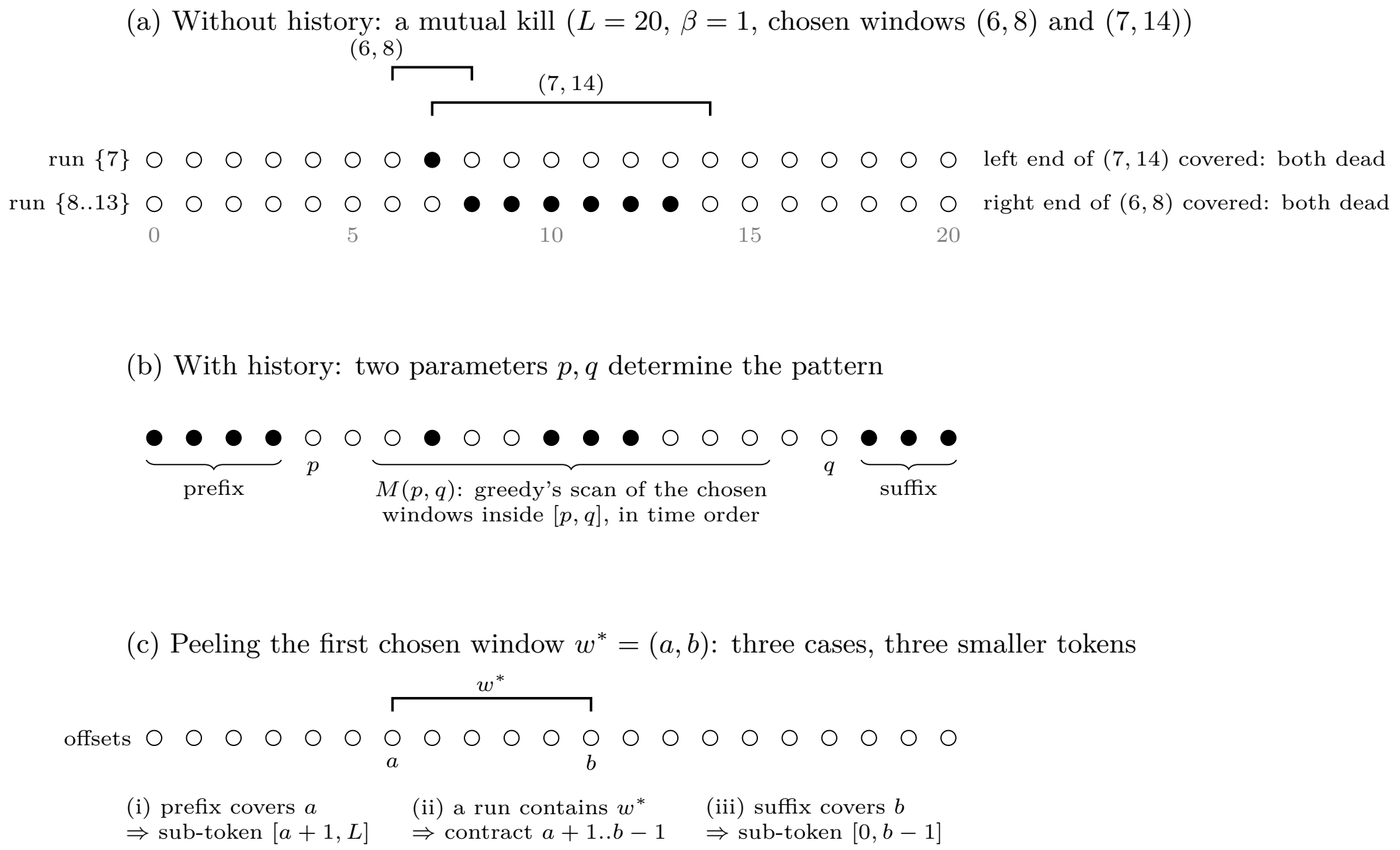

%% file: sections/warmup.tex
\section{Warm-up: the exact bound for short tokens, by hand}\label{sec:warmup}\label{sec:short}
This section proves the tight bound $\JLIM$ for optimal tokens of at most three characters. Every step except the per-step
inequality and the recursion theorem, which rest on the shift and charging lemmas (Appendices~\ref{app:gtmodel}
and~\ref{sec:proof}), can be checked by hand.

\paragraph{Windows.}
Fix an optimal set $S^*$ with $|S^*|\le k$ and, in every word, an optimal family $\mathcal P$ of disjoint occurrences of
tokens of $S^*$. For $P\in\mathcal P$ of length $L$, \emph{offset} $e\in\{0,\dots,L\}$ is the pair between the $(e-1)$-th and $e$-th
character of $P$; offsets $1,\dots,L-1$ are its interior pairs and $0,L$ its boundary pairs. A \emph{window} is $(a,b)$ with
$0\le a$, $a+2\le b\le L$, and $\ell(a,b)=b-a-1$ is its number of interior pairs. For a sub-probability distribution $D_L$
on windows let
\[
\varepsilon_e=\sum_{(a,b):\,e\in\{a,b\}}D_L(a,b)\,\ell(a,b),\qquad \pi_e=\Pr_{D_L}[a<e<b]:
\]
the expected length lost if offset $e$ invalidates the windows that start or end there, and the probability that $e$ lies
strictly inside the window. $D_L$ is \emph{$(\alpha,\beta)$-good} (Definition~\ref{def:good}) if
$\mathbb E_{D_L}[\ell]\ge\alpha(L-1)$, $\pi_e+\varepsilon_e\le\beta$ for $1\le e\le L-1$, and $\varepsilon_0,\varepsilon_L\le\beta/2$.
The weight $\frac12$ at the ends reflects that a boundary pair borders at most two optimal occurrences.

\paragraph{The recursion.}
If for each line $(\alpha_i,\beta_i)$, $\beta_i\ge1$, of a finite family and each token length there is an
$(\alpha_i,\beta_i)$-good distribution, then every step satisfies $k\,g_{t+1}\ge\alpha_i\OPT_k-\beta_iG_t$ for every $i$
(Theorem~\ref{thm:frame}, Lemma~\ref{lem:perP}). With $x_t=G_t/\OPT_k$ this reads $x_{t+1}\ge\psi_k(x_t)$ for
$\psi_k(x)=x+\frac1k\max\big(0,\max_{i:\beta_i\le k}(\alpha_i-\beta_ix)\big)$, so $\GTv_k\ge\psi_k^{\circ k}(0)\OPT_k$; and if
$k\ge\max_i\beta_i$ then $\psi_k^{\circ k}(0)\ge y(1)$, where $y'=\max_i(\alpha_i-\beta_iy)$, $y(0)=0$
(Theorem~\ref{thm:env}). The argument never uses that $S^*$ is optimal, so $\OPT_k$ may be replaced by $f(S^*)$ for any
set $S^*$ of at most $k$ strings, with good distributions needed only for the lengths in $S^*$ (Lemma~\ref{lem:compare}).

\begin{theorem}[Short tokens]\label{thm:short}
Let $\mathcal A=\{(\frac12,1),(\frac23,2),(\frac56,3),(1,4)\}$, and for $k\ge2$ let $\gamma^{\rm s}_k=\psi_k^{\circ k}(0)$ with
$\psi_k(x)=x+\frac1k\max\big(0,\max_{(\alpha,\beta)\in\mathcal A,\,\beta\le k}(\alpha-\beta x)\big)$; put $\gamma^{\rm s}_1=1$.  For every
corpus, every $k\ge1$, every set $S^*$ of at most $k$ strings of length $2$ or $3$, every tie-breaking rule, and both the
exact and the lazy loop,
\[
\GTv_k\ \ge\ \gamma^{\rm s}_k\,f(S^*).
\]
Here $\gamma^{\rm s}_2=\SrhoTwo$, $\gamma^{\rm s}_3=\SrhoThree$, $\gamma^{\rm s}_4=\SrhoFour$, and $\gamma^{\rm s}_k\ge\JLIM\approx\SyOneSixteen$
for every $k\ge4$.  In particular, if some optimal vocabulary consists of tokens of length at most $3$, then
$\GTv_k\ge(\JLIM)\OPT_k$ for every $k$.  This constant is best possible along the budgets $k=n^2$: the J4 corpora of Theorem~\ref{thm:j4} have
an optimal vocabulary of length-$3$ tokens (Theorem~\ref{thm:opt}) and ratio $\GTv_K/\OPT_K\to\JLIM$.
\end{theorem}

\begin{proof}
\emph{Window distributions.}  For $m\in[0,1]$ let $D_2=\frac{1+m}2\cdot(0,2)$ and
$D_3=m\cdot(0,3)+\frac{1-m}2\cdot(0,2)+\frac{1-m}2\cdot(1,3)$.  Both are $(\frac{1+m}2,1+3m)$-good, hence exactly good
(Lemma~\ref{lem:perP}).  For $L=2$: the only window
has $\ell=1$, so $\mathbb E[\ell]=\frac{1+m}2$; offset $1$ has $\pi_1=\frac{1+m}2\le1$ and $\varepsilon_1=0$; and
$\varepsilon_0=\varepsilon_2=\frac{1+m}2\le\frac{1+3m}2$.  For $L=3$: $\mathbb E[\ell]=2m+(1-m)=2\cdot\frac{1+m}2$; offset $1$ lies strictly
inside $(0,3)$ and $(0,2)$ and is the left end of $(1,3)$, so $\pi_1+\varepsilon_1=m+\frac{1-m}2+\frac{1-m}2=1$, and offset $2$ is
symmetric; $\varepsilon_0=2m+\frac{1-m}2=\frac{1+3m}2$, and $\varepsilon_3$ is symmetric.  The values $m=0,\frac13,\frac23,1$ give $\mathcal A$.

\emph{Recursion.}  By Lemma~\ref{lem:compare} with these distributions, Theorem~\ref{thm:env} gives
$\GTv_k\ge \gamma^{\rm s}_kf(S^*)$ for $k\ge2$; $k=1$ is Proposition~\ref{prop:easy}.  The values of $\gamma^{\rm s}_2,\gamma^{\rm s}_3,\gamma^{\rm s}_4$
are exact iterations; for instance, for $k=2$ only $(\frac12,1)$ and $(\frac23,2)$ are available, and $x_1=\frac13$,
$x_2=\frac13+\frac12(\frac12-\frac13)=\frac5{12}$.
(For $k=2$ the argument of Proposition~\ref{prop:easy}, which applies to any $S^*$, gives the better value $\frac12$.)

\emph{The limit.}  For $k\ge4=\max\beta$, Theorem~\ref{thm:env} gives $\gamma^{\rm s}_k\ge y(1)$ with $y'=\varphi(y)$, $y(0)=0$,
$\varphi(x)=\max_{(\alpha,\beta)\in\mathcal A}(\alpha-\beta x)$.  All four lines pass through $(\frac16,\frac13)$, since
$\frac{1+m}2-\frac{1+3m}6=\frac13$; so $\varphi(x)=\max(1-4x,\frac12-x)$, with the kink at $x=\frac16$.  While $y\le\frac16$,
$y=\frac14(1-e^{-4t})$, which reaches $\frac16$ at $t_1=\frac{\ln3}4<1$; afterwards $\frac12-y(t)=\frac13e^{-(t-t_1)}$.  Hence
$y(1)=\frac12-\frac13e^{t_1-1}=\frac12-3^{1/4}/(3e)$.  Since $\gamma^{\rm s}_2,\gamma^{\rm s}_3>y(1)$, the bound $\GTv_k\ge y(1)f(S^*)$ holds for
every $k$.  Tightness is Theorems~\ref{thm:j4} and~\ref{thm:opt}.
\end{proof}

\noindent In the limit the J4 run has normalized gain $e^{-4t}$ until $t_1$ and $\frac13e^{-(t-t_1)}$ afterwards, exactly the
extremal solution of this recursion (Figure~\ref{fig:traj}); so for short tokens the window argument loses nothing.

%% file: sections/uniform.tex
\section{\GT{} on every corpus}\label{sec:uniform}
For long tokens the per-offset accounting of Section~\ref{sec:warmup} is too lossy, and exact goodness against all
$2^{L+1}$ patterns of covered offsets is false: at $L=20$ two chosen windows whose covered runs kill each other push the
best line $\beta=1$ below $\frac12$ (Proposition~\ref{prop:cxL20}). The window distribution may, however, depend on the
strings greedy has chosen \emph{and on their order}. For a token $y$ of length $L$, its \emph{ordered window set} $\mathcal O$ lists, in the order of greedy's
choices, the sets of windows of $y$ that spell each chosen string. Given $\mathcal O$, the covered offsets of an occurrence lie in an
explicit family $\mathcal F(L;\mathcal O)$ of $O(L^2)$ patterns indexed by the first and last uncovered offset (the history
lemma, Lemma~\ref{lem:2P}), and $\alpha^{\rm H}(L;\mathcal O;\beta)$ denotes the best fraction of pairs a distribution can keep
against this family at price $\beta$ (Appendix~\ref{sec:main}). Everything rests on one inequality per integer $\beta$.

\begin{theorem}[Structural theorem]\label{thm:H}
For every integer $\beta\in\{1,\dots,6\}$, every $L\ge2$ and every ordered window set $\mathcal O$ of a length-$L$ token,
\[
\alpha^{\rm H}(L;\mathcal O;\beta)\ \ge\ \frac{\beta}{\beta+1}.
\]
\end{theorem}

\noindent The proof peels off the first chosen window (Figure~\ref{fig:idea}c): the case without chosen windows is a hand-checkable
base case (Theorem~\ref{thm:E0}), $\beta\ge2$ is a mixing induction (Theorem~\ref{thm:H2}) and $\beta=1$ an additive one
(Theorem~\ref{thm:PH}). For $\beta=1$ the constant $\frac12$ cannot be raised; the proof works for every integer $\beta$, but lines beyond $\beta=6$ add only about $10^{-4}$. Solving the recursion with the six lines
$(\frac\beta{\beta+1},\beta)$ in closed form gives the main result (proof in Appendix~\ref{sec:main}).

\begin{theorem}[Main lower bound]\label{thm:main}
For every corpus, every $k\ge1$, any tie-breaking and both the exact and the lazy loop, $\GTv_k\ge\gamma_k\OPT_k$, where
$\gamma_1=1$,
\[
\gamma_2=\tfrac12,\quad\gamma_3=\tfrac7{18},\quad\gamma_4=\tfrac{239}{640},\quad\gamma_5=\tfrac{1363}{3750},
\]
and, for every $k\ge6$,
\[
\gamma_k\ \ge\ y(1)\ =\ \frac12-\frac1{3e}\Big(\frac95\Big)^{1/2}\Big(\frac{10}9\Big)^{1/4}\Big(\frac{15}{14}\Big)^{1/5}
\ =\ \MAINyOneFull\ldots
\]
In particular $\GTv_k\ge\MAINconst\cdot\OPT_k$ for every $k$.
\end{theorem}
\noindent The values $\gamma_1=1$ and $\gamma_2=\frac12$ come from the elementary bound $\GTv_k\ge\OPT_k/k$
(Proposition~\ref{prop:easy}).

\paragraph{Bad instances.}
Junction corpora (Appendix~\ref{sec:upper}) use $K=n^2$ planted tokens $a_ux_jz_v$ whose cut strings $z_va_u$ appear at
the junctions between consecutive tokens of a word; one committed cut covers one pair and invalidates two planted
occurrences. The four-token family J4 has a \GT{} run (adversarial ties) that commits only cut strings, its planted tokens
are an optimal vocabulary, and the run has $\GTv_K/\OPT_K=\frac12-\frac14(1-\frac1n)^{n-R_1}\to\JLIM$ (Theorems~\ref{thm:j4}
and~\ref{thm:opt}); disjoint copies
and one tiny word per pick make the bound hold for all large $k$ and with unique choices (Theorem~\ref{thm:unique}).
So $\limsup_k\rho_k\le\JLIM\approx\JFourLimSeven$, where $\rho_k$ is the worst-case ratio, and the gap to
Theorem~\ref{thm:main} is under $3\%$. Small budgets: $\rho_2\in[\frac12,\KtwoTie]$, $\rho_3\in[\frac7{18},\frac12]$ and
$\rho_4\in[\frac{239}{640},\frac7{16}]$ (Proposition~\ref{prop:smallk}).

\paragraph{Alphabets and hardness.}
The junction corpora can be recoded into $2\sqrt k$ letters (Proposition~\ref{prop:recode}). Over $256$ symbols \GT{} provably exceeds
$0.33861$ for every $k\ge\ByteThresh$, and over $k^c$ symbols with $\frac13<c<\frac12$ the asymptotic worst case lies in
$[\frac12-\frac1{6e},\frac12]$ (Corollary~\ref{cor:byte}, Theorem~\ref{thm:half}, Proposition~\ref{prop:U};
Appendix~\ref{app:smallalpha}). Over unbounded alphabets the objective itself is APX-hard: no polynomial-time approximation scheme exists unless
P$=$NP (Theorem~\ref{thm:hard}, Appendix~\ref{app:hard}). Over a fixed alphabet it is not APX-hard unless P$=$NP, by the
EPTAS of Ion et al.\ \cite[Theorem~5.6]{ion2026}.

%% file: sections/fixed.tex
\section{Fixed alphabets: one analysis for both greedy algorithms}\label{sec:fixed}
Junction corpora need alphabets that grow with the budget, while byte-level tokenizers have about $256$ symbols and tens of
thousands of tokens or more. On a fixed alphabet both \GT{} and greedy BPE are asymptotically optimal, and the worst deficit
has order exactly $1/\log k$. Write $I_M(t)$ for the count-weighted number of $M$-short pairs at time $t$
(Section~\ref{sec:model}). The upper bound for both algorithms rests on one lemma.

\begin{lemma}[Domination]\label{lem:dom}
For every $M\ge2$ and every $t$: $P(\mathcal C)\le\frac{M+1}{M-1}\,G_t+I_M(t)$.
\end{lemma}
\begin{proof}
Work in one word.  Distribute each pair between units $U,V$ that is not $M$-short to its two units in proportion to their
covered pairs: $U$ receives $\frac{|U|-1}{|U|+|V|-2}\le\frac{|U|-1}{M-1}$.  A unit has at most two neighbouring pairs, so the number
of such pairs is at most $\frac2{M-1}$ times the number of covered pairs.  Every pair is covered, $M$-short, or neither; sum
with the counts and use $G_t=\sum_wc_w\cdot\#\text{covered pairs}$ (Lemma~\ref{lem:basic}(iii)).
\end{proof}
\noindent The proof uses only that the units partition every word and that $G_t$ counts the pairs inside units, so it
applies verbatim to the tokens of a BPE run after $t$ rules, with $G_t$ its saving.

\begin{theorem}[Fixed alphabets]\label{thm:fixed}\label{thm:P}\label{th:main}
(a) \emph{\GT{}.} For every corpus, every $M\ge2$, every $k\ge1$, any tie-breaking and both loops, with $a=\frac{M+1}{M-1}$:
\[
2N_M\,g_{t+1}\ \ge\ P(\mathcal C)-a\,G_t\quad(t<k),\qquad\text{hence}\qquad
\GTv_k\ \ge\ \frac{M-1}{M+1}\Big(1-\Big(1-\frac a{2N_M}\Big)^k\Big)\OPT_k
\]
whenever $2N_M\ge a$.

\noindent(b) \emph{Greedy BPE.} For integers $M\ge2$ and $k\ge1$, both greedy rules satisfy
\[
 P(\mathcal C)-G_k\ \le\ G_k\left(\frac{2}{M-1}+\frac{2N_M}{k}\right).
\]
If $P(\mathcal C)>0$ and $\sigma=|\Sigma|\ge2$ is fixed, then uniformly over inputs and greedy choices,
\[
 \frac{G_k}{\OPT^{\mathrm{part}}_k}\ \ge\ \frac{G_k}{P(\mathcal C)}\ \ge\ 1-\frac{2+o(1)}{\log_\sigma k}.
\]
\end{theorem}
\noindent In (a) the choice $M=\lfloor\log_\sigma(k/(4\ln k))\rfloor\ge2$ gives $N_M<k/(2\ln k)$ and
$\GTv_k/\OPT_k\ge1-2/\log_\sigma(k/(4\ln k))-1/k$. The two parts bound the $M$-short pairs differently. For \GT{},
each $M$-short pair spells a candidate whose score collects at least half of its short pairs (Lemma~\ref{lem:short}),
so $I_M(t)\le2N_Mg_{t+1}$; the recursion follows from Lemma~\ref{lem:dom}. For BPE, the current types of two adjacent
tokens are determined by their combined original spelling, so the short adjacencies at the end are at most $N_MT$, where
$T$ is the final maximum raw pair frequency; that frequency never increases and every step realizes at least half of it,
so $kT\le2G_k$. The proofs are in Appendix~\ref{app:fixed}. The BPE bound also gives convergence when
$\log\sigma=o(\log k)$, and over $k^c$ letters part (a) gives the ratio $\frac{M-1}{M+1}$ in the limit whenever $c<\frac1M$.

\paragraph{The matching lower bound.}
Encode logical bits by $E(0)=ab$ and $E(1)=abb$, extended by concatenation. For $m\ge1$ let $\lambda$ be the smallest even integer $\ge2$ with $2^{\lambda-2}\ge m$, choose $m$ distinct $\lambda$-bit strings $x_i$ starting and ending in $0$ and $m$ distinct
$y_j$ starting and ending in $1$, and set
\[
 L_i=a E(x_i)a,\qquad R_j=aa E(y_j)a .
\]
The corpus $\mathcal B_m$ contains $m$ copies of each standalone word $L_i$ and $R_j$, one copy of each of the $m^2$ grid
words $L_iR_j$, and $B$ copies of the guard word $aa$, where $B$ exceeds the total length of the other words, counted with
multiplicity, by one. Its total length is $O(m^2\log m)$.

\begin{figure}[t]
\centering
\begin{tikzpicture}[x=0.55cm,y=0.5cm,font=\small]
  \foreach \c [count=\i from 0] in {a,b,a,a,a,a,b,b} \node[minimum size=0.4cm,inner sep=0] at (\i,0) {\texttt{\c}};
  \node at (-1.2,0) {$\cdots$}; \node at (8.2,0) {$\cdots$};
  \draw[thick,fill=gray!30,rounded corners=2pt,fill opacity=0.5] (1.6,-0.45) rectangle (3.4,0.45);
  \draw[thick,fill=gray!15,rounded corners=2pt,fill opacity=0.5] (3.6,-0.45) rectangle (5.4,0.45);
  \draw[very thick,dashed] (2.5,-0.9) -- (2.5,0.9);
  \draw[decorate,decoration={brace,amplitude=3pt}] (2.6,1.0) -- (8.6,1.0) node[midway,above=3pt,font=\scriptsize] {$R_j=aa\,E(y_j)\,a$ with $E(1)=abb$};
  \draw[decorate,decoration={brace,amplitude=3pt,mirror}] (-1.6,-1.0) -- (2.4,-1.0) node[midway,below=3pt,font=\scriptsize] {$L_i=a\,E(x_i)\,a$ with $E(0)=ab$};
  \node[font=\scriptsize,anchor=west,align=left] at (9.3,0.3) {cut $L_i\mid R_j$ (dashed) lies inside the first committed $aa$:\\no later unit can end there};
\end{tikzpicture}
\caption{The lost boundary in a grid word $L_iR_j$ of $\mathcal B_m$. The cut lies after the first of four $a$'s, which
follow a $b$; the left-to-right commitment of $aa$ (\GT{}) or the full rule $aa$ (BPE) erases it.}
\label{fig:lost}
\end{figure}
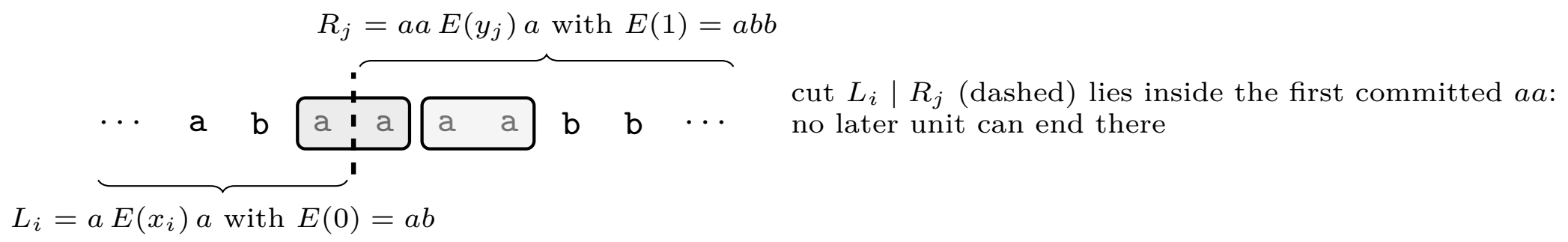

\begin{theorem}[One binary corpus for both algorithms]\label{thm:binary}\label{thm:fixed-binary}\label{th:binary}
Fix $0<\epsilon<1$.

\noindent(a) \emph{\GT{}.} For all sufficiently large $k$, a binary corpus has the following property. Once \GT{} makes its first selection and commits all its leftmost disjoint occurrences, every segmentation using at most $\lfloor(2-\epsilon)k\rfloor$ optional vocabulary entries in total and respecting that first commitment (no piece boundary at a pair it covers) saves at most
\[
 \left(1-\frac{c_\epsilon}{\log k}\right)\OPT_k
\]
for some $c_\epsilon>0$. In particular, the worst same-budget deficit of the committed \GT{} segmentation has order $1/\log k$ over every fixed alphabet of size at least two.

\noindent(b) \emph{Greedy BPE.} For every sufficiently large $k$ there is a binary corpus on which every sequence beginning with the first full merge of either greedy rule and using at most $K=\lfloor(2-\epsilon)k\rfloor$ rules in total saves at most $(1-c_\epsilon/\log k)\OPT^{\mathrm{full}}_k$, where $\OPT^{\mathrm{full}}_k$ uses budget $k$ and $c_\epsilon>0$. The sequence may use arbitrary partial replacements after the first rule. In particular, the same-budget greedy savings deficit on corpora has order $1/\log k$ on every fixed alphabet of size at least two.

\noindent In both parts the corpus is $\mathcal B_m$, with $m=\lfloor(k-1)/2\rfloor$ in (a) and $m=\lfloor k/2-30\sqrt k\rfloor$ in (b).
\end{theorem}

\begin{proof}[Proof idea]
The guard words make \texttt{aa} the unique first choice of \GT{} and of both BPE rules. In a grid word the cut
$L_i\mid R_j$ lies after the first of four consecutive $a$'s preceded by $b$ (Figure~\ref{fig:lost}), so the left-to-right commitment of
\texttt{aa} straddles it, and no later unit of that word has the cut as an endpoint. Counting the heads $ab$ of the
encoded bits shows that a token spelling a standalone word could occur in a grid word only at its nominal position,
which ends or starts at the erased cut; so no standalone spelling is usable in any grid word. The comparator uses one
token per standalone word and two per grid word: a direct vocabulary $\{L_i,R_j,aa\}$ for \GT{}, and $2m+O(\sqrt m)$ full
rules that build the codewords from halves for BPE. A continuation that misses $d\ge\epsilon m/2$ standalone spellings
pays $dm$ extra tokens; otherwise it has at most $(2-\epsilon)m$ types left for the grid, and since the string $aaba$ marks the
start of every grid word, these spell at most $v^2/4+O(v)$ two-token grid words out of $m^2$. Either way the excess is
$\Omega_\epsilon(m^2)$ tokens against savings $O(m^2\log k)$. The full proof, common to both parts, is in
Appendix~\ref{app:binary}.
\end{proof}

\paragraph{One string, and a token-length cap.}
On a single word the vocabulary can contain the whole word, so the direct model has no single-string analogue of
Theorem~\ref{thm:binary}. For merges it has one, and the first merge can cost far more than savings suggest.

\begin{theorem}[Forced first merge on one binary string]\label{thm:string}\label{th:string}
For every sufficiently large $k$, the following deterministic binary strings can be constructed in $O(k\log k)$ time on a word RAM.
\begin{enumerate}
\item For each fixed $0\le\gamma<1/3$, there is a string of length $\Theta(k\log k)$ for which any sequence beginning with the first full merge of either greedy rule and using at most $\lfloor(1+\gamma)k\rfloor$ rules in total saves at most $(1-c_\gamma/\log k)\OPT^{\mathrm{full}}_k$, where $\OPT^{\mathrm{full}}_k$ uses budget $k$ and $c_\gamma>0$.
\item There is a string of length $|s|=\Theta(k\log k)$ that $k$ full rules reduce to one token. Every sequence of at most $k$ rules beginning with the first full merge of either greedy rule leaves at least $(1/57-o(1))k$ tokens.
\end{enumerate}
Both statements allow arbitrary partial rules after the first step.
\end{theorem}

\begin{theorem}[Repair on one string]\label{thm:repair}\label{th:repair}
Let a sequence of $k$ full or partial rules reduce a nonempty binary string to $F$ tokens. After any prescribed first full pair rule, a continuation using partial rules can reach at most $F$ tokens with at most $2k+3$ rules in total, including the prescribed rule.
\end{theorem}
\noindent The cells of Theorem~\ref{thm:string} follow an Euler circuit in which every arc has a unique cut signature; an
obstruction shows that every assignment of token endpoints fails at at least a third of the cuts, which yields a budget-free inequality between the number of rules and the output length (Appendix~\ref{app:string}).
Theorem~\ref{thm:repair} simulates the prescribed merge by a two-state transducer along the comparator's rules. So after the
first greedy merge on one binary string, the asymptotic number of rules needed to recover the best $k$-rule final length
lies between $\frac43k$ and $2k$, while on corpora it is at least $2k$ (Theorem~\ref{thm:binary}). Greedy also forces the shortest
scale (proof of Corollary~\ref{cor:shortest}): for each fixed $\delta>0$ and large $k$, $k$ greedy rules reduce every binary string of length at most
$(\frac12-\delta)k\log_2k$ to one token.

A token-length cap gives an exact answer for \GT{}. For $M\ge2$, run \GT{} with all occurring substrings of lengths $2,\dots,M$ as
candidates, let $G_\infty$ be its committed savings once every remaining gain is zero (it may use up to $N_M$ types), and
let $\OPT^{\le M}_k$ be the optimal direct savings with $k$ optional types of lengths $2,\dots,M$.

\begin{theorem}[Length-capped committed savings]\label{thm:capped-exhaustion}
For either greedy loop and every tie-breaking rule,
\[
 G_\infty\ \ge\ \left(1-\frac1M\right)\OPT^{\le M}_k\qquad(k\ge1).
\]
Exhaustion occurs within $N_M$ positive selections. For every fixed $k\ge2$, the infimum of $G_\infty/\OPT^{\le M}_k$ over binary corpora with $\OPT^{\le M}_k>0$ is exactly $1-1/M$, even if the first choice must be unique. For $k=1$ the infimum is one. At a common finite budget $k\ge1$, if $N_M\ge1$, the same capped run satisfies
\[
 G_k\ \ge\ \left(1-\frac1M\right)
 \left[1-\left(1-\frac{1}{2N_M}\right)^k\right]\OPT^{\le M}_k .
\]
\end{theorem}
\noindent The sharp example is the word $ab^Ma$: greedy may commit $b^M$ first and then stops at savings $M-1$, while
$ab\mid b^{M-1}a$ saves $M$; an extra word $b^M$ makes that first choice unique.

\paragraph{An approximation scheme for merge savings.}
Theorem~\ref{thm:fixed}(b) makes greedy a $(1-\epsilon)$-approximation once $k$ exceeds a threshold depending only on
$\sigma$ and $\epsilon$, and below it rule lists can be enumerated; so optimal full and partial merge savings admit a
deterministic EPTAS on every fixed alphabet (Corollary~\ref{cor:eptas}, Appendix~\ref{app:string}).

%% file: sections/discussion.tex
\section{Related work, provenance and limitations}\label{sec:discussion}
\paragraph{Concurrent complexity results.}\label{sec:related}
Ion, Mateiu, Miclaus and Popa \cite{ion2026} study the optimization problems, not the deployed greedy algorithms. For
direct savings, our $f$, they give a deterministic EPTAS on every fixed alphabet \cite[Theorem~5.6]{ion2026} and analyse
greedy on the occurrence-coverage relaxation, $(1-1/e)/2$, and greedy by true marginal savings, $(1-e^{-3})/3$
\cite[Theorems~5.2, 5.4]{ion2026}; \GT{}'s occurrence scoring is neither. Their EPTAS selects arbitrary strings, whereas
Corollary~\ref{cor:eptas} concerns merge sequences, in which every type is one binary merge of available types; neither theorem implies the other, though for partial rules their argument \cite[Section~5.6]{ion2026} adapts with prefix merges; for full rules we use Theorem~\ref{thm:fixed}(b). Their observation that all types of lengths $2,\dots,b$ already save a $1-\epsilon$ fraction when
the budget allows them \cite[Section~5.6]{ion2026} is a statement about one vocabulary; Theorem~\ref{thm:capped-exhaustion}
is the exact guarantee for \GT{}'s own run under a cap. For the direct token count they rule out $n^{1-\epsilon}$-approximations on binary inputs of total length $n$ and prove
unary hardness for merge models. Bastan, He, Pavan and Vinodchandran \cite{bastan2026} rule out
$n^{1-\epsilon}$-approximations for binary direct tokenization with the token-count objective and give bicriteria
algorithms that exceed the budget and the optimal token count by a factor $\sqrt{M'}\log n$ ($M'$ the maximum document length). \ourname{} \cite{complexity} prove that the final token count
after $k$ full or partial merges on one binary string cannot be approximated within $cn/\log n$ unless P$=$NP
\cite[Theorem~10]{complexity}, and that two unary merge problems are strongly NP-complete. Their hardness results are conditional statements about polynomial-time algorithms for the
token count. Our results instead bound the savings of the greedy algorithms themselves, unconditionally. In that sense
Theorems~\ref{thm:binary} and~\ref{thm:string} are bicriteria lower bounds for greedy's first decision: neither almost twice
the budget on corpora, nor a third more on one string, recovers the savings, and Theorem~\ref{thm:string}(2) gives a
final-length factor of order $|s|/\log|s|$ after one forced merge.

\paragraph{Connections.}
Kozma and Voderholzer's bad BPE instances rest on the mechanism of Figure~\ref{fig:xaby}: ``we destroy two units of
utility and gain one'' \cite[App.~B]{kozma2024arxiv}. For \GT{} the same trade, organized into junction corpora, gives
$\JFourLimSeven$, and the binary corpus of Theorem~\ref{thm:binary} uses it once per grid word for both algorithms. The guarantees say the converse: a covered pair can hurt only the two optimal occurrences it borders and the windows
that start or end at it. Greedy guarantees beyond submodularity use the submodularity ratio \cite{dask2011,bian2017}, which
is $0$ here; our window distributions play the role of a factor-revealing LP \cite{jain2003}, solved in closed form for
every token length. Choosing a static dictionary and parsing optimally is a classical compression problem
\cite{storer1982}. For Re-Pair, Hucke, Je\.z and Lohrey give binary examples with grammar size a factor $\Omega(\log N/\log\log N)$
above optimal \cite{hjl}, and the pair-spelling and frequency facts behind Theorem~\ref{thm:fixed}(b) go back to Navarro and
Russo \cite{nr} and Ga\'nczorz \cite{ganczorz}; we use them for a budget-dependent guarantee, uniform in input length.

\paragraph{Verification provenance.}
Theorems~\ref{thm:main}, \ref{thm:short}, \ref{thm:fixed}, \ref{thm:binary}, \ref{thm:string}, \ref{thm:repair},
\ref{thm:capped-exhaustion} and~\ref{thm:hard}, the junction theorems and everything they use are written proofs; the main
constant is a closed form, evaluated in interval arithmetic. Computed are Proposition~\ref{prop:cxL20} (an explicit rational
certificate), the byte table and threshold of Corollary~\ref{cor:byte}, and the exact replays of explicit instances; each
was carried out by two separately written programs. Further computational tests of written proofs are not part of the
proofs. Appendix~\ref{app:comp} lists every computation.

\paragraph{Limitations.}\label{sec:limits}
The worst case of \GT{} lies in $[\MAINconst,\JFourLimSeven]$; the ends meet for optimal tokens of at most three characters,
and we know no construction with longer optimal tokens below $\JFourLimSeven$. The examples near $\JFourLimSeven$ use about
$2\sqrt k$ letters. Over bytes, our bounds improve $\MAINconst$ by more than $10^{-4}$ only from $k\approx87{,}500$ on; for $16384<k<\ByteThresh$
we know no family below $0.3386$, and the byte recursion reaches it only near the top of this range (not at $100{,}256$); common sizes such as $32{,}768$ and $50{,}000$ fall in this
range. On fixed alphabets the order $1/\log k$ is exact, but its leading constant is open for both algorithms; the
fixed-alphabet bounds are informative only at very large budgets (the ratio $\frac35$ of $M=4$ needs $k\gg256^4$ for bytes).
The guarantees concern savings, $\sum_wc_w(|w|-\#\text{tokens})$, on the training corpus; a guarantee on tokens saved is not
one on the token count. For \GT{} they hold for the committed segmentation and for optimal re-encoding with the chosen
tokens; the released encoder can segment differently and is not covered, and for the released lazy solver we give an
argument and a differential test, not a proof (Appendix~\ref{app:impl}). Theorem~\ref{thm:binary}(a) applies to
segmentations that preserve the first commitment; fresh optimal resegmentation with the first entry and the $k$ comparator
entries recovers the optimum. After one greedy BPE merge on one string, the rule factor needed for recovery is only known to
lie in $[\frac43,2]$. All bad instances are synthetic, and we make no claim about natural text.

\ifdefined\NAMEDBUILD\paragraph{Use of AI tools.}\else\paragraph{Use of AI tools.}\fi
AI-assisted tools, including Claude (Anthropic) and GPT-6.2 Astra (OpenAI), were used throughout the research process to
support mathematical reasoning, computational experimentation, proof verification, and manuscript preparation. The authors
guided the research direction, directed and evaluated the central arguments, and oversaw the computational and
verification workflows. The authors take full responsibility for the results and conclusions presented.

%% file: app/notation.tex
\section{Notation}\label{app:notation}
Table~\ref{tab:notation} lists the symbols used in more than one place.  Symbols local to one proof, and all symbols of
Appendix~\ref{app:string}, are defined where they are used.
\begin{table}[ht]\centering\footnotesize
\renewcommand{\arraystretch}{1.0}
\begin{tabular}{p{0.2\textwidth}p{0.72\textwidth}}\toprule
Symbol & Meaning\\\midrule
$w$, $c_w$ & a word of the corpus and its count\\
pair $i$ & the adjacent characters $i-1,i$ of a word; a token of length $L$ covers its $L-1$ interior pairs\\
$[s,e)$ & an occurrence of a string in a word; interior pairs $s+1,\dots,e-1$, boundary pairs $s$ and $e$\\
$T$ & candidate set (all substrings of length $\ge2$, or a capped subset, Appendix~\ref{app:impl})\\
$f(S)$ & pairs covered (tokens saved) when every word is segmented optimally with $S$\\
$k$, $\OPT_k$ & budget; optimum of $f$ over $|S|\le k$\\
$x_t$, $S_t$, $g_t$, $G_t$ & greedy's $t$-th choice, chosen set, gain of the $t$-th choice, committed value $g_1+\dots+g_t$\\
$\GTv_k$ & $=G_k$, the value of \GT{}'s committed segmentation; $f(S_k)\ge G_k$\\
$\Sc_t(x)$ & score of string $x$ at time $t$ (leftmost scan of valid occurrences)\\
$\rho_k$ & worst-case ratio $\inf\GTv_k/\OPT_k$ over corpora and tie-breaks\\
$S^*$, $\mathcal L_{S^*}$ & an optimal set, or any comparison set of at most $k$ strings (Lemma~\ref{lem:compare}); its token lengths\\
$P$, $L_P$, $c_P$ & an occurrence of a token of $S^*$ in the fixed optimal segmentation, its length, its word's count\\
offset $e$ & $e\in\{0,\dots,L\}$, the pair between characters $e-1$ and $e$ of $P$; $0,L$ are boundary pairs\\
$\Pi$, $\Pi_P$ & a pattern (set of covered offsets); the pattern of $P$ at time $t$\\
$(a,b)$, $\ell(a,b)$ & window of a token (offsets $a<b$, $b\ge a+2$) and its gain $b-a-1$\\
$X(\Pi,(a,b))$ & uncovered interior offsets of the window if both ends are uncovered, else $0$\\
$\mathrm{wt}(e)$, $\mathrm{cost}(\Pi)$ & $\frac12$ at $e\in\{0,L\}$, $1$ otherwise; $\mathrm{cost}(\Pi)=\sum_{e\in\Pi}\mathrm{wt}(e)$\\
$D$, $D_L$ & sub-probability distribution on windows\\
$(\alpha,\beta)$ & a line: fraction of pairs kept, cost per covered pair (Definitions~\ref{def:exact}, \ref{def:good})\\
$\pi_e$, $\varepsilon_e$ & per-offset quantities of Definition~\ref{def:good}\\
$\varphi$, $y(t)$, $\psi_k$, $\gamma_k$ & envelope $\max_i(\alpha_i-\beta_ix)$, solution of $y'=\varphi(y)$, one-step map and its $k$-fold iterate (Theorem~\ref{thm:env})\\
$\mathcal W$, $\mathcal Q(\mathcal W)$ & chosen windows of a token; patterns compatible with them (Appendix~\ref{sec:pce})\\
$\mathcal O=(C_1,\dots,C_r)$ & ordered window set: chosen windows grouped by choice, in time order (Appendix~\ref{sec:history})\\
$M(p,q)$, $\mathcal F(L;\mathcal O)$ & middle process; history family of patterns (Appendix~\ref{sec:history})\\
$\Phi_D(\Pi)$, $\alpha^{\rm H}$ & $\mathbb E_DX(\Pi,\cdot)+\beta\,\mathrm{cost}(\Pi)$; best constant against $\mathcal F(L;\mathcal O)$\\
$V$, $v_w$, $\mathrm{Val}$, $\lambda^{\rm L},\lambda^{\rm R}$ & target $\frac\beta{\beta+1}(L-1)$; window value $D_w\ell(w)$; total value; left and right loads (Appendix~\ref{sec:E0})\\
$w^*=(a,b)$, $\ell$ & the peeled window and its gain (Appendix~\ref{sec:peel})\\\midrule
$n$, $g$, $K$ & Appendix~\ref{sec:upper}: $g=n$, budget $K=n^2$ of the junction corpora\\
$y_j$, $a_u,z_v,x_j$ & planted tokens $a_{u}x_jz_{v}$ and their letters\\
$R_1$, $m_r$, $\mu_r$ & number of phase-1 rounds; mass of the cut strings of phase-1 round $r$ and of phase-2 round $r$\\
$Y$ & the planted tokens $\{y_0,\dots,y_{K-1}\}$ of the junction corpora (optimal by Theorem~\ref{thm:opt})\\
$\mathcal M$, $w_0$ & a token's occurrence mass next to a still-uncut junction; $w_0=\frac12\big(1-(1-\frac1n)^{n-R_1}\big)$\\\midrule
$\mathcal C$, $|\mathcal C|$ & a corpus; its total length\\
$\sigma$, $c$, $P(\mathcal C)$ & alphabet size; $\log\sigma/\log k$; total number of pairs (the savings ceiling)\\
units, $M$-short & current pieces of a word (committed occurrences and free characters, or BPE tokens); a pair between units $U,V$ with $|U|+|V|\le M$\\
$N_M$, $I_M(t)$ & distinct substrings of length $2..M$; mass of $M$-short pairs\\
$M$ & a length parameter (Section~\ref{sec:fixed}); also the token-length cap of Theorem~\ref{thm:capped-exhaustion}\\
$\OPT^{\mathrm{full}}_k$, $\OPT^{\mathrm{part}}_k$ & best savings of $k$ full, resp.\ partial, merge rules\\
$\OPT^{\le M}_k$, $G_\infty$ & best direct savings with $k$ types of length $\le M$; \GT{}'s savings at exhaustion\\
$G_k$ (BPE) & savings of $k$ greedy BPE rules (raw or saving greedy)\\
$\mathcal B_m$, $L_i,R_j$ & the binary corpus of Theorem~\ref{thm:binary}; its standalone codewords\\
$F$, $F_0$ & number of output tokens (Theorem~\ref{thm:repair}, Appendix~\ref{app:string})\\
$\Lambda$, $\Omega$ & Appendix~\ref{app:hard} only: blow-up factor and forcing weight\\
\bottomrule\end{tabular}
\caption{Notation.}\label{tab:notation}
\end{table}

%% file: app/model.tex
\section{The \GT{} model: invariants and the shift lemma}\label{app:gtmodel}\label{sec:shiftsec}
The definitions are in Section~\ref{sec:model}.  An occurrence $[s,e)$ has its interior pairs
$s+1,\dots,e-1$, left boundary pair $s$ (if $s\ge 1$) and right boundary pair $e$ (if $e\le n-1$).  Since
$|w|=\mathrm{PARTITION}(w,S\cup\Sigma)+\mathrm{cover}(w,S)$, where PARTITION is the minimum number of pieces, maximizing $f$ is
Problem~1 of \cite{lim2025}.  The candidate set $T$ is the set of all substrings of length $\ge2$ of corpus words, which is
the default of Lim et al.\ and of the implementation.  Our lower bounds, except those that use Theorem~\ref{thm:fixed}(a), use only
that $T$ is closed under taking substrings of length $\ge2$, which also holds for a length cap or a minimum frequency
(Appendix~\ref{app:impl}).  Read literally, the pseudocode of Algorithm~1 of \cite{lim2025} has a typo in the commit step;
we read it as committing the accepted token, with positions visited left to right in each word, as in the
implementation (Appendix~\ref{app:impl}).  ``The lazy loop'' always means this reading.

The first lemma collects the basic invariants of the committed state; we use it throughout.

\begin{lemma}[Basic structure]\label{lem:basic}
At every time: (i) any two committed occurrences in a word are nested or disjoint; (ii) for a valid
occurrence $q$ of a string not yet chosen, every committed occurrence meeting $q$ lies inside $q$, and
\[\text{gain}(q)=\#\text{uncovered chars of }q+\#\text{maximal committed occurrences inside }q-1;\]
(iii) committing $q$ makes exactly its uncovered interior pairs covered; covered pairs stay covered
forever; and $G_t=\sum_w c_w\cdot\#\{\text{covered pairs of }w\}$.
\end{lemma}
\begin{proof}
If a committed occurrence $c$ meets $q$ but is not contained in it, then $c$ extends past an end of $q$ that is not
an end of the word, so $c$ contains the boundary pair of $q$ at that end as an interior pair, and that pair is covered, contradicting validity.  This gives the first claim of (ii) and, by induction, (i).  Inside $q$ the maximal committed occurrences and the uncovered characters partition
$q$ into consecutive pieces, which are separated by exactly the uncovered interior pairs; so the number of pieces
minus one equals the number of uncovered interior pairs.
(iii) is immediate from (ii) and the fact that commits only add occurrences.
\end{proof}

The case $k=1$ is easy, and it already gives a (weak) bound for every $k$.

\begin{proposition}[$k=1$; elementary bound]\label{prop:easy}
$\GTv_1=\OPT_1$, and $\GTv_k\ge \OPT_k/k$ for every $k$.
\end{proposition}
\begin{proof}
On the empty state every occurrence is valid with gain $|x|-1$.  The leftmost scan of equal-length intervals
places a maximum number of pairwise disjoint occurrences (earliest-finish interval scheduling), so $\Sc_0(x)=f(\{x\})$ and $g_1=\max_x f(\{x\})=\OPT_1$.  For general $k$, restricting an optimal segmentation to the
occurrences of one token shows $f(S)\le\sum_{x\in S} f(\{x\})$, hence $\OPT_k\le k\,g_1\le k\,G_k$.
\end{proof}

\paragraph{Running example.}
Take the single word $w=\texttt{abababab}$ with count $1$ and index its characters $0,\dots,7$; pair $i$, for
$1\le i\le7$, is the pair of characters $i-1$ and $i$.  The string \texttt{abab} occurs at positions $0$, $2$
and $4$, i.e.\ on the intervals $[0,4)$, $[2,6)$ and $[4,8)$.
On the empty state every occurrence is valid.  The scan of \texttt{abab} places $[0,4)$, skips $[2,6)$ because it
overlaps $[0,4)$, and places $[4,8)$, for a score of $3+3=6$.

\subsection{The shift lemma}
The rest of the analysis of \GT{} rests on one structural fact about \GT{}'s state: overlapping occurrences of a string that greedy
has not chosen cannot see different covered pairs.

\begin{figure}[t]
\centering
\begin{tikzpicture}[x=0.62cm,y=0.62cm,font=\small]
  \begin{scope}
    \foreach \s in {1,3,5} \draw[thick,fill=gray!25,rounded corners=3pt] (\s-0.42,-0.4) rectangle (\s+1.42,0.4);
    \foreach \c [count=\i from 0] in {a,b,a,b,a,b,a,b} \node[minimum size=0.5cm,inner sep=0] at (\i,0) {\texttt{\c}};
    \foreach \r in {1,...,7} \node[font=\scriptsize,gray] at (\r-0.5,-0.85) {\r};
    \foreach \r in {2,4,6} \fill (\r-0.5,-0.55) circle (2.2pt);
    \foreach \r in {1,3,5,7} \draw (\r-0.5,-0.55) circle (2.2pt);
    \draw[thick] (-0.45,0.75) -- (-0.45,0.95) -- (3.45,0.95) -- (3.45,0.75);
    \draw[thick] (3.55,0.75) -- (3.55,0.95) -- (7.45,0.95) -- (7.45,0.75);
    \node at (1.5,1.4) {\texttt{abab}: invalid};
    \node at (5.5,1.4) {\texttt{abab}: invalid};
    \draw[->,thick] (3.5,2.0) -- (3.5,1.0);
    \node at (3.5,2.3) {pair 4};
    \node at (3.5,-2.4) {(a) after committing \texttt{ba}};
  \end{scope}
  \begin{scope}[xshift=6.3cm]
    \foreach \s in {0,2,4,6} \draw[thick,fill=gray!25,rounded corners=3pt] (\s-0.42,-0.4) rectangle (\s+1.42,0.4);
    \foreach \c [count=\i from 0] in {a,b,a,b,a,b,a,b} \node[minimum size=0.5cm,inner sep=0] at (\i,0) {\texttt{\c}};
    \foreach \r in {1,...,7} \node[font=\scriptsize,gray] at (\r-0.5,-0.85) {\r};
    \foreach \r in {1,3,5,7} \fill (\r-0.5,-0.55) circle (2.2pt);
    \foreach \r in {2,4,6} \draw (\r-0.5,-0.55) circle (2.2pt);
    \draw[thick] (-0.45,0.75) -- (-0.45,0.95) -- (3.45,0.95) -- (3.45,0.75);
    \node at (1.5,1.35) {$p$: gain 1};
    \draw[thick] (1.55,1.75) -- (1.55,1.95) -- (5.45,1.95) -- (5.45,1.75);
    \node at (3.5,2.35) {$q=p+2$: gain 1};
    \foreach \r in {1,2,3} \draw[->,gray] (\r-0.5,-1.15) to[bend right=35] (\r+1.5,-1.15);
    \node at (3.5,-2.4) {(b) after committing \texttt{ab}};
  \end{scope}
\end{tikzpicture}
\caption{The running example $w=\texttt{abababab}$.  Shaded boxes are committed occurrences; a filled dot marks a
covered pair, an open dot an uncovered one.  (a) Committing \texttt{ba} covers pairs $2,4,6$.  Pair $4$ is the
shared boundary of the two disjoint occurrences of \texttt{abab}, so both become invalid; so does
$[2,6)$.  (b) After committing \texttt{ab}, the overlapping occurrences $p=[0,4)$ and $q=[2,6)$ of \texttt{abab} are both
valid.  Shifting by $\delta=2$ maps the interior pairs $1,2,3$ of $p$ to the interior pairs $3,4,5$ of $q$ (arrows),
and covered pairs map to covered pairs.  So $p$ and $q$ have the same gain (Lemma~\ref{lem:shift}).}
\label{fig:example}
\end{figure}
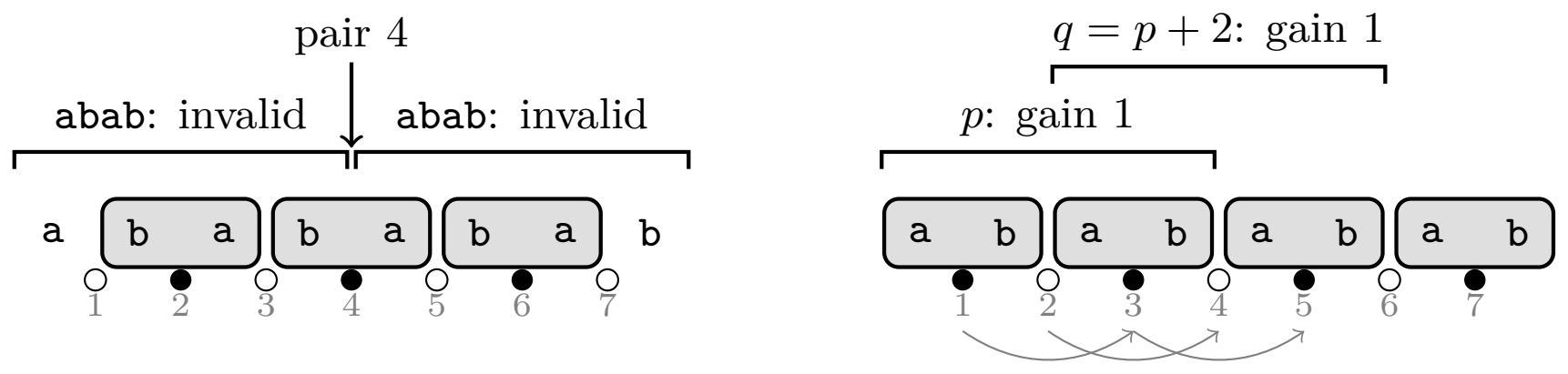

\paragraph{Why it is needed.}
Our proof will credit greedy with the total gain of a family of disjoint occurrences of some string $x$.  \GT{}'s
left-to-right scan, however, may skip one of them because an overlapping occurrence came first.  In the running
example, suppose \GT{} has committed \texttt{ab}, covering pairs $1,3,5,7$.  The occurrences of \texttt{abab} at $0$,
$2$ and $4$ are then all valid, each with gain $1$, and the scan places $[0,4)$ and $[4,8)$ but skips $[2,6)$.  Nothing
is lost, because the blocker $[0,4)$ has the same gain as the skipped $[2,6)$ (Figure~\ref{fig:example}b).  The shift
lemma says that this is always so.  Panel~(a) of the figure shows the other side of the coin: a single commit of
\texttt{ba} invalidates every occurrence of \texttt{abab}, so greedy may destroy more than it gains.

\begin{lemma}[Chosen strings]\label{lem:chosen}
If $x\in S_t$, every valid occurrence of $x$ at time $t$ has gain $0$.
\end{lemma}
\begin{proof}
Let $x$ be chosen at step $s+1\le t$ and $q$ an occurrence of $x$ valid at time $t$.  Since covered
pairs stay covered, $q$ was valid at time $s$.  Either the commit pass committed $q$, and then all its
interior pairs are covered from then on (gain $0$); or it skipped $q$ because $q$ overlapped the
previously committed occurrence $p$ of $x$, with $p_{\rm start}<q_{\rm start}<p_{\rm end}$.  In the second case the left
boundary pair of $q$ is an interior pair of $p$, covered from time $s+1$ on, and $q$ is not valid at
time $t$, a contradiction.
\end{proof}

The shift lemma is the heart of the proof.  It says that two overlapping valid occurrences of an unchosen string
see the same pattern of covered pairs (Figure~\ref{fig:example}b).  We need it to show that the leftmost scan
loses nothing when it skips an occurrence (Lemma~\ref{lem:int}).

\begin{lemma}[Shift]\label{lem:shift}
Let $x\notin S_t$, and let $p,q$ be occurrences of $x$ in the same word $w$, both valid at time $t$,
with $q_s=p_s+\delta$ for some $1\le\delta\le|x|-1$.  Then for every pair $r$ with $p_s<r<p_e$,
\[
r \text{ is covered at time } t\iff r+\delta\text{ is covered at time }t .
\]
In particular $\text{gain}_t(p)=\text{gain}_t(q)$.
\end{lemma}
\begin{proof}[Proof sketch]
The map $r\mapsto r+\delta$ is a bijection from the interior pairs of $p$ to those of $q$, which gives the
``in particular''.  Since $p$ and $q$ spell the same string, the text is $\delta$-periodic on $[p_s,q_e)$.  We
induct on $t$.  When the string $z$ is committed at step $t+1$, one shows three things: shifted copies of
valid occurrences of $z$ inside $p$ (or $q$) are valid; a placed occurrence that covers a pair inside $p$
lies inside $p$; and an occurrence $N\subseteq p$ is placed iff $N+\delta$ is placed.  The last claim is proved by
taking a counterexample with leftmost start and deriving a counterexample further left.  The full proof is in
Appendix~\ref{app:shift}.
\end{proof}

\begin{remark}
The proof never uses that the committed string $z$ is a maximizer, only that it differs from $x$.  So
Lemma~\ref{lem:shift} holds for every sequence of commits of distinct strings.
\end{remark}

With the shift lemma, a skipped occurrence can be charged to the occurrence that blocked it.

\begin{lemma}[Interference is free]\label{lem:int}
Let $x\notin S_t$ and let $Q$ be any set of pairwise disjoint occurrences of $x$ (in any words), all
valid at time $t$, and let $c_q$ be the count of the word containing $q$.  Then $\Sc_t(x)\ \ge\ \sum_{q\in Q} c_q\,\text{gain}_t(q)$.
\end{lemma}
\begin{proof}
Each $q\in Q$ is placed by the scan, or, being valid, is skipped because it overlaps the last placed
occurrence $p$ before it ($p_s<q_s<p_e$); in that case $p$ is valid, $1\le q_s-p_s\le|x|-1$, and Lemma~\ref{lem:shift}
gives $\text{gain}(p)=\text{gain}(q)$.  Such a $p$ is not in $Q$, since it overlaps $q$, and it blocks at most one
element of $Q$, since two disjoint occurrences of length $|x|$ cannot both start in $(p_s,p_e)$.  So the map
$q\mapsto$ (itself if placed, its blocker otherwise) is an injection from $Q$ into the placed occurrences that
preserves gain, while $\Sc_t(x)$ is the count-weighted total gain of the placed occurrences.
\end{proof}

{\sloppy
\begin{corollary}[Scores never increase]\label{cor:mono}
For every unchosen $x$, $\Sc_t(x)=\max_Q\sum_{q\in Q}c_q\,\text{gain}_t(q)$ over disjoint families $Q$ of occurrences
valid at time $t$; hence $\Sc_{t+1}(x)\le\Sc_t(x)$ whenever $x\notin S_{t+1}$.  Consequently the lazy loop of Algorithm~1 always accepts an exact maximizer, i.e.\ it is the exact greedy up to
tie-breaking.
\end{corollary}\par}
\begin{proof}
``$\ge$'' is Lemma~\ref{lem:int}, and ``$\le$'' holds because the placed occurrences form such a family.  Validity
and gains can only decrease over time, so the maximum can only decrease.  In a lazy loop, every unchosen candidate carries a stored score that was its true score at some earlier time, so
every stored score is at least the current one.  The accepted candidate's stored score is current and maximal among
stored scores, hence at least every current score.
\end{proof}

\subsection{Proof of the shift lemma (Lemma~\ref{lem:shift})}

\label{app:shift}
Note that $r\mapsto r+\delta$ maps the interior pairs of $p$ bijectively onto those of $q$, which gives
the ``in particular''.  Also, since $p$ and $q$ both spell $x$, the text is $\delta$-periodic on
$[p_s,q_e)$: $w_i=w_{i+\delta}$ for $p_s\le i<p_e$.  Hence for every occurrence $N\subseteq p$ of a
string $z$, $N+\delta\subseteq q$ is also an occurrence of $z$, and for $N\subseteq q$, $N-\delta\subseteq p$
is an occurrence of $z$.

\begin{proof}
Induction on $t$.  At $t=0$ nothing is covered.  Assume the statement at time $t$ (for all strings and
all such pairs of occurrences) and let $z=x_{t+1}$ be the string committed at step $t+1$.  Fix
$x\notin S_{t+1}$ and $p,q$ as in the lemma, valid at time $t+1$.  By monotonicity $p,q$ are valid at
time $t$ and $x\notin S_t$, so the induction hypothesis gives, with $\mathbf 1^{\rm cov}_t(r)\in\{0,1\}$ indicating that
pair $r$ is covered at time $t$,
\begin{equation}\label{eq:IH}
\mathbf 1^{\rm cov}_t(r)=\mathbf 1^{\rm cov}_t(r+\delta)\qquad(p_s<r<p_e).
\end{equation}
``Placed'' below refers to the scan of $z$ at time $t$ (the commit pass of step $t+1$).

\smallskip\noindent\emph{Step 1 (shifted copies are valid).}  If $N\subseteq p$ is an occurrence of $z$
valid at time $t$, then $N+\delta$ is valid at time $t$; if $N\subseteq q$ is valid at time $t$, then
$N-\delta$ is valid at time $t$.
Take $N\subseteq p$.  Left boundary: if $N_s=p_s$ then the left boundary pair of $N+\delta$ is $q_s$,
which is uncovered (or absent) because $q$ is valid; otherwise $p_s<N_s<p_e$ and
$\mathbf 1^{\rm cov}_t(N_s+\delta)=\mathbf 1^{\rm cov}_t(N_s)=0$ by \eqref{eq:IH} and validity of $N$.  Right boundary:
if $N_e=p_e$ the right boundary pair of $N+\delta$ is $q_e$, uncovered or absent since $q$ is valid;
otherwise $p_s<N_e<p_e$ and $\mathbf 1^{\rm cov}_t(N_e+\delta)=\mathbf 1^{\rm cov}_t(N_e)=0$.  The case $N\subseteq q$ is
symmetric (using validity of $p$ at the ends and \eqref{eq:IH} with $r=N_s-\delta$, $r=N_e-\delta$,
which lie in $(p_s,p_e)$ unless $N_s=q_s$, resp.\ $N_e=q_e$).

\smallskip\noindent\emph{Step 2 (a blocking occurrence that covers an interior pair of $p$ or $q$ lies
inside it).}  Let $M$ be placed and let $V$ be an occurrence with $M_s<V_s<M_e$, $V\subseteq p$, $V$
valid at time $t$.  Then $M\subseteq p$.  Indeed $V_s$ is an interior pair of $M$, so it is covered at
time $t+1$.  If $V_s=p_s$ this contradicts the validity of $p$ at time $t+1$ (and $V_s=p_s=0$ is impossible
since $M_s<V_s$).  Otherwise $p_s<V_s<p_e$, so $M$ (committed at time $t+1$) shares the characters
$V_s-1,V_s$ with $p$, and Lemma~\ref{lem:basic}(ii) at time $t+1$ gives $M\subseteq p$.  The same holds with $q$ in place of
$p$.

\smallskip\noindent\emph{Step 3 (placement is shift-invariant).}  We claim:
(i) if $N\subseteq p$ is placed then $N+\delta$ is placed; (ii) if $N\subseteq q$ is placed then
$N-\delta$ is placed.

(i) Suppose not, and among the placed $N\subseteq p$ with $N+\delta$ not placed choose one with
$N_s$ minimal.  By Step~1, $N+\delta$ is valid at time $t$; as it is not placed, the scan skipped it
because it overlaps the last placed occurrence $M$ before it: $M_s<N_s+\delta<M_e$.  By Step~2 (with
$V=N+\delta\subseteq q$), $M\subseteq q$.  Then $M-\delta\subseteq p$ is an occurrence of $z$, valid at
time $t$ by Step~1, and $M_s-\delta<N_s<M_e-\delta$, i.e.\ $M-\delta$ overlaps the placed $N$.  Placed
occurrences are pairwise disjoint, so $M-\delta$ is not placed, hence it was skipped because of the last
placed occurrence $M'$ before it: $M'_s<M_s-\delta<M'_e$.  By Step~2 (with $V=M-\delta\subseteq p$),
$M'\subseteq p$.  Now $M'+\delta$ overlaps the placed $M$ ($M'_s+\delta<M_s<M'_e+\delta$), so it is not
placed.  Thus $M'\subseteq p$ is placed, $M'+\delta$ is not placed, and $M'_s<M_s-\delta<N_s$,
contradicting the minimality of $N_s$.

(ii) Suppose $N\subseteq q$ is placed and $N-\delta$ is not.  By Step~1 $N-\delta$ is valid at time
$t$, so it was skipped because of the last placed $M$ before it, $M_s<N_s-\delta<M_e$; by Step~2,
$M\subseteq p$.  Then $M+\delta$ overlaps the placed $N$, so $M+\delta$ is not placed, contradicting (i)
applied to $M$.

\smallskip\noindent\emph{Step 4 (conclusion).}  By Lemma~\ref{lem:basic}, a pair is covered at time $t+1$ iff it was
covered at time $t$ or it is an interior pair of a placed occurrence $N$ of $z$.  Let $p_s<r<p_e$.  If $r$
is an interior pair of a placed $N$, then $N$ shares the characters $r-1,r$ with $p$, so $N\subseteq p$
by Lemma~\ref{lem:basic}(ii) at time $t+1$; by Step~3(i) $N+\delta$ is placed and $r+\delta$ is one of its interior pairs.
Conversely, if $r+\delta$ (which satisfies $q_s<r+\delta<q_e$) is an interior pair of a placed $N$, then
$N\subseteq q$, $N-\delta$ is placed by Step~3(ii), and $r$ is an interior pair of $N-\delta$.  Together
with \eqref{eq:IH}, $\mathbf 1^{\rm cov}_{t+1}(r)=\mathbf 1^{\rm cov}_{t+1}(r+\delta)$.
\end{proof}

%% file: app/windows.tex
\section{Windows and the recursion}\label{sec:proof}
This section turns the locality of the damage into a bound.  We credit greedy with random substrings (windows) of the
optimal tokens, charge every covered pair to greedy's own value, and solve the resulting recursion.  Section~\ref{sec:warmup} applies it by hand to tokens of at most three characters, and Appendix~\ref{sec:main}
handles all token lengths.

\paragraph{The plan in one inequality.}
Fix a step $t$.  Take an optimal token $y$ of length $L$ and a random window $(a,b)$ drawn from a distribution
$D_L$, and let $x=y[a{:}b]$.  By the shift lemma, the next greedy gain $g_{t+1}$ is at least the total gain of the
copies of this window inside the optimal occurrences of $y$.  Suppose $D_L$ keeps, in expectation, an $\alpha$
fraction of the pairs of $y$, and each covered pair, by being unavailable or by invalidating windows that start or end at it, reduces the expected window gain by at most $\beta$.  Summing over the at most $k$ optimal tokens, and noting that
greedy has covered $G_t$ pairs, gives
\[
k\,g_{t+1}\ \ge\ \alpha\,\OPT_k-\beta\,G_t .
\]
The gap $\frac\alpha\beta\OPT_k-G_t$ thus shrinks by a factor $1-\beta/k$ per step, which yields
$\GTv_k\ge\frac\alpha\beta(1-e^{-\beta})\OPT_k$ for $k\ge\beta$, by the same computation as in the $1-1/e$ proof.  For example, for a
three-character token, take the whole token with probability $m$ and each of its two two-character substrings with
probability $\frac{1-m}2$.  In expectation this keeps a fraction $\alpha=\frac{1+m}2$ of the token's two pairs, and
one checks that each covered pair costs at most $\beta=1+3m$ (Theorem~\ref{thm:short}); varying $m$ trades a larger
$\alpha$ against a larger $\beta$.  For longer tokens, $D_L$ will be allowed to depend on the strings greedy has already
chosen (Appendix~\ref{sec:main}).

Fix an optimal set $S^*$ with $|S^*|\le k$ and, in every word, an optimal family $\mathcal P$ of
disjoint occurrences of tokens of $S^*$, so that $\OPT_k=\sum_{P\in\mathcal P} c_P(L_P-1)$, where $L_P=|P|$
and $c_P$ is the count of the word containing $P$.  For $P\in\mathcal P$ we use \emph{offsets}
$e\in\{0,\dots,L_P\}$: offset $e$ denotes the pair between the $(e-1)$-th and $e$-th character of
$P$; offsets $1,\dots,L_P-1$ are the interior pairs of $P$, and offsets $0,L_P$ are its boundary
pairs (when they exist).  At time $t$ let $\Gamma_P\subseteq\{1,\dots,L_P-1\}$ be the covered interior
offsets, and let $\chi^\ell_P,\chi^r_P\in\{0,1\}$ indicate that the left/right boundary pair of $P$ exists
and is covered.

\subsection{Windows, patterns and the per-step inequality}\label{sec:windows}
We now make this plan precise.  A window is a substring of a token, given by its two
end offsets; what a window still offers greedy depends only on which offsets of the optimal occurrence are covered.

\paragraph{Windows and patterns.}  A \emph{window} of a length-$L$ token is $(a,b)$ with $0\le a$, $a+2\le b\le L$;
$\ell(a,b)=b-a-1$ is its number of interior pairs, and $q_P(a,b)$ is the aligned occurrence of $y[a{:}b]$ inside $P$.
The \emph{pattern} of $P$ at time $t$ is
$\Pi_P=\Gamma_P\cup\{0:\chi^\ell_P=1\}\cup\{L_P:\chi^r_P=1\}\subseteq\{0,\dots,L_P\}$, the set of its covered offsets.
For a pattern $\Pi\subseteq\{0,\dots,L\}$ and a window $(a,b)$ put
\[
X(\Pi,(a,b))=\mathbf 1[a\notin\Pi]\,\mathbf 1[b\notin\Pi]\cdot\#\{e:a<e<b,\ e\notin\Pi\},\qquad
\]
\[
\mathrm{cost}(\Pi)=|\Pi\cap[1,L-1]|+\tfrac12|\Pi\cap\{0,L\}|.
\]
So $X$ counts the uncovered pairs strictly inside the window if neither end of the window is covered, and is $0$
otherwise.  The cost charges $1$ for every covered interior offset and $\frac12$ for every covered boundary offset,
because a boundary pair can be shared by two optimal occurrences.

\begin{lemma}[Window gain]\label{lem:wingain}
For $P\in\mathcal P$, an occurrence of the token $y$, a time $t$ and a window $(a,b)$ of $y$, let
$X_P(a,b)=\mathbf 1[q_P(a,b)\text{ valid at time }t]\cdot\text{gain}_t(q_P(a,b))$.  Then $X_P(a,b)=X(\Pi_P,(a,b))$.
\end{lemma}
\begin{proof}
The interior pairs of $q_P(a,b)$ are the offsets $a+1,\dots,b-1$ of $P$.  Its left boundary pair is offset $a$ of
$P$: for $a\ge1$ this is an interior pair of $P$, covered iff $a\in\Gamma_P$; for $a=0$ it is the left boundary pair of
$P$, covered iff it exists and $\chi^\ell_P=1$.  In both cases it is covered iff $a\in\Pi_P$, and likewise at $b$.  So
$q_P(a,b)$ is valid iff $a,b\notin\Pi_P$, and then its gain is its number of uncovered interior pairs.
\end{proof}

\begin{definition}\label{def:exact}
A sub-probability distribution $D_L$ on the windows of a length-$L$ token is \emph{exactly $(\alpha,\beta)$-good} if
\[
\mathbb E_{(a,b)\sim D_L}X(\Pi,(a,b))\ \ge\ \alpha(L-1)-\beta\,\mathrm{cost}(\Pi)\qquad\text{for every }\Pi\subseteq\{0,\dots,L\}.
\]
\end{definition}
Here $\alpha$ is the fraction of the token's pairs that a random window keeps when nothing is covered, and $\beta$ is
the price of each covered pair.  The condition must hold for the worst pattern, chosen after $D_L$; there are
$2^{L+1}$ patterns.

The covered pairs charged in this way must be paid for by greedy's own value.  The next lemma does this; it is where
``a covered pair borders at most two optimal occurrences'' enters.

\begin{lemma}[Charging]\label{lem:charge}
Let $I_t=\sum_P c_P|\Gamma_P|$ and let $B_t$ be the (count-weighted) number of covered pairs that are
boundary pairs of some $P\in\mathcal P$.  Then $I_t+B_t\le G_t$ and
$\sum_P c_P(\chi^\ell_P+\chi^r_P)\le 2B_t$.  Hence $\sum_Pc_P\,\mathrm{cost}(\Pi_P)\le G_t$.
\end{lemma}
\begin{proof}
A boundary pair of $P$ is not interior to any $P'\in\mathcal P$ (that would make $P,P'$ overlap).  So
interior and boundary pairs are disjoint sets of covered pairs, and their total is at most $G_t$ by
Lemma~\ref{lem:basic}(iii).  A pair is a boundary pair of at most two members of $\mathcal P$.  Finally
$\mathrm{cost}(\Pi_P)=|\Gamma_P|+\frac12(\chi^\ell_P+\chi^r_P)$.
\end{proof}

Combining the lemmas gives the recursion and a first bound.

\begin{theorem}[Greedy step]\label{thm:frame}
Suppose that for every $L\ge2$ there is an exactly $(\alpha,\beta)$-good $D_L$, with $\beta\ge1$.  Then for every
corpus and every $k\ge\beta$, $\ \GTv_k\ \ge\ \frac{\alpha}{\beta}\big(1-e^{-\beta}\big)\OPT_k$.
\end{theorem}
\begin{proof}
Fix $t<k$, $y\in S^*$ of length $L$ and a window $(a,b)$; put $x=y[a{:}b]$.  If $x\notin S_t$, apply
Lemma~\ref{lem:int} to $Q=\{q_P(a,b): P\text{ an occurrence of }y,\ q_P\text{ valid}\}$, which is disjoint since the
$P$ are: $g_{t+1}\ge\Sc_t(x)\ge\sum_P c_PX_P(a,b)$.  If $x\in S_t$, every $X_P(a,b)=0$ by Lemma~\ref{lem:chosen}, and
$g_{t+1}\ge0$.  Average over $(a,b)\sim D_L$ (a sub-probability; $g_{t+1}\ge 0$), and use Lemma~\ref{lem:wingain} and
Definition~\ref{def:exact}: $g_{t+1}\ge\sum_{P}c_P\big(\alpha(L-1)-\beta\,\mathrm{cost}(\Pi_P)\big)$, the sum over the
occurrences $P$ of $y$.  Sum over the at most $k$ tokens of $S^*$ and use Lemma~\ref{lem:charge}:
\[
k\,g_{t+1}\ \ge\ \alpha\OPT_k-\beta \sum_Pc_P\,\mathrm{cost}(\Pi_P)\ \ge\ \alpha\OPT_k-\beta G_t .
\]
(If the scores are all $0$ at some $t<k$, this gives $G_k=G_t\ge\frac\alpha\beta\OPT_k$.)  With $\bar G=\alpha\OPT_k/\beta$ and $h_t=\bar G-G_t$ we get $h_{t+1}\le(1-\beta/k)h_t$ while $h_t>0$, and
$0\le1-\beta/k$.  Hence $h_k\le(1-\beta/k)^k\bar G\le e^{-\beta}\bar G$ (if some $h_t\le0$, then $G_k\ge G_t\ge \bar G$),
i.e.\ $G_k\ge(1-e^{-\beta})\bar G$.
\end{proof}

\subsection{A per-offset sufficient condition}
Exact goodness has one constraint per pattern.  (The word ``exactly'' refers to the exact accounting of patterns: the
per-offset condition below is the stronger, more conservative one, by Lemma~\ref{lem:perP}.)  The following condition
has one constraint per offset, and it is the accounting of the plan above.  For a distribution $D_L$ let
\[
\varepsilon_e=\sum_{(a,b):\,e\in\{a,b\}}D_L(a,b)\,\ell(a,b),\qquad \pi_e=\Pr_{D_L}[a<e<b]:
\]
the expected length lost
if offset $e$ invalidates the windows that start or end there, and the probability that $e$ lies strictly inside the
window.

\begin{definition}\label{def:good}
$D_L$ is \emph{$(\alpha,\beta)$-good} if
$\ \mathbb E_{D_L}[\ell]\ge\alpha(L-1)$, $\ \pi_e+\varepsilon_e\le\beta$ for $1\le e\le L-1$, and
$\varepsilon_0\le\beta/2$, $\varepsilon_L\le\beta/2$.
\end{definition}

\begin{lemma}[Per-offset bound]\label{lem:perP}
Every $(\alpha,\beta)$-good distribution is exactly $(\alpha,\beta)$-good.
\end{lemma}
\begin{proof}
Fix $\Pi$.  If a window has an end in $\Pi$ then $X=0\ge\ell-|\Pi\cap(a,b)|-\ell$; otherwise
$X=\ell-|\Pi\cap(a,b)|$.  Hence $X(\Pi,(a,b))\ge \ell-|\Pi\cap(a,b)|-\ell\,(\mathbf 1[a\in \Pi]+\mathbf 1[b\in \Pi])$
(also when an end is covered, as $|\Pi\cap(a,b)|\le\ell$).  Taking expectations, an interior $e\in \Pi$ costs at most
$\pi_e+\varepsilon_e\le\beta$, and offset $0$ or $L$ costs at most $\varepsilon_0,\varepsilon_L\le\beta/2$, while
$\mathbb E[\ell]\ge\alpha(L-1)$.
\end{proof}
The per-offset condition is lossy: it charges a window that dies at a covered end once for the end and again for
every covered pair inside it.  Its advantage is that it has explicit solutions, which we use for short tokens below.

\subsection{A different \texorpdfstring{$\beta$}{beta} at every step}

Early in the run $G_t$ is small, and a large $\alpha$ matters more than a small $\beta$; late in the run the reverse
holds.  Since the recursion holds for every certified pair $(\alpha,\beta)$ at every step, we may use the best one
at each step.

\begin{theorem}[Envelope]\label{thm:env}
Let $(\alpha_i,\beta_i)_{i\in I}$, $I$ finite, with $\beta_i\ge1$ be such that for every $i$ and every $L\ge2$ there is an
exactly $(\alpha_i,\beta_i)$-good $D_L$.  Put $\varphi(x)=\max_i(\alpha_i-\beta_ix)$ and let $y$ solve $y'=\varphi(y)$,
$y(0)=0$.  For $k\ge2$ let $\psi_k(x)=x+\frac1k\max\big(0,\max_{i:\beta_i\le k}(\alpha_i-\beta_ix)\big)$ and
$\gamma_k=\psi_k^{\circ k}(0)$.  Then $\GTv_k\ge \gamma_k\OPT_k$ for every $k\ge2$, and $\gamma_k\ge y(1)$ whenever
$k\ge\max_i\beta_i$.
\end{theorem}
\begin{proof}
The proof of Theorem~\ref{thm:frame} gives, at every step and for every $i$ separately,
$k\,g_{t+1}\ge\alpha_i\OPT_k-\beta_iG_t$; also $g_{t+1}\ge0$.  With $x_t=G_t/\OPT_k$ this is
$x_{t+1}\ge\psi_k(x_t)$.  Each function $x\mapsto x+(\alpha_i-\beta_ix)/k$ with $\beta_i\le k$ is
nondecreasing, so $\psi_k$ is nondecreasing, and induction gives $x_t\ge\psi_k^{\circ t}(0)$.  If
$k\ge\max_i\beta_i$, all lines are used, and we show $\psi_k^{\circ t}(0)\ge y(t/k)$ by induction.  As long as
$\varphi(y)>0$, $y$ increases and $\varphi$ is decreasing, so
$y(\tfrac{t+1}k)=y(\tfrac tk)+\int_{t/k}^{(t+1)/k}\varphi(y(s))\,ds\le y(\tfrac tk)+\varphi(y(\tfrac tk))/k
=\psi_k(y(\tfrac tk))\le\psi_k(\psi_k^{\circ t}(0))$.  (Once $\varphi(y)=0$, $y$ is constant and the
inequality persists.)
\end{proof}
In words, $y(1)$ is the area under the normalized greedy-gain curve when, at every step, the adversary leaves greedy
no more than the best certified line allows; Figure~\ref{fig:traj} shows this curve for short tokens.

The bounds transfer to the lazy loop that Lim et al.\ actually run.

\begin{corollary}[The paper's lazy Algorithm~1]\label{cor:lazy}
Theorems~\ref{thm:frame} and~\ref{thm:env}, and hence all lower bounds in this paper, hold for the lazy priority-queue loop of
Algorithm~1 of \cite{lim2025}, with any tie-breaking in the queue, read as in Appendix~\ref{app:impl} (the accepted
token is the one committed, and positions are visited left to right within each word).
\end{corollary}
\begin{proof}
By Corollary~\ref{cor:mono} the lazy loop accepts an exact maximizer at every step, so the claim follows from the
theorems, which allow any tie-breaking.  A direct argument, which does not use Corollary~\ref{cor:mono}, also works.
Lemma~\ref{lem:int} gives $\Sc_s(x)\ge\sum_Pc_PX_P(s)$ at every time $s$.  Covered pairs only accumulate, so
validity can only be lost and gains only drop: $X_P=\mathbf 1[\text{valid}]\cdot\text{gain}\ge0$ is non-increasing
in time.  In the lazy loop every unchosen $x$ carries a stored score $R(x)=\Sc_s(x)$ computed at some time $s\le t$
(time $0$ initially, later when popped), and the accepted token $\hat\tau$ has current score $g_{t+1}=R(\hat\tau)\ge R(x)$
(the first pop is accepted at time $0$, where stored and current scores agree).  Hence for every window string
$x\notin S_t$, $g_{t+1}\ge\Sc_s(x)\ge\sum_Pc_PX_P(s)\ge\sum_Pc_PX_P(t)$, while for $x\in S_t$
Lemma~\ref{lem:chosen} gives $X_P=0$.  The proofs of Theorems~\ref{thm:frame} and~\ref{thm:env} use the greedy rule
only through this per-window inequality $g_{t+1}\ge\sum_Pc_PX_P(a,b)$, from which Lemmas~\ref{lem:wingain}
and~\ref{lem:charge} give $k\,g_{t+1}\ge\alpha_i\OPT_k-\beta_iG_t$ for every certified pair $(\alpha_i,\beta_i)$; so
both proofs carry over unchanged.
\end{proof}

\subsection{Comparison sets}
The proof of Theorem~\ref{thm:frame} never uses the optimality of $S^*$.

\begin{lemma}[Comparison sets]\label{lem:compare}
Let $S^*$ be any set of at most $k$ strings (not necessarily optimal), and let $\mathcal L_{S^*}=\{|y|:y\in S^*\}$.  If
for every $i$ in a finite index set and every $L\in\mathcal L_{S^*}$ there is an exactly $(\alpha_i,\beta_i)$-good $D_L$, with
$\beta_i\ge1$, then Theorems~\ref{thm:frame} and~\ref{thm:env} and Corollary~\ref{cor:lazy} hold with $\OPT_k$ replaced by
$f(S^*)$.
\end{lemma}
\begin{proof}
Take $\mathcal P$ to be a maximum-value disjoint family of occurrences of tokens of $S^*$, so that its value is
$f(S^*)$.  Optimality of $S^*$ is never used after this point: Lemmas~\ref{lem:chosen}, \ref{lem:shift}
and~\ref{lem:int} do not mention $S^*$; Lemma~\ref{lem:wingain} concerns a single occurrence; Lemma~\ref{lem:charge} uses
only that the members of $\mathcal P$ are pairwise disjoint; and the proof of Theorem~\ref{thm:frame} uses $|S^*|\le k$
and $\sum_Pc_P(L_P-1)=f(S^*)$.  Goodness is applied only to members of $\mathcal P$, whose lengths lie in
$\mathcal L_{S^*}$, so good distributions are needed only for those lengths.  The proofs of Theorem~\ref{thm:env} and
Corollary~\ref{cor:lazy} use the per-step inequality only.
\end{proof}

\begin{remark}\label{rem:shortscope}
(i) The hypothesis concerns the comparison set only; \GT{}'s own choices are unrestricted.  (ii) Tightness holds
along the budgets $k=\nu K$ at which the J4 corpora and their disjoint copies are built (Theorems~\ref{thm:j4}
and~\ref{thm:opt}, Lemma~\ref{lem:copies}); we
do not claim that every large $k$ has a near-extremal corpus with a short optimal vocabulary.  (iii) The extremal trajectory of the recursion and the J4 run coincide step for step in the limit
(Figure~\ref{fig:traj}): both have normalized gain $e^{-4t}$ until $t_1$ and $\frac13e^{-(t-t_1)}$ afterwards.
\end{remark}

%% file: app/main.tex
\section{The bound 0.3287 for every corpus}\label{sec:main}
This appendix proves $\GTv_k\ge\MAINconst\cdot\OPT_k$ for every corpus and every $k$.  The warm-up loses for long tokens
for two reasons.  The per-offset condition charges a window that dies at a covered end twice.  Exact goodness
(Definition~\ref{def:exact}) removes the double charge, but it lets the adversary choose any pattern of covered offsets,
in every occurrence independently, and most of these patterns cannot occur.  Once the window distribution may depend on
\emph{which strings greedy has chosen and in which order}, the adversary is left with a two-parameter family of patterns
per token, and against this family one explicit distribution per line suffices.

\subsection{Chosen windows}\label{sec:pce}
Take a token of length $4$ and the pattern $\{1\}$: offset~$1$ of an occurrence is covered while offsets $0$ and $2$ are
not.  The only commit that can do this is a commit of the prefix $y[0{:}2]$ at exactly this position.  But once
$y[0{:}2]$ is chosen, Lemma~\ref{lem:chosen} says that no occurrence of $y$ can keep offsets $0$, $1$ and $2$ all
uncovered, since its window $(0,2)$ would then be a valid occurrence of a chosen string with positive gain.  So the
pattern $\{1\}$ cannot coexist with the empty pattern in occurrences of the same token at the same time.

A \emph{run window} of a pattern $\Pi$ is a window $(a,b)$ with $a,b\notin\Pi$ and $a+1,\dots,b-1\in\Pi$: a maximal covered
stretch strictly between two uncovered offsets.  For a set $\mathcal W$ of windows of a length-$L$ token let
\[
\begin{aligned}
\mathcal Q(\mathcal W)=\big\{\Pi\subseteq\{0,\dots,L\}:\ &\text{every run window of }\Pi\text{ is in }\mathcal W,\\
&\text{and }X(\Pi,w)=0\text{ for every }w\in\mathcal W\big\}.
\end{aligned}
\]

\begin{lemma}[Run windows are chosen]\label{lem:pce}
Fix a time $t$ and a token $y$ of length $L$, and let $\mathcal W=\{(a,b):y[a{:}b]\in S_t\}$ be its chosen windows.
Then $\Pi_P\in\mathcal Q(\mathcal W)$ for every occurrence $P$ of $y$.
\end{lemma}
\begin{proof}
Let $(a,b)$ be a run window of $\Pi_P$; we work in $P$'s offsets.  Offset $a+1$ is covered, so pair $a+1$ is interior to
some committed occurrence.  The committed occurrences with pair $a+1$ in their interior all contain the two characters
around that pair, so they are pairwise nested (Lemma~\ref{lem:basic}(i)); let $Q=[s,e)$ be the largest.  If $s<a$, then
pair $a$ is interior to $Q$ and hence covered.  For $a\ge1$ this contradicts $a\notin\Pi_P$.  For $a=0$, pair $0$ is the left
boundary pair of $P$: if it exists it is covered, contradicting $0\notin\Pi_P$, and if it does not exist, $P$ starts at the
beginning of the word and $s<0$ is impossible.  So $s=a$.  Every interior pair of $Q$ is covered, and pair $b$ is not
(or does not exist), so $e\le b$.  If $e<b$, then pair $e$, which is offset $e\in\{a+1,\dots,b-1\}$, is covered: it is interior
to a committed $Q'$ that contains the characters $e-1\in Q$ and $e\notin Q$.  By Lemma~\ref{lem:basic}(i), $Q\subsetneq Q'$, so
$Q'$ also has pair $a+1$ in its interior, contradicting the choice of $Q$.  Hence $Q=[a,b)$ is a committed occurrence of
$y[a{:}b]$, and $(a,b)\in\mathcal W$.  Finally let $w\in\mathcal W$.  By Lemma~\ref{lem:wingain}, $X(\Pi_P,w)$ is $0$ if the aligned
occurrence of $w$ is invalid, and otherwise its gain, which is $0$ by Lemma~\ref{lem:chosen}.
\end{proof}

With no chosen windows the adversary can only cover a prefix and a suffix of the offsets: a maximal covered block that
contains neither $0$ nor $L$ is the interior of a run window, so
\begin{equation}\label{eq:Qempty}
\mathcal Q(\emptyset)=\big\{\{0,\dots,p-1\}\cup\{q+1,\dots,L\}:\ 0\le p\le q+1\le L+1\big\}.
\end{equation}
Chosen windows forbid some patterns (each chosen window must be dead) but allow others: covered runs strictly inside
the token.  One might hope that the second effect never helps the adversary, so that it would suffice to beat the
prefix-and-suffix patterns of~\eqref{eq:Qempty}.  For a token of $20$ characters it does help.

\begin{proposition}[Two chosen windows that help the adversary]\label{prop:cxL20}
Let $L=20$, $\beta=1$ and $\mathcal W=\{(6,8),(7,14)\}$, and put
$\alpha_{\mathcal W}(L,\beta)=\max_D\min_{\Pi\in\mathcal Q(\mathcal W)}\big(\mathbb E_DX(\Pi,\cdot)+\beta\,\mathrm{cost}(\Pi)\big)/(L-1)$
over sub-probabilities $D$ on windows.  Then
\[
\alpha_{\mathcal W}(20,1)=\tfrac{300654}{602471}\approx0.49903\;<\;\tfrac12=\alpha_\emptyset(20,1).
\]
\end{proposition}
The mechanism is a \emph{mutual kill} (Figure~\ref{fig:idea}a).  A pattern with run window $(6,8)$ covers offset $7$, the
left end of $(7,14)$, so both chosen windows are dead.  A pattern with run window $(7,14)$ covers offsets $8,\dots,13$,
including the right end of $(6,8)$, so again both are dead.  Each chosen string's covered run kills the other one, and
the adversary mixes the two kinds of patterns.  The proof (Appendix~\ref{app:cx}) gives an explicit mixture of $32$
patterns and an explicit distribution on $22$ windows, both checked in exact rational arithmetic.

\subsection{Greedy's history}\label{sec:history}
The mutual kill cannot happen inside \GT{}.  Greedy chooses $(6,8)$'s string and $(7,14)$'s string at different steps.
Whichever comes first is committed first in every occurrence where both of its ends are uncovered, and its covered run
then blocks the other forever.  So in a run of \GT{} the adversary cannot mix the two kinds of patterns freely: in the order $(6,8)$ first, a run window
$(7,14)$ forces the prefix to cover offsets $0,\dots,6$, and in the other order a run window $(6,8)$ forces the suffix to
cover offsets $14,\dots,20$.
The next lemma makes this precise for every token: given the order of greedy's choices, the pattern of an occurrence is
determined by two numbers.

\paragraph{Ordered window sets.}
Fix a token $y$ of length $L$ and a time $t$.  For each step $s\le t$ let $C_s=\{(a,b):y[a{:}b]=x_s\}$ be the windows of
$y$ that spell greedy's $s$-th choice.  The nonempty $C_s$, in the order of $s$, form the \emph{ordered window set}
$\mathcal O=(C_1,\dots,C_r)$ of $y$ at time $t$ (we renumber them $1,\dots,r$), with union $\mathcal W$, the chosen windows.  The
classes are disjoint, and the windows of one class have pairwise distinct left endpoints.  Below, an \emph{ordered window
set} is any finite sequence of disjoint sets of windows with this last property; the windows of one class need not have
equal lengths.

\paragraph{The middle process.}
For $0\le p\le q\le L$ define $M(p,q)=M_r$ by $M_0=\emptyset$ and, for $j=1,\dots,r$: scan the windows $(a,b)\in C_j$ with
$p\le a$ and $b\le q$ in increasing order of $a$, and \emph{commit} $(a,b)$ iff $a\notin M_{j-1}$, $b\notin M_{j-1}$, and $a$ is
at least the right end of the window last committed in this scan (if any); then $M_j$ is $M_{j-1}$ together with the
interiors of the windows committed in class $j$.  This is \GT{}'s leftmost scan, restricted to the windows that lie
between $p$ and $q$.  The \emph{history family} is
\[
\begin{aligned}
\mathcal F(L;\mathcal O)=\Big\{\Pi\in\mathcal Q(\mathcal W):\ &\Pi=\{0,\dots,L\},\ \text{or}\\
&\Pi=[0,p)\cup(q,L]\cup M(p,q)\text{ for some }0\le p\le q\le L\Big\},
\end{aligned}
\]
where $[0,p)=\{0,\dots,p-1\}$ and $(q,L]=\{q+1,\dots,L\}$.  Since $M(p,q)\subseteq\{p+1,\dots,q-1\}$, the offsets $p$ and $q$
are uncovered, and $|\mathcal F(L;\mathcal O)|=O(L^2)$ for every $\mathcal O$, however the windows interact.

\begin{lemma}[History lemma]\label{lem:2P}
At every time $t$, every occurrence $P$ of $y$ has $\Pi_P\in\mathcal F(L;\mathcal O)$, where $\mathcal O$ is the ordered window set
of $y$ at time $t$.
\end{lemma}
\begin{proof}[Proof sketch]
Let $p$ and $q$ be the smallest and the largest uncovered offset of $\Pi_P$ (if there is none, $\Pi_P$ is the full
pattern).  A commit that covers an offset $e\in[p,q]$ must lie inside $[p,q]$: a commit reaching past the left end of $P$
covers all offsets up to $e$, including $p$; one reaching past the right end covers $q$.  So the offsets of $[p,q]$ are
covered only by commits of windows inside $[p,q]$, and an induction over the steps shows that the commits inside $[p,q]$
are exactly those of the middle process: validity at the time of the commit is the condition $a,b\notin M_{j-1}$, and the
leftmost scan within the occurrence skips exactly the windows that overlap the previously committed one.  Membership in
$\mathcal Q(\mathcal W)$ is Lemma~\ref{lem:pce}.  The full proof is in Appendix~\ref{app:history}.
\end{proof}

\noindent For the example of Proposition~\ref{prop:cxL20}, the cheap mixture of the mutual kill is not
available under either order, and by Theorem~\ref{thm:H} below the bound $\frac12$ holds for this example.

\subsection{The refined framework and the structural theorem}
In the proof of Theorem~\ref{thm:frame} the window distribution enters only through the inequality
$g_{t+1}\ge\sum_Pc_P\,\mathbb E_DX_P$ for one token $y$ at one time $t$, and this inequality holds for every
sub-probability $D$.  So $D$ may be chosen after looking at the state at time $t$, in particular after looking at the
ordered window set of $y$.  For an ordered window set $\mathcal O$ of a length-$L$ token and $\beta\ge1$ put
\[
\Phi_D(\Pi)=\mathbb E_DX(\Pi,\cdot)+\beta\,\mathrm{cost}(\Pi),\qquad
\alpha^{\rm H}(L;\mathcal O;\beta)=\max_D\ \min_{\Pi\in\mathcal F(L;\mathcal O)}\ \frac{\Phi_D(\Pi)}{L-1},
\]
with $D$ ranging over sub-probability distributions on windows.  We call $D$ a \emph{witness} for
$\alpha^{\rm H}(L;\mathcal O;\beta)\ge\alpha$ if $\Phi_D(\Pi)\ge\alpha(L-1)$ for every $\Pi\in\mathcal F(L;\mathcal O)$.

\begin{lemma}[Refined framework]\label{lem:pcerec}
Let $(\alpha_i,\beta_i)_{i\in I}$, $I$ finite, with $\beta_i\ge1$, and let $S^*$ be any set of at most $k$ strings.  If
$\alpha_i\le\alpha^{\rm H}(L;\mathcal O;\beta_i)$ for every $i$, every $L\in\mathcal L_{S^*}$ and every ordered window set
$\mathcal O$ of a length-$L$ token, then the conclusions of Theorem~\ref{thm:env} and Corollary~\ref{cor:lazy} hold with $\OPT_k$
replaced by $f(S^*)$.
\end{lemma}
\begin{proof}
Fix $t$, $i$ and $y\in S^*$, and let $\mathcal O$ be the ordered window set of $y$ at time $t$.  Take $D$ attaining
$\alpha^{\rm H}(L;\mathcal O;\beta_i)$ (a maximum of a continuous function over a compact set).  By Lemmas~\ref{lem:2P}
and~\ref{lem:wingain}, every occurrence $P$ of $y$ has $\mathbb E_DX_P+\beta_i\mathrm{cost}(\Pi_P)\ge\alpha_i(L-1)$.  The rest of
the proofs of Theorem~\ref{thm:frame}, Lemma~\ref{lem:compare}, Theorem~\ref{thm:env} and Corollary~\ref{cor:lazy} is
unchanged: they use only the resulting inequality $k\,g_{t+1}\ge\alpha_if(S^*)-\beta_iG_t$ at each step.  (The lazy loop
commits exactly as the exact loop does, so Lemma~\ref{lem:2P} applies to it as well.)
\end{proof}

Everything now rests on one inequality per integer $\beta$, about one explicit family of patterns.

Theorem~\ref{thm:H} (Section~\ref{sec:uniform}) states $\alpha^{\rm H}(L;\mathcal O;\beta)\ge\frac{\beta}{\beta+1}$ for every
integer $\beta\in\{1,\dots,6\}$, every $L\ge2$ and every ordered window set $\mathcal O$.

For $\beta=1$ the constant cannot be raised: a token of length $2$ has $\alpha^{\rm H}(2;\emptyset;1)=\frac12$ (the pattern
$\{0\}$ costs $\frac12$ and kills the only window).  The proof uses $\beta\le6$ nowhere; with all lines $\beta\le B$ the
envelope below increases with $B$, but only by about $10^{-4}$ (from $0.32874$ to about $0.32885$), so we keep six lines.
Theorem~\ref{thm:H} is proved in
Appendices~\ref{sec:E0}--\ref{sec:beta1}.  First we derive the main result, Theorem~\ref{thm:main} (Section~\ref{sec:uniform}).

\begin{proof}[Proof of Theorem~\ref{thm:main}]
$k=1$ and $k=2$ are Proposition~\ref{prop:easy} ($\GTv_k\ge\OPT_k/k$).  By Theorem~\ref{thm:H}, the six lines $(\alpha_\beta,\beta)=(\frac\beta{\beta+1},\beta)$,
$\beta=1,\dots,6$, satisfy the hypothesis of Lemma~\ref{lem:pcerec} for every token length, with $S^*$ an optimal set.  So
Theorem~\ref{thm:env} applies.  For $3\le k\le5$, $\gamma_k=\psi_k^{\circ k}(0)$ is a rational iteration with the lines
$\beta\le k$.  To see how it works, take $k=2$, where it gives only $\frac5{12}$: $x_1=\frac12\cdot\frac23=\frac13$ (line $\beta=2$), $x_2=\frac13+\frac12\max(\frac12-\frac13,\frac23-\frac23)=\frac5{12}$.

For $k\ge6=\max\beta$, Theorem~\ref{thm:env} gives $\gamma_k\ge y(1)$, where $y'=\varphi(y)$, $y(0)=0$,
$\varphi(y)=\max_\beta\big(\frac\beta{\beta+1}-\beta y\big)$.  The lines $\beta$ and $\beta+1$ cross at
$y=\frac{\beta+1}{\beta+2}-\frac\beta{\beta+1}=\frac1{(\beta+1)(\beta+2)}$, and these crossings decrease in $\beta$; so line
$6$ is the maximum on $[0,\frac1{42}]$, line $\beta\in\{2,\dots,5\}$ on $[\frac1{(\beta+1)(\beta+2)},\frac1{\beta(\beta+1)}]$, and line $1$
on $[\frac16,\frac12]$.  On a segment where line $\beta$ is the maximum, $y$ moves from $y_0$ to $y_1$ in time
$\frac1\beta\ln\frac{\alpha_\beta-\beta y_0}{\alpha_\beta-\beta y_1}$.  At the two ends of the segment of line $\beta\in\{2,\dots,5\}$,
$\alpha_\beta-\beta y$ equals $\frac\beta{\beta+2}$ and $\frac{\beta-1}{\beta+1}$, so the segment takes time
$T_\beta=\frac1\beta\ln\frac{\beta(\beta+1)}{(\beta-1)(\beta+2)}$; the first segment takes $T_6=\frac16\ln\frac65$.  The total
$\sum_\beta T_\beta=\frac12\ln\frac32+\frac13\ln\frac65+\frac14\ln\frac{10}9+\frac15\ln\frac{15}{14}+\frac16\ln\frac65\approx0.334$ is
less than $1$, and afterwards line $1$ gives $\frac12-y(t)=\frac13e^{-(t-\sum T_\beta)}$.  Hence
$y(1)=\frac12-\frac13e^{\sum_\beta T_\beta-1}$, which simplifies to the displayed closed form.  It exceeds $\MAINconst$, and so do
$\gamma_2,\dots,\gamma_5$.  Corollary~\ref{cor:lazy} gives the lazy loop.
\end{proof}

\paragraph{What is computed.}  Only the decimal value of the closed form, which we evaluated in interval arithmetic
($y(1)\in[\MAINyOneFull,\ \MAINyOneFull+10^{-16}]$); computational tests of the proof are described in
Appendix~\ref{sec:provenance}.

\paragraph{The extremal trajectory.}  As for short tokens (Figure~\ref{fig:traj}), $y'(t)$ is the normalized greedy gain of
the worst case the recursion allows: it decays at rate $6$ at first, then at rates $5,4,3,2$, and from $t=\sum T_\beta\approx0.334$
on at rate $1$.  For tokens of at most three characters only the rates $4$ and $1$ appear, and the junction corpora of
Appendix~\ref{sec:upper} realize that trajectory exactly.  Whether a corpus realizes the six-rate trajectory is the open
question behind the remaining gap (Section~\ref{sec:discussion}).

\subsection{Proof of Theorem~\ref{thm:H}, part 1: no chosen windows}\label{sec:E0}
Throughout the proof, $\beta$ is a fixed integer, $V=\frac\beta{\beta+1}(L-1)$ is the target, and $\mathrm{wt}(e)=\frac12$ for
$e\in\{0,L\}$ and $\mathrm{wt}(e)=1$ otherwise, so that $\mathrm{cost}(\Pi)=\sum_{e\in\Pi}\mathrm{wt}(e)$.  We write a sub-probability
$D$ through its \emph{values} $v_w=D_w\ell(w)$, its total value $\mathrm{Val}=\sum_wv_w=\mathbb E_D\ell$ and its mass
$\sum_wD_w=\sum_wv_w/\ell(w)$.  The left load of an offset $e$ is $\lambda^{\rm L}(e)=\sum_{w=(e,b)}v_w$, and its right load is
$\lambda^{\rm R}(e)=\sum_{w=(a,e)}v_w$.  The first part, the case $\mathcal W=\emptyset$, is a second warm-up: every step can be
checked by hand.

\begin{lemma}[Load bound]\label{lem:mono}
If $\lambda^{\rm L}(e)\le\beta\,\mathrm{wt}(e)$ and $\lambda^{\rm R}(e)\le\beta\,\mathrm{wt}(e)$ for every offset $e$, then
$\Phi_D(\Pi)\ge \mathrm{Val}$ for every $\Pi\in\mathcal Q(\emptyset)$.
\end{lemma}
\begin{proof}
Let $\Phi(p,q)$ be $\Phi_D$ of the pattern $[0,p)\cup(q,L]$; then $\Phi(0,L)=\mathrm{Val}$.  Raise $p$ one step at a time with $q=L$, then
lower $q$.  A prefix step $p\to p+1$ covers offset $p$: windows with $a=p$ die, windows with $b=p$ are already dead (they have
$a<p$), and offset $p$ is interior to no live window (live windows have $a\ge p+1$).  So $\mathbb E_DX$ drops by at most
$\lambda^{\rm L}(p)\le\beta\,\mathrm{wt}(p)$, which is the cost increase.  A suffix step covers offset $q$, which is not yet covered; it
kills exactly the live windows with $b=q$, no live window has $a=q$, and $q$ is interior to no live window.  So $\Phi$
never decreases.
\end{proof}

\begin{theorem}[No chosen windows]\label{thm:E0}
For every integer $\beta\ge1$ and every $L\ge2$ there is a sub-probability $Z=Z(L,\beta)$ with $\Phi_Z(\Pi)\ge V$ for every
$\Pi\in\mathcal Q(\emptyset)$; that is, $\alpha^{\rm H}(L;\emptyset;\beta)\ge\frac\beta{\beta+1}$.
\end{theorem}
\begin{proof}
\emph{Short tokens, $L\le\beta+1$.}  A pattern of cost at least $1$ has $\Phi_Z\ge\beta>V$, since $V<\beta$ when $L-1\le\beta$.  The
patterns of $\mathcal Q(\emptyset)$ of cost less than $1$ are $\emptyset$, $\{0\}$ and $\{L\}$.  For $L=2$ take the point mass on
$(0,2)$: $\Phi(\emptyset)=1\ge V$ and $\Phi(\{0\})=\Phi(\{2\})=\frac\beta2\ge V$.  For $3\le L\le\beta+1$ let
$u=\max\big(0,\frac{V-\beta/2}{L-2}\big)$ and $Z=(1-2u)\cdot(0,L)+u\cdot(0,L-1)+u\cdot(1,L)$.  Then
$\Phi(\{0\})=\Phi(\{L\})=u(L-2)+\frac\beta2\ge V$, and $\Phi(\emptyset)=(1-2u)(L-1)+2u(L-2)=L-1-2u\ge V$ iff
$2u\le\frac{L-1}{\beta+1}$.  If $u>0$ this reads $\beta(2L-\beta-3)\le(L-1)(L-2)$; the left side is a concave function of
$\beta$ with maximum at $\beta=L-\frac32$, so for $\beta\ge L-1$ it is at most its value $(L-1)(L-2)$ at $\beta=L-1$.  Also
$2u\le\frac{L-1}{\beta+1}\le1$, so $Z\ge0$ and has mass $1$.

\emph{Long tokens, $L\ge\beta+2$: the zigzag.}  Write $L-1=(\beta+1)m+r$ with $0\le r\le\beta$, so $m\ge1$, and put
$A=\beta m+r=L-1-m$; note $V=\beta m+\frac{\beta r}{\beta+1}$.  We give values $v_w$ with all loads at most $\beta\,\mathrm{wt}$,
$\mathrm{Val}=V$ and mass at most $1$; Lemma~\ref{lem:mono} then gives the claim.  \emph{Long} windows $(a,a+L-m)$ have $\ell=A$, and
\emph{short} windows have $\ell=A-1$.
\begin{itemize}
\item \emph{Type I ($2r\le\beta+1$).}  Put $s=\frac{\beta r}{\beta+1}\le\frac\beta2$.  Long windows $a=0,\dots,m$ get $v=\frac\beta2$
for $a\in\{0,m\}$ and $\beta-s$ otherwise; short windows $(a+1,a+L-m)$, $a=0,\dots,m-1$, get $v=s$.  Left loads: $\frac\beta2$ at
$0$, $(\beta-s)+s=\beta$ at $1,\dots,m-1$, $\frac\beta2+s\le\beta$ at $m$; the right loads are symmetric.  Value:
$\mathrm{Val}=(m-1)(\beta-s)+\beta+ms=\beta m+s=V$.  Mass: $\frac{\beta m-(m-1)s}A+\frac{ms}{A-1}\le1$ iff $r(A-1)\ge s(A+m-1)$, i.e.\ iff
$(\beta+1)(A-1)\ge\beta(A+m-1)$, i.e.\ iff $r\ge1$; for $r=0$ we have $s=0$ and mass $\beta m/A=1$.
\item \emph{Type II ($2r\ge\beta+1$).}  Put $s=\frac{\beta(\beta+1-r)}{\beta+1}\le\frac\beta2$.  Long windows $a=0,\dots,m$ get $v=s$;
short windows $(a,a+L-m-1)$, $a=0,\dots,m+1$, get $v=\beta-s$ for $1\le a\le m$ and $\frac\beta2-s$ for $a\in\{0,m+1\}$.  Left loads:
$\frac\beta2$ at $0$, $\beta$ at $1,\dots,m$, $\frac\beta2-s$ at $m+1$; the right loads are symmetric.  Value:
$\mathrm{Val}=(m+1)s+m(\beta-s)+\beta-2s=\beta m+\beta-s=V$.  Mass: $\frac{(m+1)s}A+\frac{\beta m+\beta-(m+2)s}{A-1}\le1$ iff
$s(A+m+1)\ge A(\beta+1-r)$, i.e.\ iff $\beta(m+1)\ge A$, i.e.\ iff $r\le\beta$.
\end{itemize}
Every window with positive value is a window: $A\ge1$, and short windows carry value only when $r\ge1$ (Type~I) or
$r\ge\frac{\beta+1}2$ (Type~II), and then $A-1=\beta m+r-1\ge1$.
\end{proof}

\noindent The left ends of the zigzag lie in $[0,m+1]$ and its right ends in $[L-m-1,L]$; for $\beta=1$ the two zones are the two
halves of the token.

\subsection{Proof of Theorem~\ref{thm:H}, part 2: peeling one chosen window}\label{sec:peel}
With chosen windows, the zigzag alone fails: a chosen window inside its left zone lets the adversary cover a hole where
the zigzag has its left ends, at no gain.  We remove chosen windows one at a time, building the witness for $\mathcal O$
from witnesses for two smaller instances.  Two operations produce them (Figure~\ref{fig:idea}c).

\paragraph{Sub-tokens and contraction.}
For $0\le s<s'\le L$, the \emph{sub-token} $[s,s']$ is a token of length $s'-s$, and a window $w$ of it lifts to $w+s$.  The
\emph{contraction} of a window $w^*=(a,b)$ with $\ell=b-a-1\ge1$ removes its interior: $\phi(e)=e$ for $e\le a$ and $\phi(e)=e+\ell$
for $e>a$ maps $\{0,\dots,L-\ell\}$ onto $\{0,\dots,L\}\setminus\{a+1,\dots,b-1\}$, and a window $w'=(a',b')$ of the contracted token lifts to
$(\phi(a'),\phi(b'))$.  A sub-probability $D'$ on the windows of the smaller token lifts to the windows of the token.

\begin{lemma}[Embedding]\label{lem:embed}
Let $D'$ satisfy $\Phi'_{D'}(\Pi')\ge T$ for every $\Pi'$ in a set $\mathcal F'$ of patterns of the smaller token.
\begin{enumerate}
\item[(a)] (Sub-token.)  If $\Pi'=(\Pi\cap[s,s'])-s\in\mathcal F'$, then
$\Phi_{\mathrm{lift}D'}(\Pi)\ge T+\beta\,\mathrm{ex}(\Pi)$ with
$\mathrm{ex}(\Pi)=\sum_{e\in\Pi\setminus[s,s']}\mathrm{wt}(e)+\frac12\mathbf 1[s\in\Pi,\,s>0]+\frac12\mathbf 1[s'\in\Pi,\,s'<L]$.
\item[(b)] (Contraction.)  If $\Pi'=\phi^{-1}(\Pi)\in\mathcal F'$, then $\Phi_{\mathrm{lift}D'}(\Pi)\ge T+\beta\,|\Pi\cap\{a+1,\dots,b-1\}|$.
\end{enumerate}
\end{lemma}
\begin{proof}
(a) A window inside the sub-token sees the same covered offsets in $\Pi$ and in $\Pi'$, so $X(\Pi,w+s)=X'(\Pi',w)$, and
$\mathrm{cost}(\Pi)=\mathrm{cost}'(\Pi')+\mathrm{ex}(\Pi)$: offsets outside the sub-token are not counted by $\mathrm{cost}'$, and the
ends $s,s'$ of the sub-token weigh $\frac12$ there but $1$ in the token unless they are $0$ or $L$.  (b) The ends of $w'$ and of its lift
correspond, and the interior of the lift contains $\phi$ of the interior of $w'$, so $X(\Pi,\phi(w'))\ge X'(\Pi',w')$; the
removed offsets are interior offsets, so $\mathrm{cost}(\Pi)=\mathrm{cost}'(\Pi')+|\Pi\cap\{a+1,\dots,b-1\}|$.
\end{proof}

\paragraph{Cheap windows and relevant patterns.}
Call a pattern \emph{relevant} if $\beta\,\mathrm{cost}(\Pi)<V$; an irrelevant pattern has $\Phi_D(\Pi)\ge V$ for every $D$.  Call
a window \emph{cheap} if $\ell<\frac{L-1}{\beta+1}$.  More generally, for a threshold $\tau$ let $\mathcal O^{<\tau}$ be $\mathcal O$
with all windows of gain $\ell\ge\tau$ deleted (and empty classes dropped).

\begin{lemma}[Only cheap windows matter]\label{lem:CH}
Every $\Pi\in\mathcal F(L;\mathcal O)$ with $\mathrm{cost}(\Pi)<\tau$ lies in $\mathcal F(L;\mathcal O^{<\tau})$.  In particular every relevant
pattern lies in $\mathcal F(L;\mathcal O^{<(L-1)/(\beta+1)})$.
\end{lemma}
\begin{proof}
The full pattern lies in every family, so write $\Pi=[0,p)\cup(q,L]\cup M(p,q)$.  A window committed
by the middle process has its interior in $\Pi$, so its gain is at most $\mathrm{cost}(\Pi)<\tau$.  Each commit decision depends
only on the set $M_{j-1}$ and on the last window committed in the current scan, both of which are determined by the
committed windows; so the process on $\mathcal O^{<\tau}$ commits the same windows, and $M(p,q)$ is the same.  The run windows of
$\Pi$ lie in $\mathcal W$ and have gain at most $\mathrm{cost}(\Pi)<\tau$, so they lie in $\mathcal W^{<\tau}$, and every window of
$\mathcal W^{<\tau}\subseteq\mathcal W$ is dead in $\Pi$.
\end{proof}

From now on let all windows of $\mathcal O$ be cheap, $\mathcal O\neq\emptyset$, and let $w^*=(a,b)$ be the window with the smallest
left end in the first class $C_1$; write $\ell=b-a-1$.  Since $w^*$ is chosen first among the windows of $C_1$ and nothing
was chosen before $C_1$, the middle process commits $w^*$ whenever $w^*$ lies between $p$ and $q$.  Only here does the
proof use the order of greedy's choices; without the order, the mutual kill breaks this step.

\begin{lemma}[Three cases]\label{lem:three}
Every $\Pi=[0,p)\cup(q,L]\cup M(p,q)\in\mathcal F(L;\mathcal O)$ satisfies exactly one of: (i) $p>a$ (the prefix covers $a$);
(ii) $p\le a$, $q\ge b$, and $\Pi$ has a run window $u=(a_u,b_u)$ with $a_u\le a$ and $b_u\ge b$, so that $\Pi$ covers all of
$a+1,\dots,b-1$; (iii) $p\le a$ and $q<b$ (the suffix covers $b$).
\end{lemma}
\begin{proof}
Only case (ii) needs proof.  There $w^*$ lies between $p$ and $q$.  In the scan of $C_1$ we have $M_0=\emptyset$ and no window
of $C_1$ starts before $a$, so $w^*$ is committed and its interior lies in $\Pi$.  The maximal covered block containing that
interior lies strictly between the uncovered offsets $p\le a$ and $q\ge b$, so it is the interior of a run window $u$ of $\Pi$.
\end{proof}

For the peeling we need three smaller ordered window sets: $\mathcal O'$, obtained by deleting $w^*$ and every window with an
end in $\{a+1,\dots,b-1\}$ and contracting the rest; $\mathcal O^{\rm L}$, the windows inside the sub-token $[a+1,L]$; and
$\mathcal O^{\rm R}$, the windows inside $[0,b-1]$.  Each has fewer windows than $\mathcal O$, since $w^*$ is in none of them.

\begin{lemma}[Restriction]\label{lem:restrict}
Let $\Pi\in\mathcal F(L;\mathcal O)$.  Then $\phi^{-1}(\Pi)\in\mathcal F(L-\ell;\mathcal O')$.  In cases (i) and (ii),
$(\Pi\cap[a+1,L])-(a+1)\in\mathcal F(L-a-1;\mathcal O^{\rm L})$.  In cases (ii) and (iii), $\Pi\cap[0,b-1]\in\mathcal F(b-1;\mathcal O^{\rm R})$.
\end{lemma}
\noindent The proof (Appendix~\ref{app:peel}) rests on one fact: if $u$ is an uncovered offset with $p\le u\le q$, then the
middle process on $[p,q]$ splits into independent processes on $[p,u]$ and on $[u,q]$, because no committed window can have
$u$ in its interior.  In case (ii) the contraction is exact: $w^*$ is committed first, the windows with an end in its
interior are never committed, and the process on the contracted token makes the same decisions.

\begin{lemma}[Peeling]\label{lem:peel}
Suppose $b\le L+\frac12-\frac{L-1}{\beta+1}$ (\emph{left region}).  Let $D_{\rm K}$ be a witness for
$\alpha^{\rm H}(L-\ell;\mathcal O';\beta)\ge\frac\beta{\beta+1}$ and $D_{\rm T}$ one for
$\alpha^{\rm H}(L-a-1;\mathcal O^{\rm L};\beta)\ge\frac\beta{\beta+1}$.  Put $B=a\beta+\frac{\beta-1}2$ and $\lambda=B/(\ell+B)$.
Then $D=\lambda\cdot\mathrm{lift}(D_{\rm K})+(1-\lambda)\cdot\mathrm{lift}(D_{\rm T})$ is a witness for
$\alpha^{\rm H}(L;\mathcal O;\beta)\ge\frac\beta{\beta+1}$.  Symmetrically, if $a\ge\frac{L-1}{\beta+1}-\frac12$ (\emph{right
region}), the same holds with $\mathcal O^{\rm R}$ on $[0,b-1]$ in place of $\mathcal O^{\rm L}$ and $L-b$ in place of $a$.
\end{lemma}
\begin{proof}
$D$ has mass at most $1$.  Irrelevant patterns need nothing; by Lemma~\ref{lem:CH} relevant patterns are in the family of
the cheap windows, which is $\mathcal O$.  In case (iii) the suffix covers $b,\dots,L$, so $\mathrm{cost}\ge L-b+\frac12\ge\frac{L-1}{\beta+1}$
and $\Pi$ is irrelevant.  In cases (i) and (ii), Lemmas~\ref{lem:restrict} and~\ref{lem:embed} give
\[
\Phi_D(\Pi)-V\ \ge\ \lambda\big(\beta\,\mathrm{ex}_{\rm K}-c\ell\big)+(1-\lambda)\big(\beta\,\mathrm{ex}_{\rm T}-c(a+1)\big),\qquad c=\tfrac\beta{\beta+1},
\]
since $V=c(L-1)=c(L-\ell-1)+c\ell=c(L-a-2)+c(a+1)$.  Here $\mathrm{ex}_{\rm K}=|\Pi\cap\{a+1,\dots,b-1\}|$ and
$\mathrm{ex}_{\rm T}=\mathrm{cost}(\Pi\cap[0,a])+\frac12\mathbf 1[a+1\in\Pi]$.  Divide by $c$.  In case (i), $\mathrm{ex}_{\rm K}\ge0$ and
$\mathrm{ex}_{\rm T}\ge a+\frac12$ (all of $0,\dots,a$ is covered), and the right side is at least
$-\lambda\ell+(1-\lambda)\big((\beta+1)(a+\frac12)-a-1\big)=-\lambda\ell+(1-\lambda)B=0$.  In case (ii), $\mathrm{ex}_{\rm K}=\ell$ and
$\mathrm{ex}_{\rm T}\ge\frac12$ (offset $a+1$ is covered), and the right side is at least
$\lambda\beta\ell+(1-\lambda)\big(\frac{\beta+1}2-a-1\big)=\frac{\ell(\beta^2-1)(a+\frac12)}{\ell+B}\ge0$.  The right region is the mirror
image, with case (i) irrelevant (cost at least $a+\frac12$).
\end{proof}

\begin{theorem}[Peeling, $\beta\ge2$]\label{thm:H2}
For every integer $\beta\in\{2,\dots,6\}$, every $L\ge2$ and every ordered window set $\mathcal O$,
$\alpha^{\rm H}(L;\mathcal O;\beta)\ge\frac\beta{\beta+1}$.
\end{theorem}
\begin{proof}
Induction on the number of windows, for all $L$ at once.  By Lemma~\ref{lem:CH} we may delete the expensive windows.  If no
window remains, the relevant patterns lie in $\mathcal F(L;\emptyset)=\mathcal Q(\emptyset)$ and Theorem~\ref{thm:E0} applies.
Otherwise the window $w^*$ is cheap, and for $\beta\ge2$ every cheap window lies in the left or the right region: if
$a<\frac{L-1}{\beta+1}-\frac12$, then $b=a+\ell+1<\frac{2(L-1)}{\beta+1}+\frac12\le L+\frac12-\frac{L-1}{\beta+1}$, because
$\frac{3(L-1)}{\beta+1}\le L-1$.  Lemma~\ref{lem:peel} reduces to instances with fewer windows.  The smaller tokens have length at least $2$: a cheap window
exists only if $\frac{L-1}{\beta+1}>1$, and then $L-\ell\ge2$, $L-a-1\ge L-b+1\ge2$ in the left region, and $b-1\ge a+1\ge2$ in
the right region.
\end{proof}

\subsection{Proof of Theorem~\ref{thm:H}, part 3: \texorpdfstring{$\beta=1$}{beta=1}}\label{sec:beta1}
For $\beta=1$ a cheap window can lie in neither region: it can \emph{straddle} the middle of the token ($a<\frac L2-1$ and
$b>\frac L2+1$).  Then all three cases of Lemma~\ref{lem:three} contain cheap patterns, and no convex combination of
witnesses of the type used above can work: the three cheap patterns $\{a+1,\dots,b-1\}$, $[0,a]$ and $[b,L]$ partition
the offsets, so the extra costs that the three embeddings collect add up to no more than the token itself.  The way out
is to \emph{add} three witnesses, one per case, each aiming at a smaller target, and to keep track of the total
mass.  This needs a statement with a variable target.

For $t\ge0$ call $D$ a \emph{$t$-witness} for $(L,\mathcal O)$ if $\mathbb E_DX(\Pi,\cdot)+\mathrm{cost}(\Pi)\ge t$ for every
$\Pi\in\mathcal F(L;\mathcal O)$, and put $\mu(L,t)=t/(L-1-t)$.

\begin{theorem}[Additive peeling, $\beta=1$]\label{thm:PH}
For every $L\ge2$, every ordered window set $\mathcal O$ and every $t\in\frac12\mathbb Z$ with $0\le t\le\frac{L-1}2$, there is a
$t$-witness for $(L,\mathcal O)$ of mass at most $\mu(L,t)$.  At $t=\frac{L-1}2$, where $\mu=1$, this is
$\alpha^{\rm H}(L;\mathcal O;1)\ge\frac12$.
\end{theorem}
\begin{proof}
Induction on the number of windows, for all $L$ and $t$ at once.  For $t=0$ take $D=0$.  Patterns of cost at least $t$ need
nothing, so by Lemma~\ref{lem:CH} (with $\tau=t$) we may delete every window of gain at least $t$.

\emph{Base: no window left.}  Then the patterns of cost less than $t$ lie in $\mathcal Q(\emptyset)$, and by
Lemma~\ref{lem:mono} it suffices to give values with all loads at most $\mathrm{wt}$, total value $\mathrm{Val}\ge t$ and mass at most
$\mu(L,t)$.  If $t$ is an integer, use the windows $(i,L-t+i)$, $i=0,\dots,t$, of gain $G=L-t-1$, with value $1$, except
$\frac12$ for $i\in\{0,t\}$: the loads are at most $\mathrm{wt}$, $\mathrm{Val}=t$, and the mass is $t/G=\mu(L,t)$.  If $t=j+\frac12$, use the
windows $(c,c+L-j)$, $c=0,\dots,j$, of gain $G=L-j-1$, and $(c+1,c+L-j)$, $c=0,\dots,j-1$, of gain $G-1$, all with value
$\frac12$: the left loads are $\frac12$ at $0$ and $1$ at $1,\dots,j$, the right loads are symmetric, $\mathrm{Val}=j+\frac12$, and the mass
$\frac{j+1}{2G}+\frac j{2(G-1)}$ is at most $\mu(L,t)=\frac{2j+1}{2G-1}$ because $G\ge j+1$.

\emph{Peeling.}  Otherwise let $w^*=(a,b)$, $\ell=b-a-1<t$, be as in Lemma~\ref{lem:three}.  By Lemma~\ref{lem:restrict}, in each
case the restricted pattern lies in the family of a smaller instance, and by Lemma~\ref{lem:embed} it collects the extra cost
$\mathrm{ex}_{\rm K}=|\Pi\cap\{a+1,\dots,b-1\}|$ (contraction), $\mathrm{ex}_{\rm T}$ (sub-token $[a+1,L]$) or
$\mathrm{ex}_{\rm R}=\mathrm{cost}(\Pi\cap[b,L])+\frac12\mathbf 1[b-1\in\Pi]$ (sub-token $[0,b-1]$), where
$\mathrm{ex}_{\rm K}=\ell$ in case (ii), $\mathrm{ex}_{\rm T}\ge a+\frac12$ in case (i) and $\ge\frac12$ in case (ii), and
$\mathrm{ex}_{\rm R}\ge L-b+\frac12$ in case (iii) and $\ge\frac12$ in case (ii).  Sub-witnesses with a negative target are $0$.
\begin{itemize}
\item \emph{Left: $L-b+\frac12\ge t$.}  Case (iii) has cost at least $t$.  Take a $(t-\frac\ell2)$-witness $D_{\rm K}$ for
$(L-\ell,\mathcal O')$ and a $(t-\frac{a+1}2)$-witness $D_{\rm T}$ for $(L-a-1,\mathcal O^{\rm L})$, and
$D=\lambda\,\mathrm{lift}(D_{\rm K})+(1-\lambda)\,\mathrm{lift}(D_{\rm T})$ with $\lambda=\frac a{a+\ell}$.  In case (i),
$\Phi\ge t-\lambda\frac\ell2+(1-\lambda)\frac a2=t$; in case (ii), $\Phi\ge t+\lambda\frac\ell2-(1-\lambda)\frac a2=t$.  Each
sub-mass is $0$ (if its target is not positive) or at most $f(x)=\frac{t-x}{L-1-t-x}$ (with $x=\frac\ell2$, resp.\ $x=\frac{a+1}2$), and
$f$ is non-increasing because $2t\le L-1$; so the mass is at most $f(0)=\mu(L,t)$.
\item \emph{Right: $a+\frac12\ge t$.}  The mirror image, with the sub-token $[0,b-1]$, target $t-\frac{L-b+1}2$ and
$\lambda=\frac{L-b}{L-b+\ell}$.
\item \emph{Straddling: $a+\frac12<t$ and $L-b+\frac12<t$.}  Take a $t_{\rm K}$-witness for $(L-\ell,\mathcal O')$, a
$t_{\rm T}$-witness for $(L-a-1,\mathcal O^{\rm L})$ and a $t_{\rm R}$-witness for $(b-1,\mathcal O^{\rm R})$, with
$t_{\rm K}=t-\ell$, $t_{\rm T}=t-a-\frac12$ and $t_{\rm R}=t-(L-b)-\frac12$, all positive, and let $D$ be the \emph{sum} of their lifts.
Since $X\ge0$, each case is paid by its own part: case (ii) by the contraction, $\Phi\ge t_{\rm K}+\ell=t$; case (i) by the left
trim, $\Phi\ge t_{\rm T}+a+\frac12=t$; case (iii) by the right trim, $\Phi\ge t_{\rm R}+L-b+\frac12=t$.  The three masses are
$\frac{t_{\rm K}}{L-1-t}$, $\frac{t_{\rm T}}{L-t-3/2}$ and $\frac{t_{\rm R}}{L-t-3/2}$, and $t_{\rm K}+t_{\rm T}+t_{\rm R}=3t-L$, so the mass is
at most $\frac{3t-L}{L-t-3/2}$.  Finally $t(L-t-\frac32)-(3t-L)(L-1-t)=2t^2-3tL+\frac32t+L^2-L$ is a convex quadratic in $t$ with
vertex at $\frac{3L}4-\frac38>\frac{L-1}2$, so on $[0,\frac{L-1}2]$ it is at least its value $\frac{L-1}4$ at $t=\frac{L-1}2$;
hence the mass is at most $\mu(L,t)$.
\end{itemize}
In every case the sub-targets lie in $\frac12\mathbb Z$ and are at most half the sub-token length minus $\frac12$ (this uses
$t\le\frac{L-1}2$).  The sub-tokens have length at least $2$: the contraction has length $L-\ell>\frac{L+1}2$; in the left
case $t>\ell\ge1$ forces $t\ge\frac32$, hence $L-b\ge1$ and $L-a-1\ge2$; in the straddling case $L-a-1>\frac L2$ and $b-1>\frac L2$
with $L\ge4$; the right case is the mirror image.  So the induction hypothesis applies.
\end{proof}

\noindent Theorems~\ref{thm:H2} and~\ref{thm:PH} together prove Theorem~\ref{thm:H}, and with it Theorem~\ref{thm:main}.  The
history enters only through Lemma~\ref{lem:three}, case (ii); without it, the same witnesses fail on the counterexample of
Proposition~\ref{prop:cxL20} (Appendix~\ref{sec:provenance}).

\subsection{Proof of the history lemma (Lemma~\ref{lem:2P})}\label{app:history}
Fix a word, an occurrence $P=[s_0,s_0+L)$ of the token $y$, and the run.  Offset $e$ of $P$ is pair $s_0+e$ of the word.  Let
$\Pi_t$ be $P$'s pattern after step $t$ ($\Pi_0=\emptyset$).  We use two facts from the model (Section~\ref{sec:model} and
Lemma~\ref{lem:basic}).  (B1) Committed occurrences are laminar, covered pairs stay covered, and committing an occurrence covers
its interior pairs and nothing else.  (B2) At step $t$ the occurrences of $x_t$ in the word are scanned by increasing start;
one is placed iff it is valid at time $t-1$ and does not overlap the previously placed occurrence of $x_t$; all placed
occurrences are committed, so placed occurrences are pairwise disjoint.  (Validity at time $t-1$ equals validity during the
scan: a commit in the scan covers only pairs interior to its own occurrence, never a boundary pair of a later,
non-overlapping one.)

\paragraph{Classification of commits.}  An occurrence $o=[u,v)$ of $x_t$ covers a pair of $P$ only if it contains two
consecutive characters of $[s_0-1,s_0+L]$.  Then exactly one of the following holds: (I) $s_0\le u$ and $v\le s_0+L$, and $o$ is the
window $(u-s_0,v-s_0)\in C_t$ (\emph{inside}); (II) $u<s_0<v\le s_0+L$, and $o$ covers exactly the offsets $0,\dots,v-s_0-1$ of $P$;
(III) $s_0\le u<s_0+L<v$, and it covers the offsets $u-s_0+1,\dots,L$; (IV) $u<s_0$ and $v>s_0+L$, and it covers $0,\dots,L$.

\begin{proof}[Proof of Lemma~\ref{lem:2P}]
Fix $t$.  If $\Pi_t$ is the full pattern there is nothing to show.  Otherwise let $p$ and $q$ be the smallest and the largest
uncovered offset of $\Pi_t$, so the offsets $0,\dots,p-1$ and $q+1,\dots,L$ are covered.  Let $M_j$ be the sets of the middle process
for these $p,q$ and the ordered window set at time $t$; class $j$ of that set is the class $C_{t'}$ of the $j$-th step $t'$ at
which a window of $y$ was chosen.

(a) \emph{Every offset $e\in[p,q]$ that is covered at some time $t'\le t$ was covered by a type-(I) commit of a window inside
$[p,q]$.}  A type-(II) commit covering $e$ covers $0,\dots,e\ni p$; type (III) covers $e,\dots,L\ni q$; type (IV) covers $p$.  By (B1)
$p$ or $q$ would be covered at time $t$.  So $e$ was covered by an inside commit of some $(a,b)$, which covers $a+1,\dots,b-1$
forever; since $p$ is uncovered and $p\le e$, $p\notin\{a+1,\dots,b-1\}$, so $a\ge p$, and likewise $b\le q$.

(b) \emph{For every $t'\le t$, $\Pi_{t'}\cap[p,q]$ is the set $M$ produced by the classes chosen up to $t'$.}  Induction on $t'$.  A
step whose class has no window inside $[p,q]$ changes nothing in $[p,q]$, by (a).  Let step $t'$ have class $C$ and let
$(a,b)\in C$ lie inside $[p,q]$.  Validity of $(a,b)$ at time $t'-1$ means $a,b\notin\Pi_{t'-1}$, i.e.\ (by induction) $a,b\notin M_{j-1}$.
Let $o$ be the previously placed occurrence of $x_{t'}$ in the scan.  If $o$ overlaps $(a,b)$, then $o$ contains the character
$s_0+a$, so $o$ is of type (I) with left end $a'<a$, or of type (II) ($o$ has the length of $(a,b)$, so it is not of type (III)
or (IV)); type (II) would cover $p$ (as $o$ is committed), which is impossible.  So $o=(a',b')\in C$ with $a'<a<b'$, committed; its interior contains $a$; if $a'<p$ it would
contain $p$, and if $b'>q$ it would contain $q$; so $o$ lies inside $[p,q]$, and it is the last inside window committed
earlier in this scan, with $a<b'$.  (Here and below we use, by an inner induction along the scan of step $t'$, that the
windows inside $[p,q]$ placed earlier in this scan are exactly those the middle process committed earlier in the scan of
class $j$.)  Conversely, if the last inside window $o'$ committed earlier in this scan has right end
$b_{o'}>a$, then $o=o'$: otherwise $o$ was placed after $o'$ and before $(a,b)$, so $o$ does not overlap $o'$ and starts at or after
$b_{o'}>a$, contradicting that $o$ starts before $(a,b)$.  Hence $(a,b)$ is placed iff the middle process commits it.  By (a), the
offsets of $[p,q]$ newly covered at step $t'$ are exactly the interiors of the committed inside windows, which is the update
of the middle process.

(c) By (b) at $t'=t$, $\Pi_t=[0,p)\cup(q,L]\cup M(p,q)$, and $\Pi_t\in\mathcal Q(\mathcal W)$ by Lemma~\ref{lem:pce}.
\end{proof}

\noindent Two remarks.  A tie class (several windows of $y$ spelling the same string) is scanned left to right with the
skipping rule, exactly as \GT{} does; dropping this rule, or reversing the order of the classes, gives a different family
that does not contain the true patterns (Appendix~\ref{sec:provenance}).  And $\mathcal F(L;\mathcal O)$ has at most
$\binom{L+2}2+1$ elements, whatever the windows.

\subsection{Proof of the restriction lemma (Lemma~\ref{lem:restrict})}\label{app:peel}
Let $\Pi=[0,p)\cup(q,L]\cup M(p,q)\in\mathcal F(L;\mathcal O)$, not the full pattern (the full pattern restricts to the full pattern).
Recall that $w^*=(a,b)$ has the smallest left end in $C_1$, that windows of one class have distinct left ends, and that no
window lies in two classes.

\paragraph{Fact NS (no straddling).}  \emph{If $u$ is uncovered in $\Pi$ and $p\le u\le q$, then $M(p,q)\cap[p,u]=M(p,u)$ and
$M(p,q)\cap[u,q]=M(u,q)$.}  A committed window cannot have $u$ in its interior, so committed windows lie inside $[p,u]$ or inside
$[u,q]$.  Windows inside $[p,u]$ and windows inside $[u,q]$ never interact: an interior inside $[p,u]$ contains no end $\ge u$, an
interior inside $[u,q]$ contains no end $\le u$, and within one class scan every window inside $[p,u]$ precedes every window inside
$[u,q]$, while the right end of a commit inside $[p,u]$ is at most $u$, which blocks no window starting at $u$ or later.  Windows
that straddle $u$ are never committed (they would cover $u\notin M(p,q)$) and are absent from both restricted processes.
Formally, this is an induction over the classes and, within a class, along the scan.

\paragraph{Commits are nested or meet in at most one end.}  A later commit needs both of its ends uncovered, so its ends
are not in the interior of an earlier committed window, and two commits of one class scan do not overlap.  Hence two
committed windows are nested or share at most one end, and a shared end is uncovered unless it lies in the interior of a
third committed window that contains both.  So the maximal covered blocks of $M(p,q)$ are interiors of committed windows
(the maximal ones under inclusion).

\begin{proof}[Proof of Lemma~\ref{lem:restrict}]
\emph{Contraction, case (ii).}  Here $p\le a$, $q\ge b$, and $w^*$ is committed first in the scan of $C_1$ (Lemma~\ref{lem:three}).
A window with an end in $\{a+1,\dots,b-1\}$ is never committed: in $C_1$ it starts in $\{a+1,\dots,b-1\}$ (it cannot start before $a$,
and starting at $a$ would make it $w^*$), so it is blocked by the right end $b$ of $w^*$, and every later commit in $C_1$ starts
at $b$ or after, hence after it in the scan; in a later class its end lies in $M_1\subseteq M_{j-1}$.  Every other window of $C_1$
starts at $b$ or later, so the remaining scan of $C_1$ is the contracted scan; for later classes, validity and the skipping
rule correspond under $\phi$, and so does the condition of lying inside the range, because $p\le a$ and $q\ge b$.  A committed window $w\ne w^*$ has an interior disjoint from that of $w^*$ or containing it, and in
both cases $\phi(\mathrm{int}(\phi^{-1}(w)))=\mathrm{int}(w)\setminus\mathrm{int}(w^*)$.  Hence
$M(p,q)=\mathrm{int}(w^*)\cup\phi(M'(p,q-\ell))$, where $M'$ is the middle process of $\mathcal O'$, and
$\phi^{-1}(\Pi)=[0,p)\cup(q-\ell,L-\ell]\cup M'(p,q-\ell)$.

\emph{Contraction, case (i).}  Here $p>a$.  Let $p''$ be the first uncovered offset with $p''\ge p$ and
$p''\notin\{a+1,\dots,b-1\}$; then $p''\ge b$.  If there is none with $p''\le q$, every offset outside the interior of $w^*$ is
covered and $\phi^{-1}(\Pi)$ is the full pattern.  Otherwise, by NS with $u=p''$, $\Pi\cap[p'',q]=M(p'',q)$, and the windows inside
$[p'',q]\subseteq[b,L]$ have no end in the interior of $w^*$; so $\phi^{-1}(\Pi)=[0,\phi^{-1}(p''))\cup(q-\ell,L-\ell]\cup M'(\phi^{-1}(p''),q-\ell)$
(the offsets of $[p,p'')$ outside the interior of $w^*$ are covered by the choice of $p''$).

\emph{Contraction, case (iii).}  Symmetric, with $q''$ the last uncovered offset $\le q$ outside the interior of $w^*$ ($q''\le a$)
and NS with $u=q''$.

\emph{Trims.}  For $\Pi^{\rm L}=(\Pi\cap[a+1,L])-(a+1)$: in case (i), $p\ge a+1$, and every window inside $[p,q]$ lies in $\mathcal O^{\rm L}$,
so the process is the same.  In case (ii) the offsets $a+1,\dots,b_u-1$ lie in the interior of the run window $u$ and are covered,
and $b_u$ is uncovered, so $\Pi^{\rm L}$ starts with a covered prefix of length $b_u-a-1$; by NS with $u=b_u$, the rest is
$M(b_u,q)$, computed from the windows inside $[b_u,q]\subseteq[a+1,L]$.  The trim $\Pi^{\rm R}=\Pi\cap[0,b-1]$ is symmetric, with NS at
$u=a_u$ in case (ii), and directly in case (iii), where every window inside $[p,q]$ lies inside $[0,b-1]$.

\emph{Membership in $\mathcal Q$.}  In each case the restricted pattern is $[0,p^*)\cup(q^*,L']\cup M^*(p^*,q^*)$ with $p^*,q^*$
uncovered, so its run windows are maximal covered blocks of $M^*$, which are interiors of committed windows of the smaller
ordered set; so they belong to it.  Every window $w'$ of the smaller set is dead: for the trims by Lemma~\ref{lem:embed}(a),
since $X$ is unchanged, and for the contraction because $X'(\phi^{-1}(\Pi),w')\le X(\Pi,\phi(w'))=0$.
\end{proof}

\subsection{The counterexample at \texorpdfstring{$L=20$}{L=20} (Proposition~\ref{prop:cxL20})}\label{app:cx}
\input{gt/cx_proof.tex}

%% file: gt/cx_proof.tex
\begin{proof}[Proof of Proposition~\ref{prop:cxL20}]
\emph{The mechanism.}  A pattern with run window $(6,8)$ covers offset $7$, which is an end of $(7,14)$, so
$X(\Pi,(7,14))=0$; and $X(\Pi,(6,8))=0$ because the only interior offset of $(6,8)$ is covered.  A pattern with run
window $(7,14)$ covers offsets $8,\dots,13$: offset $8$ is an end of $(6,8)$, and these are all the interior offsets of
$(7,14)$, so again both chosen windows are dead.  Thus the covered run of each chosen window kills the other one, and the
adversary may use the covered runs $\{7\}$ and $\{8,\dots,13\}$ strictly inside the token, which $\mathcal Q(\emptyset)$
forbids.  (The two runs cannot occur together, since offset $7$ would have to be both covered and uncovered.)  Filtering
all $2^{21}$ patterns gives $|\mathcal Q(\mathcal W)|=281$: $134$ prefix-and-suffix patterns, $91$ patterns with run window
$(6,8)$ and $56$ with run window $(7,14)$.

\emph{Upper bound.}  For every probability distribution $\mu$ on $\mathcal Q(\mathcal W)$ and every sub-probability $D$ on
windows,
\begin{align*}
\min_{\Pi\in\mathcal Q(\mathcal W)}\big(\mathbb E_DX(\Pi,\cdot)+\beta\,\mathrm{cost}(\Pi)\big)
&\le\sum_wD_w\,\mathbb E_\mu X(\cdot,w)+\beta\,\mathbb E_\mu\mathrm{cost}\\
&\le\max\big(0,\max_w\mathbb E_\mu X(\cdot,w)\big)+\beta\,\mathbb E_\mu\mathrm{cost},
\end{align*}
since $D\ge0$ and $\sum_wD_w\le1$ (weak linear-programming duality).  The mixture $\mu$ of
Table~\ref{tab:cxmu} has $32$ patterns, all in $\mathcal Q(\mathcal W)$ (each has run window $(6,8)$, run window $(7,14)$, or
none, and keeps both chosen windows dead).  Computed exactly, $\mathbb E_\mu\mathrm{cost}=\frac{211554}{31709}$ and
$\max_w\mathbb E_\mu X(\cdot,w)=\frac{89100}{31709}$, the maximum over all $190$ windows (attained at $32$ of them).  The sum
is $\frac{300654}{31709}$, and dividing by $L-1=19$ gives $\alpha_{\mathcal W}(20,1)\le\frac{300654}{602471}$.

\emph{Lower bound.}  The distribution $D_{\mathcal W}$ of Table~\ref{tab:cxD} ($22$ windows, total mass $1$) satisfies
$\mathbb E_{D_{\mathcal W}}X(\Pi,\cdot)+\beta\,\mathrm{cost}(\Pi)\ge\frac{300654}{31709}$ for each of the $281$ patterns of
$\mathcal Q(\mathcal W)$, checked in exact rational arithmetic.  Hence $\alpha_{\mathcal W}(20,1)=\frac{300654}{602471}$.

\emph{No chosen windows.}  Let $D_\emptyset$ put mass $\frac1{20}$ on each window $(a,a+11)$, $0\le a\le9$, and mass
$\frac1{18}$ on each window $(a,a+10)$, $1\le a\le9$; the total mass is $\frac{10}{20}+\frac9{18}=1$.  Each offset
$1\le e\le9$ is the left end of one window of each family, with load $\frac1{20}\cdot10+\frac1{18}\cdot9=1=\beta\,\mathrm{wt}(e)$;
offset $0$ is the left end of $(0,11)$ only, with load $\frac12=\beta\,\mathrm{wt}(0)$; no window starts at an offset $\ge10$.  The
right ends are symmetric (offsets $11,\dots,19$ carry load $1$, offset $20$ carries $\frac12$, offsets $\le10$ carry $0$).
So Lemma~\ref{lem:mono} applies (with $\beta=1$) and gives $\alpha_\emptyset(20,1)\ge\frac1{19}\sum_wD_w\ell(w)=\frac1{19}\big(10\cdot\frac{10}{20}+9\cdot\frac9{18}\big)=\frac12$.
A mixture on $66$ patterns of $\mathcal Q(\emptyset)$, listed in the supplement, has value exactly $\frac12$, so
$\alpha_\emptyset(20,1)=\frac12$.

The supplement (folder \texttt{counterexample\_L20}) contains a short standalone program that rebuilds $\mathcal Q(\mathcal W)$
and $\mathcal Q(\emptyset)$ from the definitions by filtering all $2^{21}$ patterns and re-verifies all four certificates in
exact arithmetic.
\end{proof}

\input{gt/cx_table.tex}

%% file: gt/cx_table.tex
\begin{table}[!htb]\centering\footnotesize
\begin{tabular}{@{}llr@{\qquad}llr@{}}\toprule
covered offsets & run window & $\mu$ & covered offsets & run window & $\mu$\\\midrule
7,14--20 & (6,8) & $\tfrac{20271}{126836}$ & 0,1,7,17--20 & (6,8) & $\tfrac{2106}{348799}$\\
0--4,7 & (6,8) & $\tfrac{13374}{348799}$ & 0,1,7,18--20 & (6,8) & $\tfrac{13374}{2441593}$\\
7,16--20 & (6,8) & $\tfrac{13374}{348799}$ & 0,1,7 & (6,8) & $\tfrac{14211}{2885519}$\\
0--5,7 & (6,8) & $\tfrac{13374}{348799}$ & 0--3,7 & (6,8) & $\tfrac{1467}{348799}$\\
7,15--20 & (6,8) & $\tfrac{13374}{348799}$ & 0--5,7,18--20 & (6,8) & $\tfrac{477}{697598}$\\
7,13--20 & (6,8) & $\tfrac{990}{31709}$ & 0--2,7,15--20 & (6,8) & $\tfrac{477}{697598}$\\
0--2,7,20 & (6,8) & $\tfrac{66870}{2885519}$ & 8--13 & (7,14) & $\tfrac{89100}{412217}$\\
0,7,17--20 & (6,8) & $\tfrac{2415}{126836}$ & 0,8--13 & (7,14) & $\tfrac{7425}{412217}$\\
0--3,7,18--20 & (6,8) & $\tfrac{80244}{4534387}$ & 8--13,20 & (7,14) & $\tfrac{7425}{412217}$\\
0--3,7,19,20 & (6,8) & $\tfrac{6777}{412217}$ & 0,1,8--13 & (7,14) & $\tfrac{414}{31709}$\\
0--2,7,18--20 & (6,8) & $\tfrac{481464}{31740709}$ & 8--13,19,20 & (7,14) & $\tfrac{414}{31709}$\\
0,1,7,19,20 & (6,8) & $\tfrac{6093}{412217}$ & 0,8--13,20 & (7,14) & $\tfrac{1350}{412217}$\\
7,17--20 & (6,8) & $\tfrac{18507}{1395196}$ & 0--7 & none & $\tfrac{489474}{2885519}$\\
0,7,14--20 & (6,8) & $\tfrac{1545}{126836}$ & 0--8 & none & $\tfrac{810}{31709}$\\
0--4,7,17--20 & (6,8) & $\tfrac{2972}{348799}$ & 0--8,20 & none & $\tfrac{180}{31709}$\\
0--3,7,16--20 & (6,8) & $\tfrac{2972}{348799}$ & 0--7,20 & none & $\tfrac{6840}{2885519}$\\
\bottomrule\end{tabular}
\caption{The mixture $\mu$ of Proposition~\ref{prop:cxL20} ($L=20$, $\beta=1$, $\mathcal W=\{(6,8),(7,14)\}$): its $32$ patterns,
each given by its covered offsets ($a$--$b$ is a range), its run window (``none'': no covered run strictly inside the token),
and its weight.  The weights sum to $1$; $\mathbb E_\mu\mathrm{cost}=\frac{211554}{31709}$ and
$\max_w\mathbb E_\mu X(\cdot,w)=\frac{89100}{31709}$.  File \texttt{counterexample\_L20/cx\_L20.json} of the supplement.}
\label{tab:cxmu}
\end{table}
\begin{table}[!htb]\centering\footnotesize
\begin{tabular}{@{}lr@{\qquad}lr@{}}\toprule
window & $D_w$ & window & $D_w$\\\midrule
$(0,12)$ & $\tfrac{1}{20}$ & $(4,15)$ & $\tfrac{1697}{31709}$\\
$(0,19)$ & $\tfrac{1}{24}$ & $(4,16)$ & $\tfrac{8218}{158545}$\\
$(1,12)$ & $\tfrac{1697}{31709}$ & $(5,16)$ & $\tfrac{1697}{31709}$\\
$(1,13)$ & $\tfrac{1621}{634180}$ & $(5,17)$ & $\tfrac{8218}{158545}$\\
$(1,14)$ & $\tfrac{2841}{63418}$ & $(6,18)$ & $\tfrac{32167}{634180}$\\
$(1,19)$ & $\tfrac{1333}{63418}$ & $(6,19)$ & $\tfrac{2841}{63418}$\\
$(1,20)$ & $\tfrac{1}{24}$ & $(8,17)$ & $\tfrac{15273}{253672}$\\
$(2,13)$ & $\tfrac{10417}{190254}$ & $(8,18)$ & $\tfrac{10417}{190254}$\\
$(2,14)$ & $\tfrac{32167}{634180}$ & $(8,19)$ & $\tfrac{1621}{634180}$\\
$(3,13)$ & $\tfrac{15273}{253672}$ & $(9,19)$ & $\tfrac{1697}{31709}$\\
$(3,15)$ & $\tfrac{8218}{158545}$ & $(9,20)$ & $\tfrac{1}{20}$\\
\bottomrule\end{tabular}
\caption{The window distribution $D_{\mathcal W}$ of Proposition~\ref{prop:cxL20} ($22$ windows, total mass $1$).  For every
one of the $281$ patterns $\Pi\in\mathcal Q(\mathcal W)$, $\mathbb E_{D_{\mathcal W}}X(\Pi,\cdot)+\beta\,\mathrm{cost}(\Pi)\ge\frac{300654}{31709}$.
File \texttt{counterexample\_L20/certificates.json} of the supplement.}
\label{tab:cxD}
\end{table}

%% file: app/upper.tex
\section{Junction corpora: where the short-token bound is tight}\label{sec:upper}\label{app:upper}
We now show that the guarantee is not far from the truth, and that for short optimal tokens it is exact.  The worst-case
ratio has limit superior at most $\JLIM\approx\JFourLimSeven$ as $k\to\infty$.  The constructions use adversarial
ties, but Lemma~\ref{lem:bonus} removes all ties at an arbitrarily small cost.

\subsection{The junction mechanism}
\begin{figure}[t]
\centering
\begin{tikzpicture}[x=0.62cm,y=0.62cm,font=\small]
  \begin{scope}
    \draw[thick,fill=gray!30,rounded corners=3pt] (1.58,-0.4) rectangle (3.42,0.4);
    \foreach \c [count=\i from 0] in {a_u,x_i,z_v,a_{u'},x_j,z_{v'}} \node[minimum size=0.5cm,inner sep=0] at (\i,0) {$\c$};
    \foreach \r in {1,...,5} \node[font=\scriptsize,gray] at (\r-0.5,-0.85) {\r};
    \fill (2.5,-0.55) circle (2.2pt);
    \draw[thick] (-0.42,0.75) -- (-0.42,0.95) -- (2.42,0.95) -- (2.42,0.75);
    \draw[thick] (2.58,0.75) -- (2.58,0.95) -- (5.42,0.95) -- (5.42,0.75);
    \node at (1,1.4) {$y_i$: invalid};
    \node at (4,1.4) {$y_j$: invalid};
    \node[font=\scriptsize] at (2.5,-1.35) {cut $z_va_{u'}$ (gain 1)};
    \draw[thick,dashed] (3.58,-1.75) -- (3.58,-1.95) -- (5.42,-1.95) -- (5.42,-1.75);
    \node[font=\scriptsize] at (4.5,-2.35) {fragment $x_jz_{v'}$};
    \draw[thick,dashed] (-0.42,-1.75) -- (-0.42,-1.95) -- (1.42,-1.95) -- (1.42,-1.75);
    \node[font=\scriptsize] at (0.5,-2.35) {fragment $a_ux_i$};
    \node at (2.5,-3.2) {(a) J2 word $y_iy_j$};
  \end{scope}
  \begin{scope}[xshift=4.9cm]
    \draw[thick,fill=gray!30,rounded corners=3pt] (1.58,-0.4) rectangle (3.42,0.4);
    \draw[thick,fill=gray!30,rounded corners=3pt] (7.58,-0.4) rectangle (9.42,0.4);
    \draw[thick,fill=gray!60,rounded corners=3pt] (4.58,-0.4) rectangle (6.42,0.4);
    \foreach \c [count=\i from 0] in {a,x_1,z,a,x_2,z,a,x_3,z,a,x_4,z} \node[minimum size=0.5cm,inner sep=0] at (\i,0) {$\c$};
    \node[font=\scriptsize] at (2.5,-0.85) {I};
    \node[font=\scriptsize] at (5.5,-0.85) {II};
    \node[font=\scriptsize] at (8.5,-0.85) {III};
    \node[font=\scriptsize] at (2.5,0.85) {phase 1};
    \node[font=\scriptsize] at (8.5,0.85) {phase 1};
    \node[font=\scriptsize] at (5.5,0.85) {phase 2};
    \draw[thick,dashed] (3.58,-1.25) -- (3.58,-1.45) -- (5.42,-1.45) -- (5.42,-1.25);
    \draw[thick,dashed] (5.58,-1.25) -- (5.58,-1.45) -- (7.42,-1.45) -- (7.42,-1.25);
    \node[font=\scriptsize,align=center] at (5.5,-2.2) {after phase 1: fragments $x_2z$ and $ax_3$;\\the phase-2 cut at II kills both};
    \node at (5.5,-3.2) {(b) J4 word $y_1y_2y_3y_4$};
  \end{scope}
\end{tikzpicture}
\caption{Junction corpora.  Shaded boxes are committed cut strings, each covering one pair (a filled dot in (a)).
(a) In a two-token word the junction pair is the right boundary of $y_i$ and the left boundary of $y_j$: one unit of
gain invalidates two planted occurrences, worth $4$ pairs, and leaves the fragments $a_ux_i$ and $x_jz_{v'}$; J2 later harvests fragments of the form $x_jz$ at
one pair each.  (b) In a four-token word (letter subscripts omitted) the junctions I and III inside the two blocks are cut in
phase~1.  The two
inner fragments sit next to the uncut junction II between the blocks, so a phase-2 cut at II again invalidates two fragments per
unit of gain, instead of harvesting them.}
\label{fig:junction}
\end{figure}
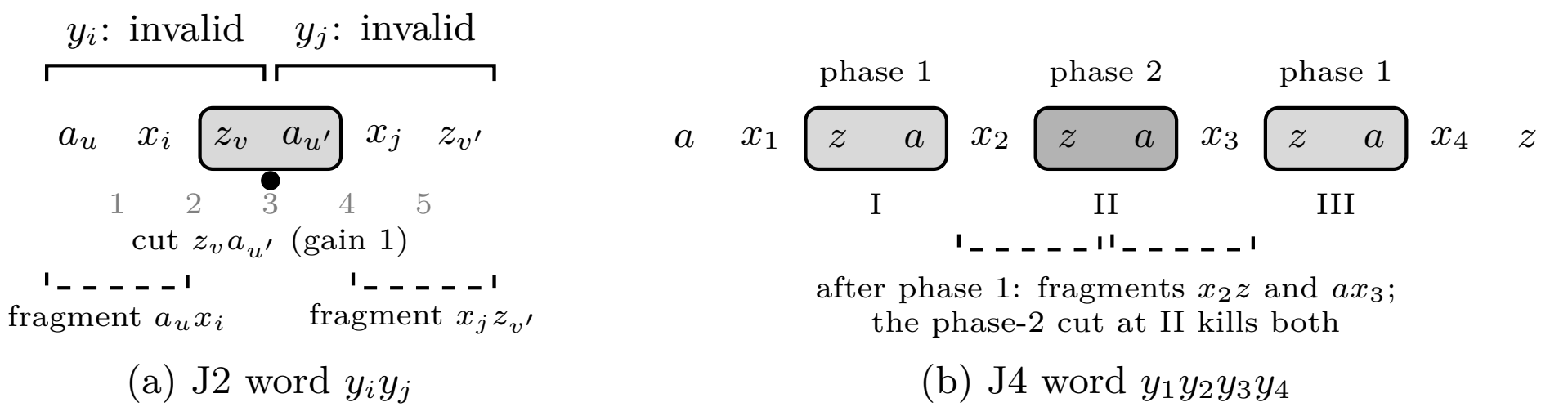

The lower-bound proof uses that a covered pair is a boundary pair of at most two disjoint optimal occurrences
(Lemma~\ref{lem:charge}).  The junction constructions realize ``two''.  Let $n\ge2$, $g=n$ and $K=n^2$.  There are $K$
\emph{planted} tokens
\[
y_j=a_{u(j)}\,x_j\,z_{v(j)}\qquad(u(j)=j\bmod n,\ v(j)=\lfloor j/n\rfloor),
\]
over pairwise distinct letters $a_0,\dots,a_{n-1}$, $z_0,\dots,z_{n-1}$ and private letters $x_0,\dots,x_{K-1}$.  Each token
covers two pairs per occurrence, each letter $a_u$ (resp.\ $z_v$) begins (resp.\ ends) exactly $g$ tokens, and each pair
$(u,v)$ belongs to exactly one token.  Every word is a concatenation of tokens.  The pair between two consecutive tokens
of a word is a \emph{junction}; its two characters are $z_va_u$, and we call the string $z_va_u$ a \emph{cut string} and say
that the junction lies on \emph{diagonal} $(u-v)\bmod n$.  If \GT{} commits the cut string $z_va_u$, the junction pair is
covered; it is the right boundary pair of the token on its left and the left boundary pair of the token on its right, so
one unit of gain invalidates two planted occurrences (Figure~\ref{fig:junction}).  We compare \GT{} with
$f(Y)$ for the planted set $Y$; since $f(Y)\le\OPT_K$, every upper bound on $\GTv_K/f(Y)$ is one on
$\GTv_K/\OPT_K$.  Theorem~\ref{thm:opt} shows more: $Y$ is an optimal vocabulary, so the ratios below are exact
ratios to the optimum.

\paragraph{The schedule in words.}  In the two-token family below, every planted token has total occurrence mass $1$,
half as the left member of a word and half as the right member.  Cutting a junction kills the occurrences on both sides of it.  Greedy keeps
cutting as long as a cut scores at least as much as its best alternative, a whole token or a fragment of one.  The
masses below are tuned so that no alternative beats the cut at any step (in the second phase of J4 the best one ties with it exactly);
adversarial tie-breaking (or a tiny perturbation, Lemma~\ref{lem:bonus}) then makes greedy cut.  We write $g=n$ for the number of tokens that share a
first letter (or a last letter), to keep this role apart from the number of letters.

\paragraph{Phase-1 masses.}  The first phase cuts junctions diagonal by diagonal.  Write $\mathcal M$ for the mass of a token's
occurrences on one side (left or right member) whose junction is still uncut.  The best competitor of a cut is then
either a whole token, valid at the occurrences whose junction is uncut, with score $2\cdot2\mathcal M=4\mathcal M$, or a
fragment $a\,x_j$ (or $x_j\,z$), which is valid at every occurrence on the side away from the junction, with score
$\frac12+\mathcal M$ when every token has total occurrence mass $1$.  So let $C(\mathcal M)=\max(4\mathcal M,\frac12+\mathcal M)$, $\mathcal M_0=\frac12$, and while
$C(\mathcal M_r)/g<\mathcal M_r$ put $m_r=C(\mathcal M_r)$ and $\mathcal M_{r+1}=\mathcal M_r-m_r/g$.  This produces $R_1$ rounds; finally scale all $m_r$ by a common
factor $\theta\ge1$ so that $\sum_rm_r=\frac g2$ (Appendix~\ref{app:masses}).  As $g\to\infty$, $R_1/g\to\tau_*=\int_0^{1/2}
\frac{d\mathcal M}{C(\mathcal M)}=\frac{\ln3}4+\ln\frac43\approx\TauStar$ (Lemma~\ref{lem:masses}).

\subsection{Two-token words: the family J2}
For round $r<R_1$ and $v<n$ let $u=v+r\bmod n$ (diagonal $r$), let $L_v$ be the tokens with $z$-letter $v$ and $R_u$ the
tokens with $a$-letter $u$, each in increasing order ($|L_v|=|R_u|=g$), and let $d=2$.  For $h<g$ and $t<d$ the corpus
contains the word $y_{L_v[h]}\,y_{R_u[(h+t)\bmod g]}$ with count $m_r/(gd)$ (scaled to integers).  In round $r$
every token is the left member of $d$ words and the right member of $d$ words, each of count $m_r/(gd)$, so each side of
each token receives $m_r/g$ per round, and every token has total occurrence mass $1$.  The run is: phase~1, the cut
strings round by round ($nR_1$ picks); phase~2, the fragments $x_jz_{v(j)}$ ($K-nR_1$ picks).

\begin{theorem}[Family J2]\label{thm:j2}
For every $n\ge10$ the run above is a valid \GT{} run with budget $K=n^2$ (every pick is a maximizer over all
candidates; ties broken adversarially), and with $Y=\{y_0,\dots,y_{K-1}\}$
\begin{gather*}
\GTv_K=\frac K2+\frac{K-nR_1}2,\qquad f(Y)=2K,\\
\frac{\GTv_K}{\OPT_K}\le\frac12-\frac{R_1}{4n}
\ \xrightarrow{\ n\to\infty\ }\ \frac12+\frac3{16}\ln3-\frac12\ln2\approx\JTwoLimSix .
\end{gather*}
\end{theorem}
The proof (Appendix~\ref{app:j2}) checks that every cut of round $r$ scores $m_r$ while every competitor scores at most
$m_r$: whole tokens and fragments are bounded by $C(\mathcal M)$, and the only strings that cross a junction and contain a private
letter occur in few words.  J2 wastes its second phase: for the last $\approx44\%$ of the budget each pick harvests a
fragment worth one pair per occurrence.

\subsection{Four-token words: the family J4}
In J4, the fragments left after phase~1 sit next to a junction that is still uncut, so phase~2 can keep cutting
junctions (Figure~\ref{fig:junction}b).  The words of J2 become \emph{blocks}; two blocks are joined into a four-token word
$y_{i_1}y_{i_2}\,y_{i_3}y_{i_4}$ whose middle junction lies on one of the diagonals $R_1,\dots,n-1$ not used by
phase~1.  Phase~2 has $R_2=n-R_1$ rounds with masses $\mu_r=\frac12(1-\frac1g)^r$: round $r$ cuts the $n$ middle junctions
on diagonal $R_1+r$.  The joining counts are chosen (Appendix~\ref{app:j4}) so that every block keeps its total usage and
every token appears in position $2$ (and in position $3$) of round-$r$ words with mass exactly $\mu_r/g$.

\begin{theorem}[Family J4]\label{thm:j4}
For every $n\ge24$ the J4 corpus (full joining) has a valid \GT{} run with budget $K=n^2$ (adversarial ties) that
commits only cut strings, and with $Y=\{y_0,\dots,y_{K-1}\}$
\begin{gather*}
\GTv_K=\frac K2+Kw_0,\qquad f(Y)=2K,\\
\frac{\GTv_K}{\OPT_K}\le\frac14+\frac{w_0}2=\frac12-\frac14\Big(1-\frac1n\Big)^{n-R_1}
\ \xrightarrow{\ n\to\infty\ }\ \JLIM\approx\JFourLimSeven,
\end{gather*}
where $w_0=\frac12(1-(1-\frac1n)^{n-R_1})$.  Its gains are non-increasing and every score after the run is smaller than the
last gain.  Consequently, for every $k\ge K$, writing $\nu=\lfloor k/K\rfloor$, there are corpora on which some \GT{} run
with budget $k$ has ratio at most $(1+\frac1\nu)(\frac14+\frac{w_0}2)$, and at most $\frac14+\frac{w_0}2$ when $K$ divides $k$;
hence $\limsup_{k\to\infty}\rho_k\le\JLIM$.
\end{theorem}
The proof is in Appendix~\ref{app:j4}.  We replayed J4 instances exactly only for $n\le20$, mostly with a smaller cyclic joining
(Appendix~\ref{app:j4}; the full joining was replayed at $n=10$ and $n=12$), because the full joining at $n=24$ already
has about $2$ million words; the statement for $n\ge24$ rests on the written proof, whose numeric side conditions were
checked exactly.
\begin{proof}[Proof sketch]
In phase~1 everything is as in J2, because the middle junctions are still uncut.  After phase~1 every whole token is
invalid, and a token $j$'s fragments $x_jz$ and $a\,x_j$ score $(\frac12-w_0)+\mathcal M$, where $\mathcal M$ is the mass of its position-$2$
(resp.\ position-$3$) occurrences whose middle junction is still uncut.  Each phase-2 round lowers $\mathcal M$ by $\mu_r/g$ for
every token at once, so the best fragment scores exactly $C_r=\frac12(1-\frac1g)^r=\mu_r$ at the start of round $r$, which
equals the score of the round-$r$ cuts.  Strings crossing junctions score at most $11/(2g)$ or $2m_0/g$, both below $\frac14\le
\mu_r$ for $n\ge24$.  The limit uses $e^{\tau_*}=\frac43\,3^{1/4}$.  The extension to all $k$ is Lemma~\ref{lem:copies}
(disjoint copies).  The full proof is in Appendix~\ref{app:j4}.
\end{proof}

\begin{remark}[Re-encoding]\label{rem:reencode}
The J4 run commits only cut strings.  A cut string $z_va_u$ occurs only at junctions, occurrences at different junctions
are disjoint, and every occurrence is valid when the string is committed (its boundary pairs are token-internal, and
only cut strings are ever committed), so the run commits every occurrence of every chosen string.  Hence re-encoding
optimally with the chosen set $S_K$ gives $f(S_K)=G_K$, and the bounds of Theorem~\ref{thm:j4} also hold for
$f(S_K)$.  The same holds on disjoint copies and for the tie-free corpora of Theorem~\ref{thm:unique} below, whose
added words are themselves chosen strings.  We do not claim this for J2, whose second phase commits fragments.
\end{remark}

\begin{figure}[t]
\centering
\begin{tikzpicture}[x=9cm,y=4cm,font=\small]
  \fill[gray!12] (0,0) rectangle (0.5623,1.05);
  \node[font=\scriptsize] at (0.28,1.0) {phase 1: cuts inside blocks};
  \node[font=\scriptsize] at (0.78,1.0) {phase 2: cuts between blocks (J4)};
  \draw[->] (0,0) -- (1.05,0) node[right] {$t$};
  \draw[->] (0,0) -- (0,1.12);
  \node[rotate=90,font=\scriptsize] at (-0.17,0.55) {gain $/\,(f(Y)/K)$};
  \foreach \x in {0,0.25,0.5,0.75,1} \draw (\x,0) -- (\x,-0.03) node[below,font=\scriptsize] {\x};
  \foreach \y in {0,0.25,0.5,0.75,1} \draw (0,\y) -- (-0.012,\y) node[left,font=\scriptsize] {\y};
  \draw[very thick,domain=0:0.2747,samples=60] plot (\x,{exp(-4*\x)});
  \draw[very thick,domain=0.2747:1,samples=60] plot (\x,{exp(-(\x-0.2747))/3});
  \draw[thick,dashed] (0.5623,0.25) -- (1,0.25);
  \draw[densely dotted] (0.2747,0) -- (0.2747,0.3333);
  \draw[densely dotted] (0.5623,0) -- (0.5623,0.25);
  \node[font=\scriptsize,anchor=west] at (0.2747,0.05) {$t_1$};
  \node[font=\scriptsize,anchor=west] at (0.5623,0.05) {$\tau_*$};
  \node[font=\scriptsize,anchor=west] at (0.62,0.11) {J4 run $=$ extremal recursion};
  \node[font=\scriptsize,anchor=west] at (0.70,0.30) {J2 run (harvest)};
\end{tikzpicture}
\caption{Normalized greedy gain $g_t/(f(Y)/K)$ against the fraction $t$ of the budget used, in the limit $n\to\infty$.
Solid: the J4 run.  Its gain coincides with the derivative $y'$ of the extremal solution of $y'=\max(1-4y,\frac12-y)$,
the lower-bound recursion of Theorem~\ref{thm:short} (the other two lines of $\mathcal A$ are dominated): $y'=e^{-4t}$ until
$t_1=(\ln3)/4$, then $\frac13e^{-(t-t_1)}$.  The switch at $t_1$ happens inside phase~1, when the best competitor of a cut
changes from a whole token (score $4\mathcal M$) to a fragment (score $\frac12+\mathcal M$).  The area under the curve is the ratio,
$\JLIM\approx\JFourLimSeven$.  Dashed: J2, which harvests fragments at a flat gain $\frac14$ after $\tau_*$, for ratio
$\approx\JTwoLimSix$.}
\label{fig:traj}
\end{figure}
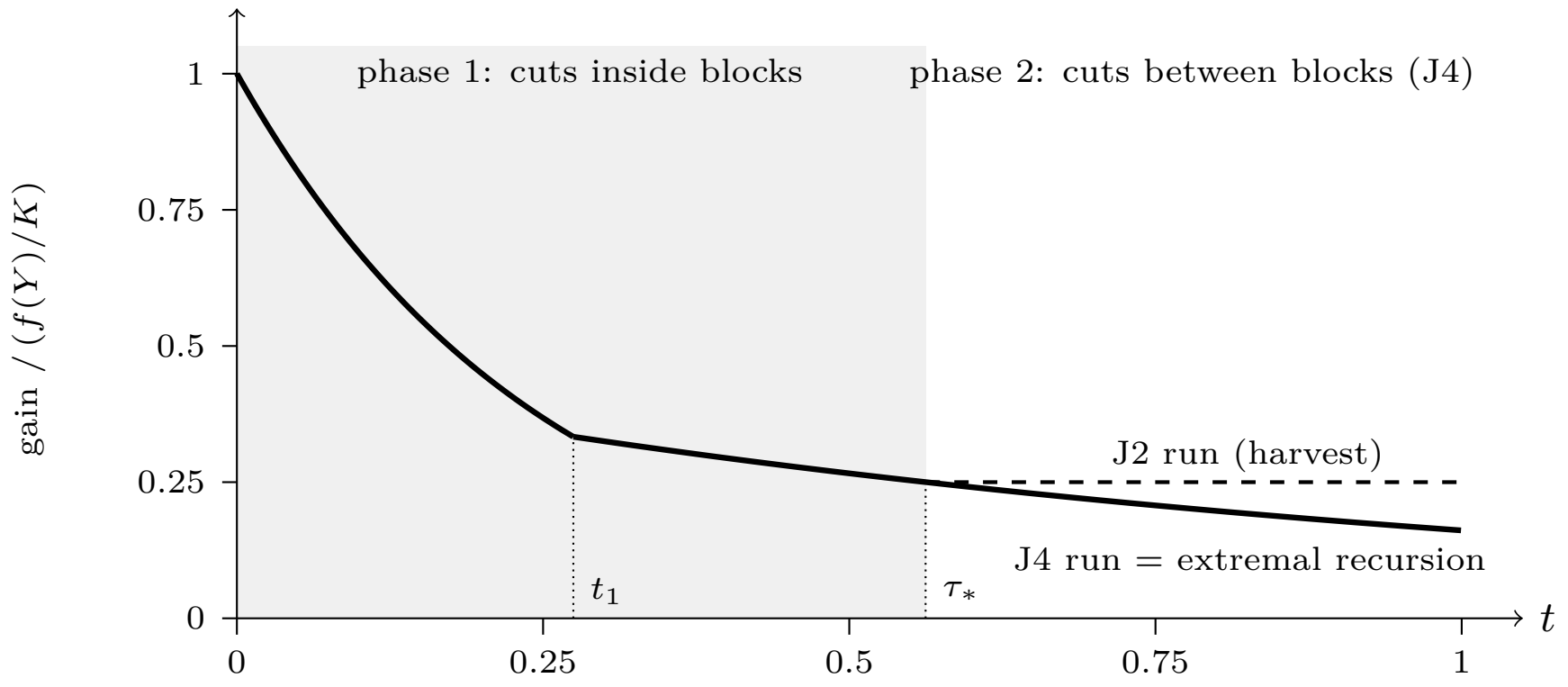

\subsection{The planted tokens are optimal}
The upper bounds above need only $\OPT_K\ge f(Y)$.  In fact the planted tokens cannot be beaten, so the bounds are not an artifact of underestimating the optimum.  A competing
vocabulary could use cut strings, which are frequent, or long strings across junctions, but each of these either
forfeits planted tokens or occurs too rarely.

\begin{theorem}[Optimality of the planted tokens]\label{thm:opt}
Let $n\ge10$.  On the J2 corpus, and on the J4 corpus with any joining that satisfies the usage identities (U1) and (U2)
of Appendix~\ref{app:j4} (in particular the full joining whenever it is defined, which includes every $n\ge16$, and the cyclic joining
with fewer words used for the replayed instances), every set of at most $K=n^2$ strings $T$ has
$f(T)\le f(Y)$; hence $\OPT_K=f(Y)$.  On $\nu$ disjoint copies with budget $\nu K$, $\OPT_{\nu K}=\nu f(Y)$.
\end{theorem}
The proof (Appendix~\ref{app:opt}) splits a competing vocabulary into substrings of planted tokens, cut strings and
strings that cross a junction, and shows that cut strings and crossing strings are worth at most a planted token
each, except for the $n$ most frequent cut strings, which are worth slightly more but each one forces $n-1$ planted
tokens out.  Consequently, at budget $K$ (and $\nu K$ on $\nu$ copies), the ratios in Theorems~\ref{thm:j2}
and~\ref{thm:j4} are exactly $\GTv_K/\OPT_K$ for the corpora with ties (Theorem~\ref{thm:opt} does not cover the
corpora with added words of Theorem~\ref{thm:unique}).  The certified
instance with $K=100$ in Table~\ref{tab:up} has $\GTv_{100}/\OPT_{100}=\JFourHundredFrac\approx\JFourHundred$.  As an independent
check, exact LP-duality certificates (rational arithmetic) prove $\OPT_K=f(Y)$ on J2 for $n\in\{\OcertJtwo\}$ and on
cyclic J4 for $n\in\{\OcertJfour\}$, including some $n<10$, which the theorem does not cover.  The statement is for budget
$K$ and its multiples; for much smaller budgets the optimum is larger than the planted value (see the remark in
Appendix~\ref{app:opt}).

\subsection{Tightness for short tokens}
Theorems~\ref{thm:short}, \ref{thm:j4} and~\ref{thm:opt} together determine the worst case for short optimal tokens.
If some optimal vocabulary has tokens of at most three characters, \GT{} keeps at least $\JLIM$ of the optimum for every
budget.  The J4 corpora have such a vocabulary (their planted tokens, by Theorem~\ref{thm:opt}), and along $k=n^2$
their ratio tends to the same constant.  The two analyses meet step by step: in the limit the J4 run has normalized gain
$e^{-4t}$ until $t_1=\frac{\ln3}4$ and $\frac13e^{-(t-t_1)}$ afterwards, which is exactly the extremal solution of the
recursion of Theorem~\ref{thm:short} (Figure~\ref{fig:traj}), so in the worst case the window-charging argument of
Appendix~\ref{sec:proof} loses nothing for such tokens.

\subsection{Unique choices}
Adversarial ties are not essential.  Every pick in J2 and J4 is a two-character string, and such runs can be made
tie-free by adding one tiny word per pick.

\begin{lemma}[Bonus words]\label{lem:bonus}
Let $\sigma_1,\dots,\sigma_m$ be a \GT{} run on a corpus $\mathcal C$ (each $\sigma_t$ a maximizer, ties allowed) in which every
$\sigma_t$ has length $2$.  Fix $\varepsilon_1>\varepsilon_2>\dots>\varepsilon_m>0$ and let $\mathcal C'$ be $\mathcal C$ together with, for each $t$, a new word
$\sigma_t$ of count $\varepsilon_t$.  Then on $\mathcal C'$ with budget $m$ the maximizer is unique at every step, the run is
$\sigma_1,\dots,\sigma_m$, its value is the value of the run on $\mathcal C$ plus $\sum_t\varepsilon_t$, and $f_{\mathcal C'}(S)\ge f_{\mathcal C}(S)$ for
every $S$.
\end{lemma}

\begin{theorem}[Unique choices]\label{thm:unique}
For every $n\ge24$, every $k\ge K=n^2$ and every $\eta>0$ there is a corpus on which \GT{} with budget $k$ has a unique
maximizer at every step and ratio at most $(1+\frac1\nu)(\frac14+\frac{w_0}2)+\eta$, with $\nu=\lfloor k/K\rfloor$, and at most
$\frac14+\frac{w_0}2+\eta$ when $K$ divides $k$, with $w_0$ as in Theorem~\ref{thm:j4}.  Hence the worst-case ratio over tie-free instances also has limit superior at
most $\JLIM$.
\end{theorem}
\noindent The proofs of Lemma~\ref{lem:bonus} and Theorem~\ref{thm:unique} are in Appendix~\ref{app:unique}.

Each added word changes the value of any vocabulary by at most its count, so when $K$ divides $k$ the optimum of the
tie-free corpus is at most $\sum_t\varepsilon_t$ above the optimum $\nu f(Y)$ of the corpus with ties
(Theorem~\ref{thm:opt}); the ratio of Theorem~\ref{thm:unique} is then exact against the optimum up to a factor $1+\eta$.

The same transfer applies to every exactly verified adversarial-tie run in Table~\ref{tab:up} whose picks have length
$2$, which includes all J2 and J4 rows.  The search-based tie-free J2/J4 instances in the table, obtained by perturbing
counts, show the same values directly.

The classical tight example for greedy maximum coverage also embeds into the problem with unique choices; its ratio tends
to $1-1/e$ (Proposition~\ref{prop:covfam}, Appendix~\ref{app:covfam}).

\paragraph{Small budgets.}
The junction idea also gives the best instances we know for small $k$; for instance $\rho_4\le\frac7{16}$
(Proposition~\ref{prop:smallk}) and, from a separate chain corpus, $\rho_2\le\KtwoTie$.  With Theorem~\ref{thm:main},
$\rho_2\in[\frac12,\KtwoTie]$, $\rho_3\in[\frac7{18},\frac12]$ and $\rho_4\in[\frac{239}{640},\frac7{16}]$.  Appendix~\ref{app:smallk-sec} gives these constructions and a
table of exactly verified instances, including a $k=100$ J4 run with ratio exactly
$\JFourHundredFrac\approx\JFourHundred$ to the optimum.

\subsection*{Proofs and further constructions}
\input{gt/smallk}
\input{gt/junction}
\input{gt/family}
\input{gt/covfam}

\input{gt/optimal_app}
\input{gt/unique_app}

%% file: gt/smallk.tex
\subsection{Small budgets and explicit instances}\label{sec:smallk}\label{app:smallk-sec}
The junction idea also gives the best instances we know for small $k$.

\begin{proposition}[Small budgets]\label{prop:smallk}
For every $k\ge2$ there is a corpus with a \GT{} run (adversarial ties) of ratio at most $(k+3)/(4k)$ for $k\le4$ (namely
$\frac58,\frac12,\frac7{16}$) and at most $(2k-1)/(4k)<\frac12$ for $k\ge4$.  Its gains satisfy $g_1\ge g_2=\dots=g_k$ and every
score after the run is at most $g_k$, so the ratio is at most $\frac7{16}$ at every multiple of $4$.
\end{proposition}
The corpus has optimal tokens $y_i=b\,c_i\,a$, all words $y_iy_jy_l$, and the singles $y_i$; \GT{} first commits the
junction string $ab$ (proof in Appendix~\ref{app:smallk}).  For $k=2$ a different corpus does better: a $\KtwoWords$-word
corpus over $\{a,b,c\}$ on which \GT{} first commits $bbb$ has worst tie-break ratio exactly $\KtwoTie$ (with
$\OPT_2=f(\{bbcb,bbab\})$ computed exactly), and a different, tie-free $15$-word corpus of the same shape has ratio $\KtwoUni$.  Hence
\[
\rho_2\in\big[\tfrac12,\ \KtwoTie\big],\qquad \rho_3\in\big[\tfrac7{18},\ \tfrac12\big],
\]
where the lower bounds are Proposition~\ref{prop:easy} and the value $\gamma_3=\frac7{18}$ of Theorem~\ref{thm:main}.  The best tie-free instances we have for $k=3,4$ are perturbations
of Proposition~\ref{prop:smallk} with ratios $\KthreeUni$ and $\KfourUni$; they are exact instances (at perturbation scale
$M=100$), and we do not claim a tie-free version of Proposition~\ref{prop:smallk} for general $k$ (its picks $abc_i$ have
length $3$, so Lemma~\ref{lem:bonus} does not apply).

\paragraph{Explicit instances.}
Table~\ref{tab:up} lists exactly verified instances.  The ratio is \GT{}'s value against the value of an explicit
$k$-token set, a lower bound on $\OPT_k$, so each entry is a certified upper bound on that instance's ratio.  For the
worst-ties J2 and J4 rows ($n\ge10$) the explicit set is the planted one, so by Theorem~\ref{thm:opt} the entry is the
exact ratio $\GTv_K/\OPT_K$ of that run.  The J4 rows
with $n<24$ use the cyclic joining of Appendix~\ref{app:j4}; the worst-ties J2 rows equal the formula of Theorem~\ref{thm:j2}, and the
worst-ties J4 rows equal the formula $\frac12-\frac14(1-\frac1n)^{n-R_1}$ of Theorem~\ref{thm:j4}, although $n<24$ lies outside
the theorem.  No instance with $n\ge24$ was replayed (the full joining at $n=24$ has about $2$ million words);
Theorem~\ref{thm:j4} for $n\ge24$ rests on its written proof, whose numeric side conditions were checked exactly.  The smallest certified ratio on a single instance is
$\JBestTieFrac\approx\JBestTie$ ($k=\JBestTieK$, $\JBestTieWords$ words).

\begin{table}[!htb]\centering\small
\input{gt/upper_table.tex}
\caption{Exactly verified upper-bound instances.  ``Worst ties'': a run in which every pick is a maximizer over all
candidates, with adversarial tie-breaking (for $k\le4$ the full tie tree was enumerated; for larger $k$ the value is that
of one exactly verified run, an upper bound on the worst tie-break).  ``Unique choices'': the maximizer is unique at
every step.  For $k=2,\dots,5$ the comparison set is optimal (exact $\OPT_k$).  J2/J4 unique-choice rows come from a
randomized count perturbation; Lemma~\ref{lem:bonus} gives tie-free instances within any $\eta>0$ of every tie row with
length-$2$ picks.  All entries bound the committed value $G_k$; for the J4 rows this equals the re-encoded value
$f(S_k)$ (Remark~\ref{rem:reencode}), while the J2 rows bound $G_K$ only.}\label{tab:up}
\end{table}

%% file: gt/upper_table.tex
\begin{tabular}{rllcr}\toprule $k$ & construction & regime & $\mathrm{GT}_k/\OPT_k\le$ & words\\\midrule
2 & chain & worst ties & $7/12\approx0.5833$ & 17\\
2 & chain & unique choices & $0.587669$ & 15\\
3 & junction & worst ties & $1/2\approx0.5000$ & 31\\
3 & junction & unique choices & $10811/21606\approx0.5004$ & 31\\
4 & junction & worst ties & $7/16\approx0.4375$ & 68\\
4 & junction & unique choices & $22419/51212\approx0.4378$ & 69\\
5 & junction & worst ties & $9/20\approx0.4500$ & 130\\
100 & J4 & worst ties & $0.367140$ & 5,600\\
100 & J2 & worst ties & $2/5\approx0.4000$ & 800\\
100 & J4 & unique choices & $0.368902$ & 5,600\\
100 & J2 & unique choices & $0.400001$ & 800\\
144 & J4 & worst ties & $0.364038$ & 11,520\\
144 & J2 & worst ties & $19/48\approx0.3958$ & 1,440\\
196 & J4 & worst ties & $0.351185$ & 21,952\\
256 & J4 & worst ties & $0.350821$ & 36,864\\
256 & J4 & unique choices & $0.352404$ & 36,864\\
400 & J4 & worst ties & $0.350316$ & 88,000\\
400 & J2 & worst ties & $3/8\approx0.3750$ & 8,000\\
900 & J2 & worst ties & $11/30\approx0.3667$ & 28,800\\
\bottomrule\end{tabular}

%% file: gt/junction.tex
\subsection{Phase-1 masses}\label{app:masses}
Recall $C(\mathcal M)=\max(4\mathcal M,\frac12+\mathcal M)$, which is non-decreasing, $4$-Lipschitz and at least $\frac12$, and the recursion of
Appendix~\ref{sec:upper}: $\mathcal M_0=\frac12$; while $C(\mathcal M_r)/g<\mathcal M_r$, put $m_r=C(\mathcal M_r)$ and $\mathcal M_{r+1}=\mathcal M_r-m_r/g$.  Let $R_1$ be the
number of rounds produced, $\mathcal M_*=\mathcal M_{R_1}$ the leftover (so $0\le \mathcal M_*\le C(\mathcal M_*)/g$), and finally multiply every $m_r$ by
$\theta=1/(1-2\mathcal M_*)$, so that $\sum_rm_r=\frac g2$.  Write $\mathcal M'_r=\frac12-\sum_{s<r}m_s/g$ (with the scaled masses) for the
per-side uncut mass at the start of round $r$, and $\tau_*=\int_0^{1/2}d\mathcal M/C(\mathcal M)=\frac{\ln3}4+\ln\frac43\approx\TauStar$.

\begin{lemma}\label{lem:masses}
For $g\ge10$:
(i) $m_0\ge m_1\ge\dots\ge m_{R_1-1}$;
(ii) $m_r\ge C(\mathcal M'_r)$ for every $r<R_1$, and $\mathcal M'_{R_1}=0$;
(iii) $R_1\le\tau_*g$; in particular $R_1<g$;
(iv) $m_0\le 2(g-1)/(g-2)\le g/4$;
(v) $R_1\ge(g-4)\big(\tau_*-\frac1{g-1}\big)$; hence $R_1/g\to\tau_*$.
\end{lemma}
\begin{proof}
(i) $\mathcal M_r$ decreases and $C$ is non-decreasing; scaling preserves the order.
(ii) Scaling only increases the masses, so $\mathcal M'_r\le \mathcal M_r$ and $C(\mathcal M'_r)\le C(\mathcal M_r)=m_r/\theta\le m_r$; and
$\mathcal M'_{R_1}=\frac12-\frac{g/2}g=0$.
(iii) Since $C$ is non-decreasing, $\int_{\mathcal M_{r+1}}^{\mathcal M_r}d\mathcal M/C\ge(\mathcal M_r-\mathcal M_{r+1})/C(\mathcal M_r)=1/g$ for every $r<R_1$; summing,
$R_1/g\le\int_{\mathcal M_*}^{1/2}d\mathcal M/C\le\tau_*$.  The integral is $\int_{1/6}^{1/2}\frac{d\mathcal M}{4\mathcal M}+\int_0^{1/6}\frac{d\mathcal M}{1/2+\mathcal M}
=\frac{\ln 3}4+\ln\frac43$.
(iv) If $\mathcal M_*\ge\frac16$, termination reads $\mathcal M_*\le4\mathcal M_*/g$, impossible for $g>4$.  So $C(\mathcal M_*)=\frac12+\mathcal M_*$ and
$\mathcal M_*\le\frac1{2(g-1)}$, hence $\theta\le\frac{g-1}{g-2}$ and $m_0=\theta C(\frac12)=2\theta$.  Finally
$2(g-1)/(g-2)\le g/4$ is $g^2-10g+8\ge0$, true for $g\ge10$.
(v) $C(\mathcal M_{r+1})\ge C(\mathcal M_r)-4(\mathcal M_r-\mathcal M_{r+1})=C(\mathcal M_r)(1-4/g)$, so
$\int_{\mathcal M_{r+1}}^{\mathcal M_r}d\mathcal M/C\le(\mathcal M_r-\mathcal M_{r+1})/C(\mathcal M_{r+1})\le1/(g-4)$.  Summing over $r<R_1$ and adding
$\int_0^{\mathcal M_*}d\mathcal M/C\le2\mathcal M_*\le\frac1{g-1}$ gives $\tau_*\le R_1/(g-4)+\frac1{g-1}$.
\end{proof}

\subsection{Proof of Theorem~\ref{thm:j2} (family J2)}\label{app:j2}
Index the six characters of a word $y_iy_j$ as $0,\dots,5$; its pairs are $1,\dots,5$, and pair $3$ is the junction.
During the run only cut strings and harvest strings $x_jz$ are committed; cut strings cover junction pairs only, and
harvest strings cover pair~$5$ (see below).  Every substring of length $\ge2$ of a word is of exactly one of three
kinds, because the letters $a_u$, $x_j$, $z_v$ are pairwise distinct and are adjacent only as $a x$, $xz$ and $za$:
\begin{itemize}
\item[(K1)] a cut string $z_va_u$;
\item[(K3)] a substring of a token: $a_{u(j)}x_j$, $x_jz_{v(j)}$, or $y_j$;
\item[(K4)] a \emph{crossing string}: it contains the junction pair and at least one private letter ($x_i$ of the left
token or $x_j$ of the right token).
\end{itemize}

\emph{Crossing strings.}  A crossing string containing $x_iz_{v(i)}a_u$ occurs only in words whose left token is $i$
and whose junction is $(v(i),u)$, which belong to a single round $r$; so its total count is at most $d\cdot\frac{m_r}{gd}
=\frac{m_r}g$.  If it contains no other private letter, it has at most $3$ interior pairs ($a\,x_i\,z\,a$).  Otherwise it
contains both $x_i$ and $x_j$ and occurs only in the single word $(i,j)$, of count $\frac{m_r}{gd}$, with at most $5$
interior pairs.  Strings containing $z_va_ux_j$ are symmetric.  Hence, with $d=2$, the score of a crossing string at a
round-$r$ junction is at most $3m_r/g$ while the junction is uncut, and at most $2m_r/g$ once it is cut (the junction pair
is then covered and contributes no gain).

\emph{Token substrings.}  For token $j$ let $\lambda_j$ ($\lambda'_j$) be the mass of its occurrences as a right (left)
member whose junction is uncut.  While no harvest string of $j$ is committed: $y_j$ scores $2(\lambda_j+\lambda'_j)$ (as a left
member it is valid iff the junction is uncut, since its left end is the word start, and symmetrically as a right
member); $a_{u(j)}x_j$ scores $\frac12+\lambda_j$ (valid in every left-member occurrence; in a right-member occurrence its left
boundary pair is the junction); $x_jz_{v(j)}$ scores $\frac12+\lambda'_j$.

\emph{Phase 1, round $r$.}  At the start of round $r$ every token has $\lambda_j=\lambda'_j=\mathcal M'_r$, since each completed round $s$
lowered each side of each token by $m_s/g$.  During round $r$ scores can only decrease (Corollary~\ref{cor:mono}), so it
suffices to bound the competitors at the start of the round: token substrings score $\le C(\mathcal M'_r)\le m_r$
(Lemma~\ref{lem:masses}(ii)); unchosen cut strings of round $s\ge r$ score $m_s\le m_r$ (Lemma~\ref{lem:masses}(i));
crossing strings at uncut junctions score $\le 3m_s/g\le m_r$; crossing strings at cut junctions score
$\le2m_0/g\le\frac12\le m_r$ (Lemma~\ref{lem:masses}(iv) and $m_r\ge C\ge\frac12$).  The cut string being picked scores exactly
$m_r$: it gains $1$ in each of its $gd$ words of count $m_r/(gd)$, its junction is uncovered there, and its boundary pairs
(pairs $2$ and $4$) are never covered during phase 1.  So every phase-1 pick is a maximizer, and phase 1 gains
$\sum_r n\,m_r=n\frac g2=\frac K2$.

\emph{Phase 2.}  After phase 1 every junction is covered, since the words use only the diagonals $0,\dots,R_1-1$.  Now
$y_j$ scores $0$; $a_{u(j)}x_j$ and $x_jz_{v(j)}$ score exactly $\frac12$ ($\lambda_j=\lambda'_j=0$); crossing strings score $\le
2m_0/g\le\frac12$; all cut strings are chosen.  The harvest string $x_jz_{v(j)}$ still scores $\frac12$ when it is picked: its
occurrences at pairs $4$--$5$ of right-member words are valid (pair $4$ could only be covered by a commit containing $a\,x_j$,
and none is made; the right end is the word end), and those at pairs $1$--$2$ of left-member words are invalid (their right
boundary is the covered junction).  Earlier harvest commits $x_{j'}z$ cover only pair $5$ of words whose right token is
$j'\ne j$.  By Corollary~\ref{cor:mono} every score stays $\le\frac12$.  So each harvest pick is a maximizer of gain
$\frac12$; there are $K-nR_1\ge1$ of them (Lemma~\ref{lem:masses}(iii)).

\emph{Values.}  $\GTv_K=\frac K2+\frac{K-nR_1}2$.  Every token has total occurrence mass $1$ and every word contains two tokens,
so the total word count is $K/2$; $Y$ covers pairs $1,2,4,5$ of every word, so $f(Y)=4\cdot\frac K2=2K$, and
$\GTv_K/f(Y)=\frac12-\frac{nR_1}{4K}=\frac12-\frac{R_1}{4n}$.  The limit follows from Lemma~\ref{lem:masses}(v):
$\frac12-\frac{\tau_*}4=\frac12+\frac3{16}\ln3-\frac12\ln2$.  Gains are non-increasing, and after the run every score is
$\le\frac12$, the last gain, so Lemma~\ref{lem:copies} applies.\qed

\subsection{Proof of Theorem~\ref{thm:j4} (family J4)}\label{app:j4}
\paragraph{Construction.}  Take the J2 word list (rounds $\xi<R_1$, masses $m_\xi$, $d=2$) and call each listed word a
\emph{block} $B=(i_1,i_2)$ of round $\xi$, with weight $c(B)=m_\xi/(gd)$.  Put $R_2=g-R_1$, $\mu_r=\frac12(1-\frac1g)^r$
for $r<R_2$, and $w_0=\sum_{r<R_2}\mu_r/g=\frac12\big(1-(1-\frac1g)^{R_2}\big)$.  A four-token word
$(B,B')=y_{i_1}y_{i_2}y_{i_3}y_{i_4}$, with $B=(i_1,i_2)$ and $B'=(i_3,i_4)$, is \emph{of round $r$} if its middle junction
$(v(i_2),u(i_3))$ lies on diagonal $R_1+r$; since phase-1 junctions use diagonals $0,\dots,R_1-1$ and phase-2 junctions use
$R_1,\dots,n-1$, all junction letter pairs of the two phases are distinct.

\emph{Full joining.}  For every $(r,\xi',\xi,v_2)$ match bijectively the $gd$ blocks of round $\xi'$ whose second token
has $z$-letter $v_2$ with the $gd$ blocks of round $\xi$ whose first token has $a$-letter $v_2+R_1+r$ (both sets have $gd$
elements: a token is the second, resp.\ first, member of $d$ blocks per round, and $g$ tokens have a given letter).  The
matched pair gets the four-token word $(B,B')$ with count $c(B,B')=c(B)c(B')\omega_r$, $\omega_r=4d\mu_r/g$, and each block
$B$ additionally appears as a standalone two-token word with count $s(B)=c(B)(1-4w_0)$.

\emph{Usage identities.}
(U1) Block $B$ meets, as a first half, exactly one partner for every $(\xi,r)$, of total count
$c(B)\sum_\xi\frac{m_\xi}{gd}\sum_r\omega_r=c(B)\frac{g/2}{gd}\cdot4dw_0=2w_0c(B)$, and the same as a second half; with
$s(B)$ its total usage is $c(B)$.  We need $s(B)\ge0$, i.e.\ $w_0\le\frac14$ (shown below).
(U2) For every token $j$ and round $r$, the total count of round-$r$ words with $i_2=j$ is
$\sum_{\xi'}d\frac{m_{\xi'}}{gd}\sum_\xi\frac{m_\xi}{gd}\omega_r=\frac12\cdot\frac1{2d}\cdot\omega_r=\mu_r/g$, and the same holds for
$i_3=j$.
Consequently, for every token $j$: its mass as the left member of a phase-1 junction is $m_\xi/g$ in round $\xi$ (the $d$
blocks with first member $j$, each with total usage $c(B)$), and likewise as a right member, so each side has total mass
$\frac12$; its mass at position $2$ and at position $3$ of four-token words is $w_0$ each, by (U2); hence its mass at
position $4$ plus standalone-right is $\frac12-w_0$, and at position $1$ plus standalone-left is $\frac12-w_0$.

\paragraph{The run.}  Phase 1: the phase-1 cut strings in the J2 order ($nR_1$ picks).  Phase 2: for $r=0,\dots,R_2-1$,
the $n$ cut strings on diagonal $R_1+r$ ($nR_2$ picks).  Total $n(R_1+R_2)=K$.

\paragraph{Proof of Theorem~\ref{thm:j4}.}  Number the pairs of a four-token word
\[1,2\,|\,3\,(J_1)\,|\,4,5\,|\,6\,(J_2)\,|\,7,8\,|\,9\,(J_3)\,|\,10,11.\]  Only cut strings are committed, so token-internal pairs are never covered.  The substrings are:
phase-1 cut strings; phase-2 cut strings; token substrings $a x_j$, $x_jz$, $y_j$; and crossing strings, which contain a
junction pair and an adjacent private letter.

\emph{Cut strings.}  A phase-1 cut string of round $\xi$ occurs exactly at the junctions of the blocks of its junction,
wherever these blocks appear, with total count $gd\cdot\frac{m_\xi}{gd}=m_\xi$ by (U1).  A phase-2 cut string
$z_va_{v+R_1+r}$ occurs exactly at $J_2$ of round-$r$ words whose second token has $z$-letter $v$; by (U2) its count is
$g\cdot\mu_r/g=\mu_r$.  Its boundary pairs are token-internal, so it gains $1$ in each occurrence until chosen.

\emph{Crossing strings.}  A crossing string at a phase-1 junction of round $\xi$ containing $x_iz\,a$ occurs only in
blocks with first member $i$ at that junction, of total usage $\le m_\xi/g$.  With no other private letter it has $\le3$
interior pairs ($\le2$ uncovered once the junction is cut); if it also contains the right member's $x_j$ and crosses no
other junction, it occurs only in block $(i,j)$, of usage $c(B)=m_\xi/(gd)$, with $\le5$ pairs.  Any crossing string that
crosses $J_2$ contains $x_{i_2}z\,a$ or $z\,a\,x_{i_3}$ at $J_2$ (letters on a phase-2 diagonal), so by (U2) its count is
$\le\mu_r/g\le\frac1{2g}$, and it has $\le11$ interior pairs.  With $d=2$: phase-1 crossing strings score $\le3m_\xi/g$
(junction uncut) or $\le2m_\xi/g$ (cut), and $J_2$-crossing strings score $\le\frac{11}{2g}$.

\emph{Token substrings, phase 1} ($J_2$ uncut).  As in J2: $y_j$ scores $2(\lambda_j+\lambda'_j)\le4\mathcal M'_\xi$, where $\lambda_j,\lambda'_j$
are the uncut right/left-member masses at phase-1 junctions; $x_jz$ scores $\frac12+\lambda'_j$ (all right-member occurrences
are valid: at position $2$ the right boundary is the uncut $J_2$, at position $4$ and standalone it is the word end) and
$a\,x_j$ scores $\frac12+\lambda_j$.  So the maximum token-substring score at the start of round $\xi$ is $C(\mathcal M'_\xi)$.

\emph{Token substrings, phase 2} (all phase-1 junctions cut).  $y_j$ scores $0$, since every occurrence has a phase-1
junction as a boundary pair.  $x_jz$ scores $(\frac12-w_0)+\mathcal M^{(2)}_j$ (position $4$ and standalone-right, plus the
position-$2$ occurrences whose $J_2$ is uncut, of mass $\mathcal M^{(2)}_j$), and $a\,x_j$ scores $(\frac12-w_0)+\mathcal M^{(3)}_j$.  By (U2) every
completed phase-2 round $s$ lowers $\mathcal M^{(2)}_j$ and $\mathcal M^{(3)}_j$ of every token by $\mu_s/g$, so at the start of round $r$
both equal $w_0-\sum_{s<r}\mu_s/g$, and $C_r=\frac12-w_0+w_0-\sum_{s<r}\mu_s/g$ satisfies $C_0=\frac12$ and
$C_{r+1}=C_r-\mu_r/g$; with $\mu_r=C_r$ this gives $C_r=\mu_r=\frac12(1-\frac1g)^r$.

\emph{Phase 1.}  Exactly as in J2, the round-$\xi$ pick scores $m_\xi\ge C(\mathcal M'_\xi)$; other phase-1 cut strings score $m_s\le
m_\xi$; phase-2 cut strings score $\mu_r\le\frac12\le m_\xi$; phase-1 crossing strings score $\le3m_s/g\le m_\xi$ (uncut,
$s\ge\xi$) or $\le2m_0/g\le\frac12$ (cut); $J_2$-crossing strings score $\le\frac{11}{2g}\le\frac12$.  Phase 1 gains $\frac K2$.

\emph{Phase 2, round $r$.}  The pick scores $\mu_r=C_r$.  Competitors at the start of the round: token substrings $\le C_r$;
other phase-2 cut strings $\mu_s\le\mu_r$ ($s\ge r$); $J_2$-crossing strings $\le\frac{11}{2g}$; phase-1 crossing strings (all
cut) $\le2m_0/g\le\frac{4(g-1)}{g(g-2)}$.  Moreover $\mu_r\ge\mu_{R_2-1}\ge\frac12(1-\frac1g)^{R_2}=\frac12-w_0\ge\frac14$, using
$w_0\le\frac14$, which we now show: $(1-\frac1g)^{R_2}\ge e^{-R_2/(g-1)}$, and by Lemma~\ref{lem:masses}(v)
$R_2=g-R_1\le g(1-\tau_*)+4\tau_*+1\le(g-1)\ln2$ for $g\ge16$.  For $g=n\ge24$, $\frac{11}{2g}\le\frac{11}{48}<\frac14$ and
$\frac{4(g-1)}{g(g-2)}<\frac14$.  So every phase-2 pick is a maximizer, and phase 2 gains $\sum_rn\mu_r=n\,g\,w_0=Kw_0$.

\emph{Values.}  $\GTv_K=\frac K2+Kw_0$; $Y$ covers both interior pairs of every token occurrence, so $f(Y)=2K$ and
$\GTv_K/f(Y)=\frac14+\frac{w_0}2=\frac12-\frac14(1-\frac1n)^{n-R_1}$.  As $R_1/n\to\tau_*$,
$(1-\frac1n)^{n-R_1}\to e^{-(1-\tau_*)}=\frac43\,3^{1/4}e^{-1}$, which gives the limit $\frac12-3^{1/4}/(3e)$.  Gains are
non-increasing ($m_{R_1-1}\ge\frac12=\mu_0$), and after the run every score is at most
$\max\big(C_{R_2},\frac{11}{2g},\frac{2m_0}g\big)=C_{R_2}<\mu_{R_2-1}$, so Lemma~\ref{lem:copies} applies.\qed

\paragraph{Smaller instances.}  For $n<24$ the crude bounds above are not claimed.  The instances of
Table~\ref{tab:up} with $n=10,\dots,20$ use a \emph{cyclic} joining with fewer words: a block $B$ of round $\xi'$ is joined only with blocks $B'$ of round
$\xi=\xi'+r\bmod R_1$ (matched as in the full joining), with count $c(B,B')=c(B)c(B')\mu_r\,g\,d/A(r)$, where
$A(r)=\sum_{\xi}m_\xi m_{\xi+r\bmod R_1}$, and each block is also a standalone word with count $c(B)\sigma_{\xi'}$,
$\sigma_{\xi'}=1-\sum_r\mu_r(m_{\xi'+r}+m_{\xi'-r})/A(r)$ (indices mod $R_1$).  (U2) holds because
$\sum_{\xi'}m_{\xi'}m_{\xi'+r}=A(r)$, and (U1) holds by the choice of $\sigma_{\xi'}$; the construction needs every
$\sigma_{\xi'}\ge0$, which was checked exactly for each replayed instance, and every step of each run was checked exactly against all
candidates (Appendix~\ref{app:comp}).

\subsection{Proof of Proposition~\ref{prop:smallk} (small budgets)}\label{app:smallk}
Let $y_i=b\,c_i\,a$ for $i=1,\dots,k$.  The corpus $\mathcal C_k$ has the \emph{singles} $y_i$ with count $k^2$, the
\emph{triples} $y_iy_jy_l$ for all $(i,j,l)\in[k]^3$ with count $1$, and the \emph{gap word} $ab$ with count
$G=\max(0,8k^2-2k^3)$.  The set $S^*=\{y_1,\dots,y_k\}$ gives $f(S^*)=k\cdot k^2\cdot2+k^3\cdot6=8k^3$.

\emph{Step 1.}  On the empty state $\Sc_0(z)=f(\{z\})$ (proof of Proposition~\ref{prop:easy}).  The only substring of
length $\ge2$ without a token letter is $ab$, with $\Sc_0(ab)=2k^3+G=\max(2k^3,8k^2)$.  A string with exactly one token
letter $c_i$ is a substring of $a\,b\,c_i\,a\,b$; the letter $c_i$ occurs with total count $4k^2$ ($k^2$ in singles, $3k^2$ in
triples), and the scores are $8k^2$ for $b\,c_i\,a$, $4k^2$ for $bc_i$, $c_ia$, $abc_i$, $c_iab$ and $abc_iab$, and at most
$6k^2$ for $abc_ia$, $bc_iab$.  A string with two token letters contains $c_iabc_j$, occurs at most $2k$ times and has at
most $7$ pairs, so it scores $\le14k\le8k^2$; a string with three token letters occurs once and scores $\le8$.  So $ab$ is a
maximizer (unique for $k\ge5$, tied with the $y_i$ for $k\le4$), and committing it covers the two junction pairs of every
triple and the pair of the gap word; $g_1=\max(2k^3,8k^2)$.

\emph{Steps $2,\dots,k$.}  By Corollary~\ref{cor:mono} it suffices to bound the scores right after step $1$.  An occurrence's
left boundary pair is now covered iff it starts at the $b$ of token $2$ or $3$ of a triple, and its right boundary pair iff
it ends at the $a$ of token $1$ or $2$.  A one-letter string (substring of $abc_iab$) has gain at most $2$ (the pairs
$bc_i,c_ia$), and each of the eight such strings is valid only at token positions of total count $k^2$ (gain $2$) or
$2k^2$ (gain $1$): for example $bc_ia$ only in singles, $abc_i$ at triple positions $2,3$; every one scores exactly $2k^2$.
Two-letter strings live in triples at positions $(1,2)$ or $(2,3)$ and score $\le4k\le2k^2$; three-letter strings score
$\le6\le2k^2$.  The string $a\,b\,c_j$ scores $2k^2$ until chosen, since committing $abc_i$ ($i\ne j$) covers only pairs
$(b,c_i)$.  So $ab,abc_1,\dots,abc_{k-1}$ is a valid run with $g_2=\dots=g_k=2k^2$, value
$\Phi_k=\max(2k^3,8k^2)+2k^2(k-1)$, and after it $abc_k$ still scores $2k^2=g_k$.  Hence
$\GTv_k/f(S^*)\le\Phi_k/(8k^3)$, which is $(k+3)/(4k)$ for $k\le4$ and $(2k-1)/(4k)$ for $k\ge4$.\qed

%% file: gt/family.tex
\subsection{Disjoint copies}\label{app:copies}
The next two lemmas turn one bad instance at budget $k$ into bad instances at every larger budget.  In both, $\Phi$
denotes the value of a given run and $O$ the optimum of the base instance.
\begin{lemma}[Disjoint copies, adversarial ties]\label{lem:copies}
Let $(\mathcal C,k)$ be an instance with a GreedTok run (each choice a maximizer) $\sigma_1,\dots,\sigma_k$ whose
gains satisfy $g_1\ge g_2\ge\dots\ge g_k>0$, such that every score in the state after $\sigma_k$ is $\le g_k$
(write $g_{k+1}$ for the maximum such score), value $\Phi=\sum g_s$, and let $\mathrm{OPT}_k(\mathcal C)=O$. Let
$\mathcal C^{(c)}$ be $c$ copies of $\mathcal C$ on pairwise disjoint alphabets. Then for every $k'\ge k$, writing
$\nu=\lfloor k'/k\rfloor$, the worst tie-break of GreedTok on $(\mathcal C^{(\nu+1)},k')$ has value at most $(\nu+1)\Phi$, and
$\mathrm{OPT}_{k'}\ge \nu O$; for $k'=\nu k$, $(\mathcal C^{(\nu)},k')$ gives value $\le \nu\Phi$ against $\mathrm{OPT}\ge \nu O$. Hence the
worst-case ratio is $\le\frac{\nu+1}{\nu}\cdot\frac\Phi O$ for all $k'\ge k$, and $\le \Phi/O$ for all multiples of $k$.
\end{lemma}
\begin{proof}
Copies share no letter, so a token lives in one copy and its score depends only on that copy's state.  Run the copies
in lockstep: in round $s=1,\dots,k$, choose $\sigma_s$ in copy $1,2,\dots$ in turn.  When a copy is chosen in round
$s$, every copy is in the state after $s-1$ or $s$ steps of $\sigma$; the maximum score in a copy after $s-1$ steps is
$g_s$, and after $s$ steps it is $g_{s+1}\le g_s$ (for $s=k$ this is the extra hypothesis).  So the chosen token is a
global maximizer and this is a valid (adversarial) GreedTok run, in which every copy contributes exactly $\Phi$ after $k$
rounds.  Truncating it after $k'$ choices is still a valid run, because choices do not depend on the budget; its value
on $\nu+1$ copies is at most $(\nu+1)\Phi$.  For OPT, take an optimal $k$-set in each of $\nu$ copies, since $f$ is
additive over copies.
\end{proof}
\noindent The junction families satisfy the hypotheses (Appendices~\ref{app:j2} and~\ref{app:j4}), and so do the
run of Proposition~\ref{prop:smallk} and every adversarial-tie instance of Table~\ref{tab:up} (Appendix~\ref{app:comp}).

\begin{lemma}[Disjoint copies, unique choices]\label{lem:copies-strict}
Let $(\mathcal C,k)$ have a tie-free GreedTok run with gains $g_1>g_2>\dots>g_k>g_{k+1}$, where $g_{k+1}$ is the maximum
score after the run, value $\Phi$, $\mathrm{OPT}_k=O$, and let $\eta=\min_{s\le k} g_s/g_{s+1}>1$. For $c\ge1$ and an integer
$N>(c-1)/(\eta-1)$, let copy $i\in\{0,\dots,c-1\}$ of $\mathcal C$ (own alphabet) have all counts multiplied by $M_i=N+i$.
Then for every budget $k'\le ck$ GreedTok is tie-free on this corpus and follows the lockstep schedule (round $s$:
copies in decreasing order of $M_i$). In particular, for $k'=ck$, $\mathrm{GT}/\mathrm{OPT}\le \Phi/O$, and for
$\nu k\le k'<(\nu+1)k$ (using $c=\nu+1$ copies) the ratio is
$\le\frac{\sum_{i\le \nu}M_i}{\sum_{i<\nu}M_{i+1}}\cdot\frac\Phi O\le\bigl(1+\frac1\nu\bigr)\frac\Phi O$.
\end{lemma}
\begin{proof}
Scores in copy $i$ are $M_i$ times the scores of $\mathcal C$ in the same state.  Suppose round $s$ is in progress and the
copies with $M$ larger than $M_i$ have already taken $\sigma_s$.  Copy $i$ offers $M_ig_s$, while the other copies still in
round $s$ offer at most $M_jg_s<M_ig_s$, since their $M_j<M_i$.  Runners-up inside any copy $j$ score strictly below that
copy's maximum.  The copies already past $\sigma_s$ offer at most $M_{\max}g_{s+1}$, and
$M_{\max}g_{s+1}<M_{\min}g_s\le M_ig_s$ because $M_{\max}/M_{\min}=(N+c-1)/N<\eta$.  So $\sigma_s$ in copy $i$ is the
unique maximizer.  The value is $\sum_iM_i\Phi$ after $ck$ steps, and $\mathrm{OPT}\ge\sum_iM_iO$ by additivity over copies
(for $\nu k\le k'<(\nu+1)k$ use the $\nu$ largest copies).
\end{proof}
\noindent The unique-choice instances of Proposition~\ref{prop:smallk} ($M=100$) have strictly decreasing gains and a
strictly smaller maximum score after the run (Appendix~\ref{app:comp}), so Lemma~\ref{lem:copies-strict} extends them to
every multiple of their budget.  For the junction families, Lemma~\ref{lem:bonus} is simpler.

%% file: gt/covfam.tex
\subsection{A max-coverage family with unique choices}\label{app:covfam}
\begin{proposition}[Max-coverage family]\label{prop:covfam}
For every $k\ge 2$ and every $R_1\ge 20k^2$ there is a corpus $\mathcal C_{k,R_1}$ (all counts $1$, $k$ words of length
$R_1+1$, alphabet of size $R_1+1$) on which every greedy choice of GreedTok is unique, and
\[
  \frac{\mathrm{GT}_k(\mathcal C_{k,R_1})}{\mathrm{OPT}_k(\mathcal C_{k,R_1})}\;\le\;1-\Bigl(1-\frac1k\Bigr)^{k}+\frac{k}{R_1}.
\]
In particular, taking $R_1=20k^3$, the worst-case ratio of GreedTok is at most $1-(1-1/k)^k+1/(20k^2)\to 1-1/e\approx 0.6321$.
\end{proposition}

\paragraph{Construction.}
Define $R_{t+1}=R_t-x_t$ with $x_t=\lfloor R_t/k\rfloor+1$ for $t=1,\dots,k$.
Let $p$ be a prime with $2k<p\le 4k$ (Bertrand). Choose integers $0\le y_{k+1},\dots,y_p\le x_k-1$ with
$\sum_{b>k} y_b=R_{k+1}-(p-1)$ (feasibility below). Let $c_b=x_b$ for $b\le k$ and $c_b=y_b$ for $b>k$.
For $b=1,\dots,p$ let $\omega_b$ be a string of $c_b+1$ letters, all letters of all blocks pairwise distinct.
Index blocks by $\mathbb Z_p$ (block $b$ has residue $b \bmod p$) and for $j=1,\dots,k$ let
\[
  W_j=\omega_{0}\,\omega_{j}\,\omega_{2j}\cdots\omega_{(p-1)j}\qquad(\text{indices mod }p),
\]
each with count $1$. As $j$ is invertible mod $p$, every word contains every block exactly once, so every letter
occurs exactly once in every word and $|W_j|=\sum_b(c_b+1)=\sum_t x_t+\sum_b y_b+p=R_1+1$.
We call $\omega_1,\dots,\omega_k$ the \emph{greedy blocks}.

\paragraph{Feasibility.} Since $x_t\le R_t/k+1$, $R_{t+1}\ge(1-\frac1k)R_t-1$, hence
$R_{t}\ge(1-\frac1k)^{t-1}R_1-k$; in particular $R_k\ge R_1/e-k\ge 7k^2-k$ and
$R_{k+1}\ge(1-\frac1k)R_k-1\ge 4k-1\ge p-1$, so the leftover mass $R_{k+1}-(p-1)$ is nonnegative.
It is at most $(p-k)(x_k-1)$: indeed $p-k\ge k$ gives $(p-k)(x_k-1)\ge R_k-(p-k)$, while
$R_{k+1}-(p-1)\le(1-\frac1k)R_k-(p-1)$, and the difference is $\ge R_k/k+k-1>0$.
Finally $x_t-x_{t+1}\ge (R_t-R_{t+1})/k-1=x_t/k-1>0$ because $x_t>R_t/k\ge R_k/k>k$; so $x_1>\dots>x_k>y_b$ for all $b>k$.

\paragraph{Scores.}
Fix $t\in\{1,\dots,k\}$ and the state after GreedTok committed $\omega_1,\dots,\omega_{t-1}$, each as one interval in each word
(we show inductively this is the run). Recall that the score of a candidate is the sum over words of the gains of its
valid occurrences, the gain of an occurrence being (number of pieces inside it)$-1$, where an uncovered letter and a
committed interval each count as one piece. Let $s$ be a candidate (a substring of length $\ge2$, not yet chosen).
\begin{enumerate}
\item[(a)] \emph{$s$ lies inside one block $\omega_b$.} Then $s$ occurs exactly once in every word, at the same offset in
$\omega_b$. If $\omega_b$ is committed, $s\neq\omega_b$ (already chosen), so one endpoint of $s$ lies strictly inside the
committed interval $\omega_b$ and the occurrence is invalid: score $0$. If $\omega_b$ is uncommitted, all letters of $s$ are
uncovered and its boundary pairs are not inside one committed interval, so the score is $k(|s|-1)\le k\,c_b$, with
equality only for $s=\omega_b$.
\item[(b)] \emph{$s$ contains a junction.} Then $s$ contains the last letter of some $\omega_u$ followed by the first
letter of $\omega_v$, which are adjacent only in the word $W_j$ with $v-u\equiv j\pmod p$; the residues $1,\dots,k$ are
distinct since $k<p$. So $s$ occurs only in $W_j$, and only once. The pieces inside one occurrence are among the pieces
of the whole word, so its gain is at most (pieces of $W_j$)$-1$. Each committed block $\omega_s$ ($s<t$) turned
$x_s+1$ uncovered letters into one piece, so this equals $|W_j|-1-\sum_{s<t}x_s=R_t$. Hence the score of $s$ is $\le R_t$.
\end{enumerate}
By construction $k x_t\ge k(\lfloor R_t/k\rfloor+1)>R_t$, $x_t>x_{t'}$ for $t'>t$, and $x_t>y_b$. So by (a) and (b),
$\omega_t$ is the \emph{unique} maximizer at step $t$, with score $k x_t$. Committing it creates one interval in each word
(the occurrence is valid: its letters are uncovered and its boundary pairs join different pieces), which completes the
induction. Hence GreedTok is tie-free and
\[
  \mathrm{GT}_k=k\sum_{t=1}^k x_t=k\,(R_1-R_{k+1}).
\]

\paragraph{Optimum.} $S^*=\{W_1,\dots,W_k\}$ has $f(S^*)=\sum_j(|W_j|-1)=kR_1$, so $\mathrm{OPT}_k\ge kR_1$ and
\[
  \frac{\mathrm{GT}_k}{\mathrm{OPT}_k}\le 1-\frac{R_{k+1}}{R_1}\le 1-\Bigl(1-\frac1k\Bigr)^k+\frac{k}{R_1},
\]
using $R_{k+1}\ge(1-\frac1k)^kR_1-\sum_{i=0}^{k-1}(1-\frac1k)^i\ge(1-\frac1k)^kR_1-k$. \qed

\paragraph{Remarks.}
(i) The instance is the classical tight example for greedy maximum coverage: the ``elements'' are the internal pairs of
the blocks; greedy block $\omega_t$ takes a $1/k$ fraction of every optimal word's remaining marginal.
(ii) The ties variant ($x_t=\lceil R_t/k\rceil$) has GreedTok's worst tie-break follow the same run and gives slightly
smaller instances. (iii) The alphabet grows with $R_1$; we did not try to bound it.
(iv) Small explicit instances of this family were verified exactly (Appendix~\ref{app:comp}).

%% file: gt/optimal_app.tex
\subsection{Proof of Theorem~\ref{thm:opt} (the planted tokens are optimal)}\label{app:opt}
Normalize counts so that every planted token has total occurrence mass $1$ (as in the constructions), so
$f(Y)=2K$.  Let $g=n\ge10$, $\theta$ be the scaling factor of Appendix~\ref{app:masses}, and $\delta=2/g$.  For a string
$t$ let $\mathrm{cnt}(t)$ be the count-weighted number of its occurrences and $\mathrm{val}(t)=(|t|-1)\,\mathrm{cnt}(t)$, an
upper bound on the pairs covered by pieces equal to $t$ in any disjoint family.

\paragraph{Facts.}  In every word, $a$ is always followed by a private letter $x$, $x$ by $z$, and $z$ by $a$ or the word
end.  We use the following facts, which hold for J2 and for J4 with any joining satisfying (U1) and (U2).
\begin{itemize}
\item[(F1)] Every token occurs with total mass $1$.
\item[(F2)] A cut string $z_va_u$ occurs only at junctions with these letters.  If $(v,u)$ lies on the diagonal of
phase-1 round $\xi$, its count is $m_\xi$, and for each of the $g$ tokens $i$ with $v(i)=v$, the occurrences whose left
token is $i$ have total mass exactly $m_\xi/g$ (token $i$ is the first member of $d$ blocks of that junction, each of total
usage $c(B)=m_\xi/(gd)$ by (U1); in J2 the blocks are the words).  If $(v,u)$ lies on a phase-2 diagonal $R_1+r$ (J4 only),
its count is $\mu_r\le\frac12$.
\item[(F3)] A string that contains $x_j$ and a character outside that occurrence of $y_j$ contains a junction pair next
to $y_j$, together with $x_j$; this fixes the side and the junction letters, so its count is at most the mass of $y_j$ next
to such a junction, which is $m_\xi/g$ (phase 1) or $\mu_r/g$ (phase 2, by (U2)), hence at most $m_0/g$.  If its only
private letter is $x_j$, it lies inside $z\,a\,x_j\,z\,a$ and has at most $4$ interior pairs.
\item[(F4)] A string with two or more private letters contains $x_iz\,a\,x_j$ at some junction.  At a phase-1 junction
only the block $(i,j)$ contains it, so its count is at most $c(B)\le m_0/(2g)$; at a phase-2 junction its count is at
most $\mu_r/g\le\frac1{2g}$ by (U2).  Every word has at most $12$ characters, so the string has at most $11$ interior pairs.
\item[(F5)] By the proof of Lemma~\ref{lem:masses}(iv), $m_0=2\theta$ with $1\le\theta\le\frac{g-1}{g-2}$.  By
Lemma~\ref{lem:masses}(i), $m_\xi\le m_1$ for $\xi\ge1$, and $m_1=\theta\,C(\frac12-\frac2g)=\theta(2-\frac8g)$ for $g\ge6$.
\end{itemize}

\paragraph{Value bounds} (for $g\ge10$).  A string with at least two private letters has
$\mathrm{val}\le11\theta/g\le\frac{11(g-1)}{g(g-2)}\le2-\delta$, since $11/(g-2)\le2$.  A string with exactly one private letter
that is not a substring of its token has $\mathrm{val}\le4m_0/g=8\theta/g\le2-\delta$.  A cut string of a phase-1 round
$\xi\ge1$ has $\mathrm{val}=m_\xi\le\theta(2-\frac8g)\le\frac{(g-1)(2g-8)}{g(g-2)}\le2-\delta$, and a phase-2 cut string has
$\mathrm{val}\le\frac12$.  The substrings $a\,x_j$ and $x_jz$ of a token have $\mathrm{val}=1$, and $y_j$ has
$\mathrm{val}=2$.  Only the round-$0$ cut strings, with count $m_0=2\theta$, can have value above $2$.

\paragraph{Proof of Theorem~\ref{thm:opt}.}  Fix $T$ with $|T|\le K$ and a family of pairwise disjoint occurrences of
strings of $T$ (\emph{pieces}) that attains $f(T)$.  Every string of length $\ge2$ is of exactly one of three kinds: a
\emph{core} string $a\,x_j$, $x_jz$ or $y_j$; a cut string $z_va_u$ (the only strings without a private letter, by the
letter order above); or a \emph{crossing} string, which contains a private letter and a character outside that token.
Let $T_{\rm core},T_{\rm cut},T_{\rm cr}$ be the corresponding parts of $T$ and $\mathcal J=\{j:T_{\rm core}\text{ contains a
core string of }y_j\}$.  Distinct $j$ need distinct core strings, so $|\mathcal J|+|T_{\rm cut}|+|T_{\rm cr}|\le K$.

Every covered pair is a token-internal pair ($a$--$x$ or $x$--$z$) or a junction pair ($z$--$a$).
\begin{enumerate}
\item A token-internal pair of an occurrence of $y_j$ with $j\notin\mathcal J$ contains $x_j$, so the piece covering it is
neither a core string nor a cut string; it is a crossing string.
\item Junction pairs may be covered by crossing pieces.
\item When a cut piece $z_va_u$ is used at a junction between occurrences of $y_i$ (left) and $y_j$ (right), the pairs
$x_i$--$z_v$ and $a_u$--$x_j$ are piece boundaries, hence uncovered.  Each token-internal pair is next to exactly one
junction or word end, so different cut uses force different pairs.
\item The token-internal pairs of occurrences of $y_j$, $j\in\mathcal J$, number at most $2$ in total, by (F1), minus
those forced uncovered in item~3.
\end{enumerate}
Items 1 and 2 are bounded by $\sum_{t\in T_{\rm cr}}\mathrm{val}(t)$.  A cut use with both neighbours in $\mathcal J$
covers one pair and forces two pairs of $\mathcal J$-tokens uncovered; with exactly one neighbour in $\mathcal J$ it covers
one and forces one.  Hence
\[
f(T)\ \le\ 2|\mathcal J|+\sum_{t\in T_{\rm cr}}\mathrm{val}(t)+\sum_{c\in T_{\rm cut}}N(c),
\]
where $N(c)$ is the mass of the occurrences of $c$ whose left token is not in $\mathcal J$.  For a phase-1 cut string
$c=z_va_u$ of round $\xi$, (F2) gives $N(c)\le m_\xi(1-\kappa_v/g)$ with $\kappa_v=|\{i\in\mathcal J:v(i)=v\}|$; for a phase-2 cut
string $N(c)\le\frac12$.

Let $V_0$ be the set of round-$0$ cut strings in $T$ with $\kappa_v\le1$, and $\zeta=|V_0|$.  Every other member of
$T_{\rm cut}\cup T_{\rm cr}$ contributes at most $2-\delta$: for a round-$0$ cut string with $\kappa_v\ge2$,
$N\le2\theta(g-2)/g\le2(g-1)/g$, and the others are covered by the value bounds.  Each member of $V_0$ contributes at most
$m_0=2\theta$ and keeps at least $g-1$ of the $g$ tokens with $z$-letter $v$ out of $\mathcal J$.  The round-$0$ cut strings
are $z_va_v$, one for each $v$, so these sets of tokens are disjoint, and $|\mathcal J|\le K-(g-1)\zeta$.  With
$R=|T_{\rm cut}|+|T_{\rm cr}|-\zeta$ we have $|\mathcal J|+\zeta+R\le K$ and
\[
f(T)\ \le\ 2|\mathcal J|+2\theta\zeta+(2-\delta)(K-|\mathcal J|-\zeta).
\]
The right side increases with $|\mathcal J|$, so it is at most its value at $|\mathcal J|=K-(g-1)\zeta$, which is
$2K+\zeta\,(2\theta-2-\delta(g-2))$.  Here $2\theta-2\le\frac2{g-2}$ and $\delta(g-2)=\frac{2(g-2)}g$, and $\frac2{g-2}\le\frac{2(g-2)}g$
because $(g-2)^2\ge g$.  So $f(T)\le2K=f(Y)$.

For $\nu$ disjoint copies on disjoint alphabets, all facts hold per copy and no string spans two copies.  The sets
$\mathcal J$ and $V_0$ are unions over the copies, the constraint $|\mathcal J_c|\le K-(g-1)\zeta_c$ holds in each copy, and the only
global constraint is $\sum_c(|\mathcal J_c|+\zeta_c+R_c)\le\nu K$.  The same computation gives $f(T)\le2\nu K$ for $|T|\le\nu K$.\qed

\paragraph{Remark.}  The budget matters: for $k$ much smaller than $K$, the round-$0$ cut strings, of value $2\theta>2$,
beat whole tokens, and $\OPT_k>2k$.  The theorem is about $k=K$ and its multiples, which are the budgets used in
Theorems~\ref{thm:j2}--\ref{thm:unique}.

%% file: gt/unique_app.tex
\subsection{Proofs of Lemma~\ref{lem:bonus} and Theorem~\ref{thm:unique}}\label{app:unique}
\begin{proof}[Proof of Lemma~\ref{lem:bonus}]
Each $\sigma_t$ is a substring of a word of $\mathcal C$, so $\mathcal C$ and $\mathcal C'$ have the same candidates.  The only substring of
length $\ge2$ of the bonus word $\sigma_t$ is $\sigma_t$ itself; it has no boundary pairs, so it is valid with gain $1$ until
$\sigma_t$ is committed, and only a commit of $\sigma_t$ covers it.  Hence, as long as the states on the words of $\mathcal C$ agree,
$\Sc'(x)=\Sc(x)+\sum_s\varepsilon_s\mathbf 1[x=\sigma_s,\ \sigma_s\text{ not yet chosen}]$ for every candidate $x$.  At step $t$, let $M_t$ be the
maximum score on $\mathcal C$.  Then $\Sc'(\sigma_t)=M_t+\varepsilon_t$; an unchosen $\sigma_s$ with $s>t$ has $\Sc'(\sigma_s)\le M_t+\varepsilon_s<M_t+\varepsilon_t$;
every other candidate has $\Sc'(x)=\Sc(x)\le M_t$.  So $\sigma_t$ is the unique maximizer, and committing it changes the words of
$\mathcal C$ exactly as on $\mathcal C$.  Induction completes the proof; adding words cannot decrease $f$.
\end{proof}

\begin{proof}[Proof of Theorem~\ref{thm:unique}]
The lockstep run on disjoint copies in the proof of Lemma~\ref{lem:copies}, truncated after $k$ picks, commits only cut
strings of the copies, which have length $2$.  Apply Lemma~\ref{lem:bonus} with $\sum_t\varepsilon_t\le\eta\,\nu f(Y)$; the optimum is
at least $\nu f(Y)$.  Rational $\varepsilon_t$ and scaling make all counts integers.
\end{proof}

%% file: app/alphabet.tex
\section{\GT{} over small alphabets}\label{sec:alphabet}\label{app:smallalpha}
Over small alphabets the worst case disappears at large budgets: over $256$ symbols, \GT{}'s ratio exceeds the
large-alphabet limit $0.33861$ for every $k\ge\ByteThresh$.  The junction corpora give every planted token a private middle
letter, so with budget $K=n^2$ they use $n^2+2n$ letters; byte-level tokenizers work the other way round, with about $256$
symbols and tens of thousands of tokens or more.  Theorem~\ref{thm:fixed}(a) and Lemma~\ref{lem:dom} (Section~\ref{sec:fixed}) give the per-step inequality
$2N_Mg_{t+1}\ge P(\mathcal C)-\frac{M+1}{M-1}G_t$; here $c=\log\sigma/\log k$, and $N_2\le\sigma^2$.  The quantities $N_M$ belong
to the corpus, so the bounds below can be evaluated on a given corpus; we use only the worst case over $\sigma$ letters, and
replacing $N_M$ by a larger number keeps every bound valid.

\subsection{Combining with the main theorem}
Theorem~\ref{thm:fixed}(a) alone stops at $\frac13$ for $c<\frac12$.  Its per-step inequality bounds the same $g_{t+1}$ as the six
lines of Theorem~\ref{thm:main}, so the two can be combined step by step.  With $x_t=G_t/\OPT_k$ and $\OPT_k\le P$, every
step satisfies $x_{t+1}\ge\psi(x_t)$ with
\[
\psi(x)=x+\max\Big(0,\ \max_{\beta\le\min(6,k)}\frac{\frac\beta{\beta+1}-\beta x}k,\ \frac{1-3x}{2N_2},\ \frac{1-2x}{2N_3}\Big),
\]
and $\psi$ is non-decreasing: the lines have slope $1-\beta/k\ge0$, and the last two branches have slopes
$1-\frac3{2N_2}$ and $1-\frac1{N_3}$, which are non-negative after replacing $N_2$ and $N_3$ by $\max(N_2,2)$ and $\max(N_3,2)$.  So
$\GTv_k\ge\psi^{\circ k}(0)\,\OPT_k$, as in Theorem~\ref{thm:env}.

\begin{corollary}[Byte alphabets]\label{cor:byte}
Let the corpus have at most $256$ letters, so $N_2\le2^{16}$ and $N_3\le2^{16}+2^{24}$.  Then for every $k$, any tie-breaking and
both loops, $\GTv_k\ge\gamma^{(256)}_k\OPT_k$ with $\gamma^{(256)}_k=\psi^{\circ k}(0)$, and
\begin{center}\small
\setlength{\tabcolsep}{4pt}\begin{tabular}{@{}lccccccccc@{}}\toprule
$k$ & $2^{14}$ & $2^{15}$ & $2^{16}$ & $2^{17}$ & $2^{18}$ & $2^{19}$ & $2^{20}$ & $2^{21}$ & $2^{22}$\\\midrule
$\gamma^{(256)}_k\ge$ & \ByteGa & \ByteGb & \ByteGc & \ByteGd & \ByteGe & \ByteGf & \ByteGg & \ByteGh & \ByteGi\\\bottomrule
\end{tabular}
\end{center}
Moreover, for every $k\ge\ByteThresh$, $\GTv_k/\OPT_k>\frac12-\frac{3^{1/4}}{3e}\approx\JFourLimSeven$, the limit of the junction
corpora.
\end{corollary}
\begin{proof}
The table is $\psi^{\circ k}(0)$ evaluated in integers scaled by $10^{18}$, rounding every branch down (valid because $\psi$ is
non-decreasing), and truncated.  For the threshold, fix $k\ge K=\ByteThresh$ and $\theta=\ByteTheta$, and let $t_0=\lfloor\theta k\rfloor\ge\theta K-1$.  On
the first $t_0$ steps use only the branch $\frac{1-3x}{2N_2}$: then $x_{t_0}\ge z=\frac13\big(1-e^{-3(\theta K-1)/(2N_2)}\big)$ or $x$
has already passed $z$.  On the remaining $k-t_0\ge(1-\theta)k$ steps use only the line $\beta=1$:
$\frac12-x_{t+1}\le(1-\frac1k)(\frac12-x_t)$.  Hence $x_k\ge\frac12-(\frac12-z)e^{-(1-\theta)}$, which increases with $K$; at
$K=\ByteThresh$ interval arithmetic gives a value that exceeds $\frac12-\frac{3^{1/4}}{3e}$ by $\ByteMargin$.
\end{proof}

\noindent At the vocabulary sizes $100{,}256$, $128{,}256$ and $200{,}019$ the recursion gives $0.33581$, $0.35253$ and $0.37942$; at
$32{,}768$ and $50{,}257$ it gives nothing beyond Theorem~\ref{thm:main}.  The recursion itself crosses $\frac12-\frac{3^{1/4}}{3e}$ a little earlier (at $k=104{,}729$ it already exceeds it); the threshold
$\ByteThresh$ is where the closed-form argument, valid for all larger $k$ at once, takes over.  Below $k\approx87{,}500$ the
recursion improves Theorem~\ref{thm:main} by less than $10^{-4}$: there $N_2$ is comparable to $k$.

\begin{theorem}[Alphabets of size $k^c$, $c<\frac12$]\label{thm:half}
If $N_2=o(k)$ (for instance $\sigma=k^c$ with $c<\frac12$), then $\liminf_k\GTv_k/\OPT_k\ge\frac12-\frac1{6e}=0.43868\ldots$
\end{theorem}
\begin{proof}
Let $\varepsilon>0$ and $t_0=\lfloor\varepsilon k\rfloor$.  The per-step inequality of Theorem~\ref{thm:fixed}(a) with $M=2$, iterated over
the first $t_0$ steps, gives $G_{t_0}\ge\frac P3\big(1-(1-\frac3{2N_2})^{t_0}\big)$, so $x_{t_0}\ge\frac13(1-e^{-3(\varepsilon k-1)/(2N_2)})=\frac13-o(1)$.
From then on the line $\beta=1$ of Theorem~\ref{thm:main} gives $\frac12-x_{t+1}\le(1-\frac1k)(\frac12-x_t)$, so
$\frac12-x_k\le(\frac16+o(1))e^{-(1-\varepsilon)}$.  Let $\varepsilon\to0$.
\end{proof}

The bound $\frac12$ is the natural barrier.  Over a small alphabet the cut strings of a junction construction must be
shared by many junctions, and such a construction reaches $\frac12$.

\begin{proposition}[A family at $\frac12$]\label{prop:U}
For every $c\in(\frac13,\frac12)$ there are corpora over $\sigma=\Theta(k^c)$ letters on which every \GT{} run, with any
tie-breaking, has $\GTv_k/\OPT_k\le\frac12+O(k^{1-3c}+k^{2c-1}+k^{-c})$.
\end{proposition}
\noindent The tokens are $a\,\xi\,z$ over three disjoint letter sets, chosen so that every cut string $za$ is very
frequent; the words follow an Eulerian circuit of the complete digraph on the tokens, so every ordered pair of tokens meets
at exactly one junction.  Greedy first commits all $\Theta(\sigma^2)=o(k)$ cut strings, which covers every junction and
kills every interior token at a cost of one covered pair per token; afterwards every string that crosses a junction is
rare.  The proof is in Appendix~\ref{app:realistic}.  For optimal tokens of at most three characters, the class where
$0.33861$ is tight for large alphabets, small alphabets raise the bound to $\frac12$:

\begin{proposition}[Tokens of at most three characters]\label{prop:S}
Let $S^*$ be any set of strings of length $2$ or $3$.  Then $f(S^*)\le2G_t+2I_2(t)$ at every $t$, and hence
$\GTv_k\ge\frac{f(S^*)}2\big(1-(1-\frac1{2N_2})^k\big)$; if $N_2=o(k)$ the ratio to $f(S^*)$ tends to at least $\frac12$.
\end{proposition}

\subsection{How large must the alphabet be for the worst case?}
The private letters of the junction corpora can be recoded.  Map $a_u$ and $z_u$ to one new letter $l_u$, and the private
letter of the token $a_ux_jz_v$ to a new letter $\xi_{(u+v)\bmod n}$.  This uses $2n=2\sqrt K$ letters, and it is injective on all
substrings of length at least $2$ of the corpus, so it preserves every greedy run and the optimum (Appendix~\ref{app:recode}).

\begin{proposition}[Junction corpora over $2\sqrt k$ letters]\label{prop:recode}
For every $n\ge24$ there is a corpus over an alphabet of $2n$ letters on which, with budget $k=n^2$, \GT{} (adversarial ties)
has a run with $\GTv_k/\OPT_k=\frac12-\frac14(1-\frac1n)^{n-R_1}$ against the true optimum, where $R_1$ is as in
Theorem~\ref{thm:j4}.  As $n\to\infty$ this tends to $\JLIM$.
\end{proposition}

\paragraph{The regimes.}  With $\sigma=k^c$: for $c<\frac13$ the ratio tends to at least $\frac12$ (Theorem~\ref{thm:fixed}(a)); for
$\frac13<c<\frac12$ the asymptotic worst case lies in $[\frac12-\frac1{6e},\frac12]$ (Theorem~\ref{thm:half} and
Proposition~\ref{prop:U}); at $\sigma=2\sqrt k$ ($c=\frac12+o(1)$) it is at most $\JLIM$ (Proposition~\ref{prop:recode}); and for
every alphabet it is at least $\MAINconst$ (Theorem~\ref{thm:main}).  For bytes, the recoded corpora fit into $256$ letters up to
$k=128^2=16384$, with ratio $\ByteRatio$ there (Proposition~\ref{prop:recode}), and Corollary~\ref{cor:byte} rules out ratios at
or below $0.33861$ from $k=\ByteThresh$ on.  What happens in between is discussed in Section~\ref{sec:discussion}.

\subsection{Units, and proofs of Propositions~\ref{prop:S} and~\ref{prop:U}}\label{app:realistic}

\paragraph{Units.}  In each word at time $t$, a \emph{unit} is either a maximal committed occurrence (maximal under inclusion in
the laminar family of committed occurrences, Lemma~\ref{lem:basic}(i)) or an uncovered character, one that lies in no
committed occurrence.  The units partition the characters into consecutive intervals.  \emph{The pairs inside a unit are
exactly the covered pairs, and the pairs between consecutive units are exactly the uncovered pairs.}  A pair inside a
unit is an interior pair of a committed occurrence.  Suppose the pair between consecutive units $U,V$ is covered, i.e.\
interior to a committed occurrence $Q$; then $Q$ contains the last character of $U$ and the first of $V$.  If $U$ is an
uncovered character this is impossible.  Otherwise $U$ is a maximal committed occurrence meeting $Q$, so $Q\subseteq U$ or
$U\subsetneq Q$ by laminarity; the first is false because $Q$ contains a character of $V$, and the second contradicts the
maximality of $U$.  In Lemma~\ref{lem:short}, the occurrences $UV$ and $U'V'$ of two short pairs whose positions among the
pairs between units differ by at least $2$ use disjoint units, so they are disjoint.

\subsubsection*{Proof of Proposition~\ref{prop:S}}
Fix a word with $m$ characters; let $D$ be its number of covered pairs, $I$ its number of $2$-short pairs, and $O$ the
number of pairs covered by an optimal segmentation with $S^*$.  An occurrence of length $L\le3$ uses $L$ characters and
covers $L-1\le\frac{2L}3$ pairs, so $O\le\lfloor\frac{2m}3\rfloor$.  By Lemma~\ref{lem:dom} for this word ($M=2$),
$m-1\le3D+I$.  If $I\ge1$, then $2D+2I\ge\frac{2(m-1-I)}3+2I\ge\frac{2m+2}3>O$.  If $I=0$, then $D\ge\lceil\frac{m-1}3\rceil$, and
$2\lceil\frac{m-1}3\rceil\ge\lfloor\frac{2m}3\rfloor$ (check $m=3j,3j+1,3j+2$).  Summing with the counts gives $f(S^*)\le2G_t+2I_2(t)$,
and Lemma~\ref{lem:short} gives $4N_2g_{t+1}\ge f(S^*)-2G_t$; the recursion of Theorem~\ref{thm:fixed}(a) with $a=2$ finishes the proof.

\subsubsection*{Proof of Proposition~\ref{prop:U}}
\paragraph{Construction $U(s,x,r)$.}  Take pairwise disjoint letter sets $A=\{a_0,\dots,a_{s-1}\}$, $Z=\{z_0,\dots,z_{s-1}\}$ and
$\Xi=\{\xi_0,\dots,\xi_{x-1}\}$, so $\sigma=2s+x$, with $x\ge2s+1$ and $2\le r\le s$.  The tokens are
$Y=\{a_u\xi_vz_w:\ w-u-v\bmod s\in\{0,\dots,r-1\}\}$, so $K=|Y|=sxr$.  Each token determines $(u,v,w)$; every prefix $a_u\xi_v$ and every
suffix $\xi_vz_w$ lies in exactly $r$ tokens, and exactly $K/s=xr$ tokens start with a given $a_u$ (end with a given $z_w$).  The
complete digraph with loops on $Y$ has in-degree and out-degree $K$ at every vertex, so it has an Eulerian circuit
$y_0y_1\cdots y_{K^2}$ ($y_{K^2}=y_0$) that uses every ordered pair $(y,y')$ exactly once.  Put $m=sx$ (so $m$ divides $K^2$),
$W=K^2/m$, and let word $j$ (count $1$) be $y_{jm}y_{jm+1}\cdots y_{jm+m}$: $m+1$ tokens and $m$ junctions.  Every ordered pair of tokens is
a junction of exactly one word, and every token occurs at most $2K$ times.

\paragraph{Claim.}  On $U(s,x,r)$, for every $k\ge K$ and every tie-breaking rule: (a) the first $s^2$ picks are exactly the
$s^2$ cut strings $z_wa_u$, each with score $K^2/s^2=x^2r^2$, so $G_{s^2}=K^2$; (b) $\GTv_k\le K^2+4W+(k-s^2)\mu^*$ with
$\mu^*=\max(rK/s,\,2r^2,\,6K/s,\,3r,\,2(m+1))$; (c) $\OPT_k\ge f(Y)=2W(m+1)$, hence
$\GTv_k/\OPT_k\le\frac12+\frac3{2(m+1)}+\frac{k\mu^*}{2K^2}$.

\begin{proof}
Letter roles are intrinsic (the three sets are disjoint), so every substring sits at a fixed offset relative to the tokens.
Let $j(x)$ be the number of junction pairs that are interior pairs of a string $x$.  Initially $\Sc_0(x)\le\mathrm{count}(x)(|x|-1)$, and
by Corollary~\ref{cor:mono} the scores of unchosen strings never increase.
\begin{itemize}
\item $j=0$ ($x$ inside a token): $a\xi$ and $\xi z$ occur at most $2Kr$ times; a whole token at most $2K$ times, with $|x|=3$.
\item $j=1$: $x$ is a suffix of length $l_1$ of some $y$ followed by a prefix of length $l_2$ of some $y'$, $l_1,l_2\in\{1,2,3\}$.
Every junction $(y,y')$ occurs once, so $\mathrm{count}(x)\le N(l_1)N(l_2)$ with $N(1)=K/s$, $N(2)=r$, $N(3)=1$.  The case $(1,1)$ is a
cut string, with count exactly $K^2/s^2$.  The others: $(2,1),(1,2)$: at most $rK/s$, length $3$; $(2,2)$: $r^2$, length $4$;
$(3,1),(1,3)$: $K/s$, length $4$; $(3,2),(2,3)$: $r$, length $5$; $(3,3)$: $1$, length $6$.
\item $j=2$: $x$ contains a full token $y''$ and both its junctions; its count is at most the number of predecessors of $y''$ with
the given suffix, at most $K/s$; its length is at most $9$.
\item $j\ge3$: $x$ contains two consecutive full tokens, an ordered pair that occurs once; count at most $1$, length at most $3(m+1)$.
\end{itemize}
(a) Every non-cut string starts with score at most
\[\max(2Kr,4K,2rK/s,3r^2,3K/s,4r,5,8K/s,3(m+1)-1)<x^2r^2,\]
using $x>2s$,
$r\ge2$ (hence $s\ge2$, $x\ge5$) and $m=sx$.  A cut string $z_wa_u$ occurs exactly at the $K^2/s^2$ junctions of its type; its
occurrences never overlap, and they stay valid as long as no pair inside a token is covered.  By induction over the first
$s^2$ steps only cut strings are committed, they cover only junction pairs, every unchosen cut string keeps score $x^2r^2$,
and every other string stays below it.  So each of the first $s^2$ picks is a cut string, whatever the tie rule, and then
every junction pair is covered: $G_{s^2}=Wm=K^2$.

(b) The covered pairs at the end are the $K^2$ junction pairs plus covered pairs inside tokens.  Pairs inside the first and the
last token of a word number at most $4W$.  An occurrence lying inside an interior token is invalid after step $s^2$ (a
boundary pair is a covered junction), so a pair inside an interior token can be covered only by a string with $j\ge1$.
After step $s^2$ the gain of such a string is at most its count times its uncovered interior pairs: for $j=1$ at most
$rK/s\cdot1$, $r^2\cdot2$, $K/s\cdot2$, $r\cdot3$ or $1\cdot4$; for $j=2$ at most $K/s\cdot6$; for $j\ge3$ at most $2(m+1)$ (two pairs
per token).  So each of the $k-s^2$ later picks adds at most $\mu^*$.

(c) Every word is tiled by tokens of $Y$, each covering $2$ pairs, and $|Y|=K\le k$; and
$\frac{K^2+4W}{2W(m+1)}=\frac{m+4}{2(m+1)}$.
\end{proof}

\paragraph{Asymptotics.}  Given $\sigma$ and $k$, take $s=\lfloor\sigma/5\rfloor$, $x=\sigma-2s$ and $r=\lfloor k/(sx)\rfloor$.  Then
$K=sxr>k-sx$, $r/s=\Theta(k/\sigma^3)$ and $1/r=\Theta(\sigma^2/k)$; for $\sigma=k^c$ with $\frac13<c<\frac12$ the conditions $2\le r\le s$ hold for
large $k$, and $\mu^*/K\le\max(r/s,\,2r/(sx),\,6/s,\,3/(sx),\,2(m+1)/K)$ with $(m+1)/K\le2/r$.  So the bound of (c) is
$\frac12+O(k^{1-3c}+k^{2c-1}+k^{-c})$.  Convergence is slow: at $\sigma=256$ and $k=2^{17}$ the bound of (c) is only $0.6638$.

\input{gt/recode_app}

%% file: gt/recode_app.tex
\subsection{Proof of Proposition~\ref{prop:recode}}\label{app:recode}
Let $\mathcal C$ be a corpus and $\mathrm{Sub}(\mathcal C)$ the set of substrings of length at least $2$ of its words.  Let
$\phi$ be a map on letters, applied letter by letter to words, with $\phi(\mathcal C)=\{(\phi(w),c_w)\}$.

\begin{lemma}[Relabelling]\label{lem:relabel}
If $\phi$ is injective on $\mathrm{Sub}(\mathcal C)$, then:
(i)~for every $t\in\mathrm{Sub}(\mathcal C)$, the occurrences of $\phi(t)$ in $\phi(\mathcal C)$ are exactly the
occurrences of $t$ in $\mathcal C$ (same words, same intervals $[s,e)$);
(ii)~$\mathrm{Sub}(\phi(\mathcal C))=\phi(\mathrm{Sub}(\mathcal C))$, and $\phi$ is a bijection between them;
(iii)~$x_1,x_2,\dots$ is a \GT{} run on $\mathcal C$ (each pick a maximizer, any tie rule) iff $\phi(x_1),\phi(x_2),\dots$
is a \GT{} run on $\phi(\mathcal C)$, with the same gains at every step and the same tie sets (mapped by $\phi$);
(iv)~$f_{\phi(\mathcal C)}(\phi(T))=f_{\mathcal C}(T)$ for every $T\subseteq\mathrm{Sub}(\mathcal C)$, and
$\OPT_k(\phi(\mathcal C))=\OPT_k(\mathcal C)$ for every $k$.
\end{lemma}
\begin{proof}
(i)~If $\phi(t)$ occurs at $[s,e)$ in $\phi(w)$, then $\phi(w[s{:}e])=\phi(t)$, and $w[s{:}e]\in\mathrm{Sub}(\mathcal C)$,
so by injectivity $w[s{:}e]=t$.  The converse is immediate.  (ii)~Every substring of $\phi(w)$ is $\phi$ of the substring
of $w$ at the same place.  (iii)~In the model, the state is a set of covered pairs per word.  Validity, gain, the leftmost
scan and the commit of a string $x$ depend only on the list of occurrence intervals of $x$ in each word and on the counts.
By (i) and (ii), these lists agree for $x$ and $\phi(x)$, and the candidate sets correspond bijectively.  Induction on the
step shows that the states agree, so the scores agree and the maximizers correspond.  (iv)~$\mathrm{cover}(w,T)$ is a
maximum over families of disjoint occurrences of strings of $T$, so it depends only on the occurrence intervals; strings
that are not in $\mathrm{Sub}(\mathcal C)$ (resp.\ $\mathrm{Sub}(\phi(\mathcal C))$) have no occurrence and contribute
nothing; so $\OPT$ is a maximum over the same values.
\end{proof}

\paragraph{The recoding.}  In J2 and J4 (any joining, any counts), every word is a concatenation $y_{i_1}\cdots y_{i_m}$
of planted tokens $y_j=a_{u(j)}x_jz_{v(j)}$, where $j\mapsto(u(j),v(j))=(j\bmod n,\lfloor j/n\rfloor)$ is a bijection
$[K]\to[n]^2$.  The alphabet is $\{a_u\}\cup\{z_v\}\cup\{x_j\}$, of size $2n+K$.  With $2n$ new letters
$l_0,\dots,l_{n-1}$, $\xi_0,\dots,\xi_{n-1}$, define
\[
\phi(a_u)=l_u,\qquad\phi(z_v)=l_v,\qquad\phi(x_j)=\xi_{(u(j)+v(j))\bmod n}.
\]

\emph{Claim: $\phi$ is injective on $\mathrm{Sub}(\mathcal C)$.}  Let $t\in\mathrm{Sub}(\mathcal C)$ and $t'=\phi(t)$.  In
every word the roles of the letters cycle $a,x,z,a,x,z,\dots$, so the role sequence of $t$ is a window of length at least
$2$ of this cycle.  The $\xi$-letters of $t'$ are exactly its $x$-role positions.  If $t'$ contains a $\xi$, the roles of
all positions follow (the left neighbour of an $x$ is an $a$, the right neighbour is a $z$, and the cycle is periodic).  If
$t'$ contains no $\xi$, then $|t|=2$ (every window of length at least $3$ contains an $x$), and the window is $za$.  So
$t'$ determines the role sequence.  Then each $a$-role letter $l_u$ is $a_u$ and each $z$-role letter $l_v$ is $z_v$.  An
$x$-role letter $\xi_s$ has, inside $t$, a left neighbour $a_u$ or a right neighbour $z_v$ (because $|t|\ge2$).  In the
first case $v=s-u\bmod n$; in the second, $u=s-v\bmod n$.  So $(u,v)$, and hence the token $j$ with $u(j)=u$, $v(j)=v$, are
determined, and the letter is $x_j$.  Thus $t'$ determines $t$.

\paragraph{Proof of Proposition~\ref{prop:recode}.}  Take the J4 corpus with the full joining (Theorem~\ref{thm:j4},
$n\ge24$) and apply $\phi$.  By Lemma~\ref{lem:relabel}(iii) the run of Theorem~\ref{thm:j4} is a run on $\phi(\mathcal C)$
with the same value $\GTv_K=\frac K2+Kw_0$.  By Lemma~\ref{lem:relabel}(iv) and Theorem~\ref{thm:opt},
$\OPT_K(\phi(\mathcal C))=\OPT_K(\mathcal C)=f(Y)=2K$.  So the ratio is $\frac14+\frac{w_0}2=\frac12-\frac14(1-\frac1n)^{n-R_1}$,
which tends to $\JLIM$.  The alphabet has $\sigma=2n=2\sqrt k$ letters, i.e.\ $\log\sigma/\log k=\frac12+\log2/\log k$.
The same applies to J2 (Theorem~\ref{thm:j2}, $n\ge10$, limit $\JTwoLimSix$).

\paragraph{Remarks.}  (i)~The proposition is stated along $k=n^2$ only: the extension to all $k$ in
Theorem~\ref{thm:j4} uses disjoint copies (Lemma~\ref{lem:copies}), which need disjoint alphabets, so $\nu$ copies need
$2n\nu$ letters.  (ii)~$2n$ letters is close to the minimum for an injective letter recoding: J4 has about $3K$ distinct
two-letter substrings, so $\sigma\ge\sqrt3\,n$.  (iii)~For a byte alphabet, $n=128$ gives $256$ letters, $k=16384$ and
ratio $\ByteRatio$; at $k\approx5\cdot10^4$ the recoding needs about $448$ letters ($n=224$, ratio $\ByteRatioBig$).  These
values are members of the proven family; they were computed from the formula, not replayed.  (iv)~As a computational check, for the J4
certificates with $n=10,12,14,16,20$ (cyclic joining, and the full joining at $n=12$) and the J2 certificate with $n=30$,
$\phi$ was verified to be injective on all substrings, and the recoded runs were replayed exactly with a separately
written simulator that checks every pick against all candidates; the gains and $f(\phi(Y))$ agree with the original
certificates.  A second, independently written replay of the J2 ($n=10,30$) and J4 ($n=10$ with both joinings, $n=12$)
certificates agreed as well, including the tie counts.

%% file: app/fixed.tex
\section{Proofs for fixed alphabets (Section~\ref{sec:fixed})}\label{app:fixed}
\subsection{\GT{}: short pairs, and Theorem~\ref{thm:fixed}(a)}
Recall that the units of a word at time $t$ are its maximal committed occurrences and its uncovered characters; by
Lemma~\ref{lem:basic}(i) they partition the word into consecutive intervals, and the covered pairs are exactly the pairs inside
units (Appendix~\ref{app:realistic}).

\begin{lemma}[Short pairs are cheap]\label{lem:short}
For every $M\ge2$ and $t<k$: $I_M(t)\le2N_M\,g_{t+1}$.
\end{lemma}
\begin{proof}
For an $M$-short pair between $U$ and $V$, the occurrence $UV$ of a string of length $|U|+|V|\in[2,M]$ is valid (its boundary
pairs lie between units, so they are uncovered or absent) and has gain exactly $1$.  By Lemma~\ref{lem:chosen} its string is
not yet chosen.  Number the pairs between consecutive units of a word from left to right.  The occurrences $UV$ of short pairs with
indices of the same parity are pairwise disjoint, so
for each string $x$, Lemma~\ref{lem:int} gives $\Sc_t(x)\ge$ half the count-weighted number of short pairs whose occurrence
spells $x$.  Summing over the at most $N_M$ strings, $I_M(t)\le2\sum_x\Sc_t(x)\le2N_Mg_{t+1}$.
\end{proof}

\begin{proof}[Proof of Theorem~\ref{thm:fixed}(a)]
The per-step inequality is Lemmas~\ref{lem:dom} and~\ref{lem:short}.  With $h_t=P-aG_t$ it reads
$h_{t+1}\le(1-\frac a{2N_M})h_t$ while $h_t\ge0$, and $h$ stays negative once it is negative; so
$h_k\le P(1-\frac a{2N_M})^k$.  Finally $\OPT_k\le P$.  (For the lazy loop, Corollary~\ref{cor:mono} makes every accepted
token a maximizer.)
\end{proof}

\noindent If $\sigma^M=o(k)$, the bracket tends to $1$, so $\liminf_k\GTv_k/\OPT_k\ge\frac{M-1}{M+1}$: at least $\frac13$ for
$c<\frac12$, $\frac12$ for $c<\frac13$, $\frac35$ for $c<\frac14$, and $1-\frac2{M+1}$ for $c<\frac1M$.  Over a fixed alphabet, the
choice $M=\lfloor\log_\sigma(k/(4\ln k))\rfloor\ge2$ gives $N_M<k/(2\ln k)$ and
\[
\frac{\GTv_k}{\OPT_k}\ \ge\ 1-\frac2{\log_\sigma\big(k/(4\ln k)\big)}-\frac1k\ \xrightarrow{\ k\to\infty\ }\ 1 .
\]
So over any fixed alphabet greedy is asymptotically optimal, and asymptotic ratios below $\frac12$ need
$\sigma\ge k^{1/3-o(1)}$.  For comparison, the counting bound of the earlier preprint needed $\sigma<k^{1/8}$ for the same conclusion and left
$\frac18\le c\le\frac12$ open.  For $\sigma=256$, however, these limits bite
only at enormous budgets (the ratio $\frac35$ of $M=4$ needs $k\gg256^4\approx4\cdot10^9$); the useful byte-level statement
is Corollary~\ref{cor:byte}.

\subsection{Greedy BPE: Theorem~\ref{thm:fixed}(b)}
\begin{proof}[Proof of Theorem~\ref{thm:fixed}(b)]
Fix the actual ordered list of full rules, including type identities. An original interval whose endpoints are surviving token boundaries can be parsed in isolation with exactly the same rules and output types. Its endpoints survived every earlier step. Inductively, each global scan reaches the first interior token with that token available, and none of its replacements crosses the right endpoint. It therefore makes exactly the replacements of the isolated scan, including on runs of identical tokens. In particular, if two current adjacent token pairs have the same concatenated original spelling, replay on that spelling returns the same two types. This establishes the pair-spelling property across all words.

Let $T$ be the maximum raw frequency at the end, or zero if there are no pairs. Maximum raw frequency is nonincreasing under any full merge. Old pairs gain no occurrences, and a new pair contains the fresh type on a specified side, so its frequency is at most the number of replacements. In a run of $h$ equal tokens one self-pair has raw count $h-1$ and gives $\lfloor h/2\rfloor$ replacements. Unequal-token occurrences are disjoint. At every step either greedy rule consequently realizes at least half the current maximum raw frequency. Thus $kT\le2G_k$, also when the run finishes early and $T=0$.

The output has $P(\mathcal C)-G_k$ adjacencies. At most $N_MT$ have combined original length at most $M$, by pair-spelling injectivity. An adjacency of tokens of lengths $u,v$ with $u+v>M$ contains $u+v-2\ge M-1$ removed original gaps internal to these tokens. Each removed gap belongs to one final token and is counted at most twice. This bounds the number of long adjacencies by $2G_k/(M-1)$ (this is Lemma~\ref{lem:dom} for the final tokens) and proves the finite inequality. Finally, set $h=\log_\sigma k$ and $M=\lfloor h-2\log_\sigma h\rfloor$ for sufficiently large $k$. Then $N_M/k\le k^{-1}\sum_{j=2}^M\sigma^j\le2/h^2$ and $M=h-O(\log h)$. Substitution proves the ratio, since $\OPT^{\mathrm{part}}_k\le P(\mathcal C)$.
\end{proof}

This proof also gives convergence to one when $\sigma$ varies with $k$ and $\log\sigma=o(\log k)$. For a unary corpus, $N_M\le M-1$, so $M=1+\lceil\sqrt{k}\rceil$ gives $(P(\mathcal C)-G_k)/G_k\le 4/\sqrt{k}+2/k$ when $G_k>0$.

\subsection{Proof of Theorem~\ref{thm:binary}}\label{app:binary}
The two parts share the corpus $\mathcal B_m$ of Section~\ref{sec:fixed}, the first greedy step, the lost boundary and the
count; only the comparator and the value of $m$ differ. Its total length is $N=O(m^2\lambda)=O(m^2\log k)$, and in both parts
$k=2m+O(\sqrt m)$.

\paragraph{The first step.} The guard words contain $B$ disjoint occurrences of $aa$, and $B$ exceeds the total length of
all other words. For BPE, the disjoint occurrences of $aa$ in the $B$ copies alone outnumber the raw occurrences of every
rival pair, so both greedy rules must begin with $aa$. For \GT{}, in the empty state the candidate $aa$ earns at least $B$ on
the guards alone, while every other candidate scores only on nonguard words and therefore earns less than $B$; the unique
first choice is $aa$.

\paragraph{The comparators.} (a) For \GT{}, $m=\lfloor(k-1)/2\rfloor$ and the direct vocabulary consisting of $L_i,R_j,aa$
costs $2m+1\le k$ entries. (b) For BPE, $m=\lfloor k/2-30\sqrt k\rfloor$, we write $h=\lambda/2$, and the following full sequence is used.
First use $ab\to P$ and $Pb\to Q$. These full rules turn all logical zeros into $P$ and ones into $Q$, leaving every external marker $a$ intact: each payload bit begins with its own $a$, and the following run of $b$'s specifies the bit. Build the left halves of the $x_i$ from their initial marker, using the full rules that extend $a$ followed by a prefix one bit at a time, in increasing prefix length up to $h$. Only prefixes starting in $P$ are used. Their first rule $(a,P)$ can apply only at the beginning of an $x_i$ payload, since every remaining $a$ is a marker and every $y_j$ starts in $Q$. Next build suffixes of length $h$ backwards from the final marker of each $x_i$. Their first rule $(P,a)$ distinguishes these ends because each $y_j$ ends in $Q$. The two halves have disjoint spans. One full rule per $i$ now makes $L_i$ a single type wherever it occurs. At this point apply $aa\to X$. It collects every isolated $aa$ and the two initial markers of every $R_j$. Assemble the $y_j$ prefixes forward from $X$ and their suffixes backwards from $a$, then use at most one root rule per $j$.

Each of the four half-building phases uses at most $\sum_{s=1}^h2^s=2^{h+1}-2$ rules. Since $2^h<4\sqrt m$ for $m>1$, the entire list uses at most $2m+32\sqrt m+3\le k$ rules for sufficiently large $k$. In both cases the comparator leaves one token per standalone or guard word and two per grid word $L_iR_j$. Let $V$ be its saving, so $\OPT_k\ge V$ in (a) and $\OPT^{\mathrm{full}}_k\ge V$ in (b); also $V\le N$.

\paragraph{The lost boundary.} In every grid word the two payloads are separated by a physical run of four $a$'s: the last marker of $L_i$, the two initial markers of $R_j$, and the head of the first bit of $y_j$. The character preceding this run is $b$, so the first greedy step (the commitment of the leftmost disjoint occurrences of $aa$, or the full rule $aa$) pairs its first two $a$'s, erasing the cut $L_i\mid R_j$. For \GT{}, a segmentation respecting the first commitment cannot have this cut as a boundary; by Lemma~\ref{lem:basic}, pairs covered on the first step remain covered throughout \GT{}. For BPE, every future token has boundaries surviving the first rule. We show that no token spelling any standalone $L_s$ or $R_t$ can occur in a grid word after that step. Each encoded payload contains $\lambda$ heads $ab$, with runs of one or two $b$'s between successive bit starts. The grid word has two such lists of exactly $\lambda$ heads, separated by the four-$a$ run, which no standalone word contains. Thus any candidate standalone occurrence must align all of its heads with one complete list. Its trailing $a$ forces the last $b$-run to be used in full. The lengths of the $b$-runs now recover the logical bits uniquely. The guard bits distinguish the two lists: $L_s$ can use only $x_i$, and $R_t$ only $y_j$. The required markers then place its endpoint at the erased cut. Such an occurrence would have the erased cut as an endpoint, so it is impossible.

\paragraph{Counting.} Set $K=\lfloor(2-\epsilon)k\rfloor$; a continuation (a respecting segmentation for \GT{}, or a rule sequence for BPE) has a dictionary of at most $K$ optional entries or rules in total. Let $d$ be the number of the $2m$ distinct standalone spellings absent from the continuation's dictionary. The $m$ copies of each such word cost at least $dm$ extra tokens altogether compared with the comparator. This remains true when the continuation chooses different joins on different copies. At most $K$ grid words can cost one token, since each needs a distinct one-token entry or compound type. All other grid words cost at least the comparator's two. If $d\ge\epsilon m/2$, then the excess is at least $\epsilon m^2/2-K$.

Otherwise the dictionary contains at least $2m-d$ different standalone spellings unusable in every grid word. At most $v=2+K-2m+d$ types, including terminals, can appear there. Since $k=2m+O(\sqrt m)$ and $d<\epsilon m/2$, we have $v\le(2-\epsilon)m$ for sufficiently large $k$.

The four-letter string $aaba$ occurs only at the beginning of any grid word. It starts the $L_i$ because the first $x_i$ bit is zero and there is another bit. Inside either encoded list the successive $a$'s are separated by $b$'s. The only other place with successive $a$'s is the middle run of four: its only candidate $aab$ is followed by $b$, because the first $y_j$ bit is one. Thus a first grid token of length at least four cannot spell a proper suffix of any grid word. The sets of long first-token types and proper suffix types are disjoint, so their ordered pairs spell at most $v^2/4$ distinct two-token grid words. The $14$ nonempty binary strings of lengths at most three account for at most $14v$ others. All remaining grid words cost at least three tokens, apart from at most $K$ one-token words. In total the excess over the comparator is at least
\[
 m^2-v^2/4-14v-2K=\Omega_\epsilon(m^2).
\]
In both cases division by the comparator saving $V\le N=O(m^2\log k)$ proves the claim. For the equal-budget conclusions, use any fixed $\epsilon<1$ and combine with Theorem~\ref{thm:fixed}; a binary corpus embeds into every larger alphabet.

\paragraph{Remark.} Fresh optimal resegmentation can abandon the first commitment. With that freedom and arbitrarily chosen further entries, the first entry together with the $k$ comparator entries already permits an optimal segmentation. Also, on a single word the unrestricted candidate set contains the whole word and the first step can achieve the full savings ceiling. Theorem~\ref{thm:binary}(a) concerns the committed parse of a collection.

\subsection{Proof of Theorem~\ref{thm:capped-exhaustion}}
\begin{proof}[Proof of Theorem~\ref{thm:capped-exhaustion}]
Consider a partition of a word of length $n\ge1$ into $t$ pieces for which adjacent lengths sum to at least $M+1$. Compare it with any capped partition into $p$ pieces, so $p\ge\lceil n/M\rceil$. If $t=2q$, pairing consecutive pieces gives $n\ge(M+1)q$, hence $Mt-n\le(M-1)q\le(M-1)p$. If $t=2q+1$, the remaining piece gives $n\ge(M+1)q+1$ and $p\ge q+1$. This gives $Mt-n\le(M-1)(q+1)\le(M-1)p$. In either case the savings satisfy
\[
 n-t\ \ge\ (1-1/M)(n-p).
\]

Starting from the current committed units, join neighboring pieces whose total length is at most $M$ until all adjacent lengths sum to at least $M+1$. Each join deletes one $M$-short boundary. The neighboring pieces only grow, so further boundaries cannot become short. These auxiliary partitions therefore add at most $I_M(i)$ to the corpus savings, counting multiplicities. Apply the preceding inequality to each resulting partition and to the optimal capped segmentation of that word under a dictionary attaining $\OPT^{\le M}_k$. Summing over the corpus gives $(1-1/M)\OPT^{\le M}_k\le G_i+I_M(i)$.

Every $M$-short boundary spells an eligible candidate, so Lemma~\ref{lem:short} gives $I_M(i)\le 2N_M(G_{i+1}-G_i)$ for both loops. At exhaustion this implies $I_M(i)=0$ whenever $N_M\ge1$; if $N_M=0$, all savings vanish. Each positive selection uses a new eligible type, proving the bound on the number of selections. For the finite-budget inequality, set $H=(1-1/M)\OPT^{\le M}_k$. Then $H-G_{i+1}\le(1-1/(2N_M))(H-G_i)$ (with $G_{i+1}=G_i$ after exhaustion), and iteration from $G_0=0$ proves the claim.

For sharpness take the single word $w=ab^Ma$. One type of length $\ell\le M$ used $r$ times saves $r(\ell-1)$. One occurrence saves at most $M-1$, and $r\ge3$ saves at most $|w|-3=M-1$. Two occurrences could save $M$ only by tiling $w$ with two identical strings. That would put its two letters $a$ at distance $|w|/2$, whereas their distance is $M+1$. Hence $b^M$ maximizes the first gain (tied with $ab^{M-1}$ and $b^{M-1}a$). Choosing it leaves lengths $1,M,1$ and exhausts the run at savings $M-1$. The two pieces $ab\mid b^{M-1}a$ attain the optimal capped savings $M$.

To force that first choice, use $C$ copies of $w$ and one extra word $b^M$. Only $b^M$ saves $M-1$ on the extra word. The run therefore commits $b^M$ and exhausts at $(C+1)(M-1)$. For every $k\ge2$ the two main-word types already save $CM$, so the ratio is at most $(C+1)(M-1)/(CM)$, which tends to $1-1/M$. For $k=1$, $G_\infty\ge g_1=\OPT^{\le M}_1$ by the argument of Proposition~\ref{prop:easy} on capped candidates, and a single two-character word attains ratio one.
\end{proof}

%% file: app/string.tex
\section{One binary string: proofs of Theorems~\ref{thm:string} and~\ref{thm:repair} and Corollary~\ref{cor:eptas}}\label{app:string}

The following theorem gives two ways to measure the cost of choosing the first rule greedily. Its recovery algorithm may select all later rules and all their replacement sites with complete knowledge of the input.
Symbols in this appendix are local to it. Theorem~\ref{thm:string} is stated in Section~\ref{sec:fixed}.

\begin{corollary}\label{cor:shortest}
The minimum length of a binary string on which greedy can save strictly less than a $k$-rule optimum is $\Theta(k\log k)$. The same order suffices to force a deficit $\Omega(1/\log k)$ on every run. In the regime $k=\Theta(|s|/\log|s|)$ the worst-case final-length factor has order $|s|/\log|s|$ along the examples of Theorem~\ref{th:string}(2).
\end{corollary}
\begin{proof}
Run $t=\lfloor k/\log_2 k\rfloor$ greedy rules. By Theorem~\ref{th:main}(b), at most $(2+o(1))|s|/\log_2 t$ adjacencies remain, uniformly over nonempty binary strings. Each additional nonempty rule removes at least one. Thus every string of length at most $(1/2-\delta)k\log_2 k$ reaches one token for each fixed $\delta>0$ and sufficiently large $k$. Theorem~\ref{th:string}(1) gives the matching length order. When $k=\Theta(|s|/\log|s|)$, Theorem~\ref{th:main}(b) also bounds greedy output by $O(|s|/\log|s|)$ tokens and the optimum is at least one.
\end{proof}

\paragraph{A family of cell identities.}
Take integers $H,d$ with $H,d\to\infty$ and $H/d\to\infty$. Write $D=Hd$, $m=9D$ and fix $r_0=6$. Put $\alpha(a)=2$, $\alpha(b)=1$, $\beta(a)=1$, $\beta(b)=2$. An identity
\[
 i=(s,p,u,e,q,v)
\]
has a starting letter $s$ and an ending letter $e$ in $\{a,b\}$, indices $0\le p<\alpha(s)$ and $0\le q<\beta(e)$, and coordinates $0\le u,v<H$ satisfying $t=(v-u)\bmod H<d$. There are $m$ identities. We construct a single text consisting of $n=3r_0m=18m$ cells in which every identity occurs exactly $3r_0$ times.

\paragraph{A graph giving the order of the cells.}
Take the $9r_0$ slots $(s,e',A,B,j)$ with $0\le A<\alpha(s)$, $0\le B<\beta(e')$ and $0\le j<r_0$. Number the slots from zero, putting first the three with $s=e'=a$, $A=B=0$ and $j=0,2,4$ in that order. For slot number $r$ put $h_r=rd+\lfloor r/2\rfloor$. Each predecessor coordinate pair $(u,v)$ determines a successor coordinate pair $(u',v')$ by
\[
 x=vd+t,\qquad y=(x+h_r)\bmod D,\qquad
 u'=y\bmod H,\quad t'=\lfloor y/H\rfloor,\quad v'=(u'+t')\bmod H.
\]
This is a bijection since $x=vd+t$ and $y=Ht'+u'$ each enumerate $\{0,\ldots,D-1\}$. Form the vertices
\[
 P_z=(s,A,u,z,j\bmod\beta(z),v),\qquad
 Q_z=(z,j\bmod\alpha(z),u',e',B,v')
\]
for $z\in\{a,b\}$, and insert the three directed arcs
\[
 P_a\to Q_a,\qquad P_b\to Q_a,\qquad P_b\to Q_b .
\]

Every vertex has indegree and outdegree $3r_0$. For an outgoing arc from $(s,p,u,e,q,v)$, the slot must have $A=p$ and $j\bmod\beta(e)=q$. There are three choices across $(e',B)$ and $r_0/\beta(e)$ possible $j$'s, with $\beta(e)$ outgoing arcs from each. For an incoming arc $B=q$, there are three choices across $(s_{\rm prev},A)$, and $r_0/\alpha(s)$ possible $j$'s, each supplying $\alpha(s)$ arcs. Inverting the slot's coordinate bijection supplies its predecessor.

The graph is weakly connected. In slots $0,1,2$ each common predecessor $P_a$ is adjacent to the three targets in marker class $(a,a)$ with $p=q=0$ and target coordinates indexed by $x,x+d,x+2d+1$ modulo $D$. Since $d$ and $2d+1$ are coprime, all these target vertices belong to one component. Every vertex of marker class $(a,a)$ can reach this component directly by taking an even $j$ in a slot with $e'=a$ and its own $A=p$. Every vertex can reach some vertex of marker class $(a,a)$ by taking $e'=a$, $A=p$, a compatible $j$ and the arc to $Q_a$. Thus we can traverse a single directed Euler circuit. Write the $n$ tails in circuit order as a linear list, omitting the closing arc.

The signature of an arc from $i$ to $i'$ is $((e_i,v_i),(s_{i'},u_{i'}))$. Every arc on the circuit has a different signature. The marker pair specifies one of the three arcs in its triple. Given $v$, the second coordinate is $u'=(vd+t+h_r)\bmod H$ since $H\mid D$. For all sufficiently large $H/d$ the $9r_0$ intervals $[h_r,h_r+d-1]$ are disjoint within $[0,H-1]$. Thus $r,t$ are unique and the slot and predecessor are determined.

\paragraph{An obstruction at one third of the cuts.}
Give each identity $i$ two integer offsets $S_i,E_i$ and two letters $A_i,B_i$. An arc $i\to i'$ passes if $E_i=S_{i'}=w$ and the corresponding condition holds:
\[
 \begin{cases}
 A_i=s_{i'} & w>0,\\
 B_{i'}=e_i & w<0,\\
 e_i s_{i'}\ne aa & w=0 .
 \end{cases}
\]
If all three arcs of a triple passed, their four endpoint offsets would coincide. Zero fails on $P_a\to Q_a$. A positive value would require $P_b$ to predict both starts, and a negative value would require $Q_a$ to predict both ends. Thus at least $n/3$ arcs of the circuit fail under every assignment.

\paragraph{Binary spelling and full comparator.}
Let $U$ and $V$ be fixed-width encodings of $(s,p,u)$ and $(e,q,v)$, each using $\ell=\lceil\log_2(3H)\rceil$ bits. Write $P=ab$, assign five logical symbols $L,R,c,0,1$ distinct exponents $j_t\in\{4,5,6,7,8\}$, and set $\mathcal E(t)=Pa^{j_t}$, extended by concatenation. The physical cell is
\[
 s\,PP\,a\,\mathcal E\big((LU)^3c(RV)^3\big)\,PPPP\,e.
\]
Concatenate the cells in the specified order, obtaining a text of length $N=\Theta(n\log H)$ (the $|s|$ of Theorem~\ref{thm:string}; here $s$ is a marker letter). If a cell has $f=6\ell+7$ logical symbols, the $a$-runs inside their encodings supply at least $2f$ disjoint $aa$ matches. The cell has at most $f+8$ letters $b$, including its two markers. Every competing pair contains a $b$ on a specified side, hence has raw count at most the total number of $b$'s. Since $f>8$, both greedy rules uniquely select $aa$. Its full left-to-right application erases every $aa$-labelled nominal cut, as the previous cell's closing frame ends in $b$ immediately before its final marker.

A full comparator first uses $ab\to P$. All physical $ab$ occurrences are designated copies because no cut is labelled $ab$ and the frames begin with $a$ and end with $b$. The rules $PP\to X$, $XX\to Y$ distinguish the initial and final frames. Extend each remaining $P$ by one $a$ at a time up to $\max j_t$ to construct the logical symbols, then apply $(X,a)$ to complete the initial frames. From the $L$ and $R$ tags build all occurring fields by prefix rules in increasing length. Both prefix trees cost $O(2^\ell)=O(H)$ rules. Two rules per left field join its three successive copies, and then attach the initial frame and its marker $s$. Likewise join the three copies of each right field and attach $Y$, its marker $e$, and the center tag $c$ on the left. The fields specify these markers. One root rule per identity now produces the $n$ cell tokens at cost $m+O(H)$ rules. The first two cells have a unique ordered pair by signature uniqueness. Their merge creates a token occurring once. Merging the growing prefix with its next cell therefore uses $n-1$ further full rules to reduce the whole input to one token.

\paragraph{Separating the cell centers.}
The witness of a cut extends from the start of the third $\mathcal E(RV)$ in its preceding cell through the first $\mathcal E(LU)$ in its following cell, including the initial $P$ of the next $\mathcal E(L)$. Its spelling locates a closing frame $PPPP$; all such frames are genuine because all $ab$ occurrences are designated. The gaps between successive $P$'s then recover both encoded fields. Every occurrence of a witness has the same cut signature, so each witness has one location in the text. The center pattern $Pa^{j_c}P$ likewise occurs exactly at centers, since nonlogical all-$a$ gaps have length zero, one or two. Use its first character as the cell center.

Consider any parse obtained by $K$ rules beginning with the full $aa$ step and write $F_0$ for its output length. Recursively split every token containing an entire witness until no token does. Each split node has a unique location in the text, because its spelling contains a witness. Distinct split nodes carry distinct types: another occurrence of that type would have the same location, and proper ancestors have strictly greater lengths. If there were $t_*$ splits, at most $K+2-t_*$ types, counting the two letters, are available to the surviving tokens. None of these splits restores an erased $aa$ cut. In a binary parse tree the sole node whose two children meet at that original cut is the $aa$ node created on the first step, and its two-letter spelling contains no witness.

Each center now lies in a separate token since a token covering consecutive centers contains their witness. Write $h$ for the number of remaining tokens without a center, so
\[
 F_0+t_*=n+h .
\]
Call an interior center good if its token touches the preceding and following center tokens directly. At most $b\le 2h+2$ centers are nongood, including the two extreme centers. A token preceding a good center cannot reach the end of the current cell's first $\mathcal E(LU)$ plus the next $P$, since it would contain the witness from the preceding center's cell. The following center token must start after the beginning of the third $\mathcal E(RV)$ for the same reason. Thus a good token contains the entire third $\mathcal E(LU)$, the center, and the entire first $\mathcal E(RV)$ with its next delimiter. Its spelling identifies its cell identity and the position of its center within the token.

\paragraph{Accounting for types and output tokens.}
For each identity having any good occurrence, designate its most frequent good token type, resolving ties arbitrarily. Let $z$ be the number of identities with no good occurrence. Every identity occurs $3r_0$ times, so $z\le b/(3r_0)$. Since a good type identifies its identity, at most $K+2-t_*-(m-z)$ good types are additional to the designated ones. If an identity has $l\ge2$ good types, at most $3r_0(l-1)/2$ of its good occurrences use additional types. For $l=2$ the less frequent type appears at most half the time; for $l\ge3$ the bound is at least the total frequency. Discard nongood cells and good cells using additional types. At most
\[
 b+\tfrac{3r_0}{2}(K+2-t_*-m+z)
 \ \le\ \tfrac32 b+\tfrac{3r_0}{2}(K+2-t_*-m)
\]
cells are discarded.

For each designated type the decoded identity and center position determine its left and right endpoint offsets $S_i,E_i$ from the nominal cell. When $E_i>0$ the type fixes the letter $A_i$ immediately after the nominal right cut, and when $S_i<0$ it fixes $B_i$ immediately before the nominal left cut. Complete unused entries arbitrarily. At every linear-text cut between retained cells the designated tokens touch. The offsets are equal, and their sign condition holds: a positive or negative offset exposes the prescribed letter, while zero could not survive on an erased $aa$ cut. The failing arcs of the Euler circuit therefore comprise at most two per discarded cell and the one omitted arc. Substitution gives
\begin{align*}
 n/3 &\le 3b+3r_0(K+2-t_*-m)+1\\
     &\le 6h+7+3r_0(K+2-n-m+F_0-h).
\end{align*}
Since $3r_0=18>6$ and $n=3r_0m$, we obtain the budget-independent inequality
\[
 \boxed{\quad K+F_0\ \ge\ n+\tfrac43m-2-\tfrac{7}{3r_0}.\quad}
\]
It applies to any choice of partial or full rules following the forced full first step.

\paragraph{Choosing the parameters.}
Put $H=\lfloor k^{3/4}\rfloor$ and choose $d=\lfloor(k-H\lceil\log_2 k\rceil^2)/(9H)\rfloor$ for part~(1). Then $m=k-o(k)$ and the comparator making the cells costs at most $k$. For $K\le(1+\gamma)k$, the boxed inequality gives $F_0-n\ge(1/3-\gamma-o(1))k$. The comparator saves $N-n=\Theta(k\log k)$, proving the relative loss. For part~(2) use $d=\lfloor(k-H\lceil\log_2 k\rceil^2)/(171H)\rfloor$. Now $n+m=19m=k-o(k)$, and the full comparator down to one token fits within $k$. The boxed bound gives $F_0\ge(1/57-o(1))k$. In both cases $H/d\to\infty$, $d\to\infty$, and $N=\Theta(k\log k)$. Constructing the graph and a deterministic Euler traversal takes $O(k)$ time. Writing the physical text takes $O(k\log k)$ time. Integer fourth roots and binary lengths suffice to set the parameters. This proves Theorem~\ref{th:string}.

\subsection{Recovering from a prescribed first merge}

The lower bounds ask how much a first greedy decision can cost even when every subsequent choice is optimal. On a single binary string there is a universal recovery construction.

Theorem~\ref{thm:repair} is stated in Section~\ref{sec:fixed}; we write $F$ for the comparator's number of tokens.
\begin{proof}[Proof of Theorem~\ref{thm:repair}]
Suppose the prescribed pair is $xy$ and its replacement token is $Z$. A left-to-right scan has two states, empty and holding one pending $x$. From the empty state, hold a newly read $x$ and emit any other letter. When holding $x$, reading $y$ emits $Z$ and clears the state. On every other read, emit the pending $x$ and process the new letter by the empty-state rule. At the end emit any pending $x$. This produces exactly the token string after the prescribed full rule, including the case $x=y$.

For each type $A$ of the comparator and incoming state $q$, represent the string $T_q(A)$ emitted while scanning the spelling of $A$, together with the resulting state. Postpone the final flush. For a terminal these outputs have length zero, one or two. Over a binary alphabet the only possible two-token output occurs for $x=y$ on reading the other letter $\bar x$ while holding $x$; it is $x\bar x$. Thus at most one preliminary binary rule represents all terminal outputs.

Process comparator rules in their original order. For a comparator rule $A\gets BC$ and state $q$, let $q'$ be the outgoing state for $B$ entered in state $q$. Then $T_q(A)=T_q(B)T_{q'}(C)$. Each of the two states needs at most one binary rule. When a child output is empty, use the other child's representation; if both are empty, record an empty output. In total the comparator rules need at most $2k$ new binary rules.

Run the states through the comparator's $F$ final tokens in order. Their nonempty outputs partition the prescribed first-rule token string except for a possible pending $x$. If one remains, attach it to the last nonempty output by one more binary rule; if every output was empty, the single letter $x$ suffices. The result has at most $F$ pieces. Realize the indicated derivation tree for each piece on its own span, applying only its prescribed joins in topological rule order. Occurrences of one rule in these trees are disjoint, and each child was constructed by an earlier rule or was already present. This gives a partial continuation on the actual first-rule output. Adding the initial rule and the two possible preliminary and final rules gives $2k+3$.
\end{proof}

The proof is a two-state specialization of the use of finite-state transducers on compressed words. Finite-state representations of BPE and transformations of straight-line programs by transducers are established~\cite{fstat,compressed}. The quantitative conclusion here counts binary rules and final tokens under a fixed budget. It is existential given the comparator. On a corpus the flush occurs separately at each word end; the argument above may leave one additional token per word if those end corrections are omitted.

For precision, define the asymptotic recovery factor on single binary strings to be the infimum of the $a$ for which $ak+o(k)$ total rules always suffice, after the first full greedy merge, to attain the final token count of the best $k$ full rules using a partial continuation. Theorem~\ref{th:string}(1) and Theorem~\ref{th:repair} place this factor in $[4/3,2]$. Theorem~\ref{th:binary} gives a lower bound of $2$ for the corresponding corpus factor.

\subsection{The approximation scheme}
\begin{corollary}\label{cor:eptas}
For each fixed alphabet, optimal full and optimal partial merge savings admit deterministic efficient polynomial-time approximation schemes. On an explicitly written corpus the running time is $f(\sigma,1/\epsilon)|\mathcal C|$ on a word RAM.
\end{corollary}
\noindent Here $|\mathcal C|$ is the total length of the corpus.
\begin{proof}[Proof of Corollary~\ref{cor:eptas}]
Given $0<\epsilon<1$, put $M=1+\lceil4/\epsilon\rceil$, $S=\sum_{j=2}^M\sigma^j$, and $K=\lceil4S/\epsilon\rceil$. If $k\ge K$, run greedy for at most $K$ steps. Theorem~\ref{th:main}(b) gives $G_K\ge P(\mathcal C)/(1+\epsilon)\ge (1-\epsilon)P(\mathcal C)$. The $K$ scans take parameter-dependent linear time using a frequency table on at most $\sigma+K$ types.

For $k<K$, enumerate the at most $(k+1)(\sigma+k)^{2k}$ topologically ordered lists of at most $k$ binary rules. For a full optimum, simulate each list. For a partial optimum, expand its types to strings and use a shortest-path segmentation dynamic program on each word, allowing the terminal and constructed types. Every partial execution gives a feasible segmentation. Conversely put the listed derivation tree for each chosen segment on its span. At each rule perform just the joins bearing that rule in these trees. Nodes bearing the same rule have disjoint spans and their children have already been formed. This realizes the segmentation even when several IDs spell the same string. Backpointers recover the choices. All expansions have length at most $2^k$, so evaluation of each list takes parameter-dependent linear time. The best list gives an exact answer in the small-budget branch.
\end{proof}

%% file: app/hard.tex
\section{Hardness of approximation}\label{app:hard}
Write \textsc{Tok}$_\ast$ for the maximization of $f$ with an unrestricted vocabulary,
\textsc{Tok}$_T$ for the version of \cite{lim2025} in which a candidate set $T$ is part of the input, and Max $k$-VC for
choosing $k$ vertices of an edge-weighted graph to maximize the weight of covered edges.

\begin{theorem}[Hardness of approximation]\label{thm:hard}
(i) For every $\delta>0$, \textsc{Tok}$_\ast$ is NP-hard to approximate within $\frac{31579}{31580}+\delta$; in particular it
is APX-hard and has no PTAS unless P$=$NP.  (ii) \textsc{Tok}$_T$ is at least as hard as Max $k$-VC: the construction of
\cite{lim2025} satisfies $f(S)=2\cov(C_S)$ for every $S\subseteq T$, so \textsc{Tok}$_T$ is UG-hard to approximate within
$\approx0.929$ (for the polynomially bounded weights of \cite{as19}; Appendix~\ref{app:hard}).
\end{theorem}

\paragraph{Overview.}
For (i) we reduce from Max-2SAT with every variable in exactly three
clauses (Berman--Karpinski \cite[\S7]{bk}), via weighted Max $k$-VC on the literal graph.  Each vertex $u$ gets $\Lambda$ fresh
copy characters $u^i$, and an edge $\{u,v\}$ becomes the words $Z u^i Z v^j Z$.  Heavy \emph{forcing words} $Zu^i$ and $u^iZ$ make every bigram $Zx$, $xZ$ worth its slot in the budget.  With all
bigrams present, the remaining budget $\Lambda k$ can buy trigrams $Zu^iZ$, and each edge word gains one pair iff an
endpoint copy is chosen; pipage rounding shows that the blown-up graph's optimum is $\Lambda^2$ times the original
one.  Other tokens (edge tokens, whole words) can appear in optimal vocabularies, but the upper-bound argument
charges them at most $w_{\max}$ per edge word, $O(\Lambda k)$ in total, against $\Theta(\Lambda^2)$.  This gives
$B+\Lambda^2\OPT_{\mathrm{VC}}\le\OPT_{\mathrm{tok}}\le B+\Lambda^2\OPT_{\mathrm{VC}}+O(\Lambda k)$, and the
Berman--Karpinski gap survives with the stated constant, which we have not tried to optimize.  Part (ii) is a quantitative reading of the NP-hardness
construction of \cite{lim2025}.

The theorem has the following consequence for the approximability of the problem.

\begin{corollary}
Unless P$=$NP, the best approximation ratio achievable in polynomial time for \textsc{Tok}$_\ast$ lies in
$[\MAINconst,\ 1-1/31580]$; the lower end is guaranteed by \GT{} (Theorem~\ref{thm:main}).  In particular \textsc{Tok}$_\ast$
has a polynomial-time constant-factor approximation but no PTAS unless P$=$NP.  (\textsc{Tok}$_\ast$ allows an unbounded
alphabet; over a fixed alphabet an EPTAS exists \cite[Theorem~5.6]{ion2026}.)
\end{corollary}

\begin{remark}[Relation to the compressed-length hardness]
Kastreva et al.\ \cite{kastreva2025} define both the compressed length and the compression reduction; maximizing our $f$
is optimizing the latter (the number of symbols saved).  Their APX-hardness theorems are for the compressed length,
and they point out that the two objectives rank tokenizers identically but differ for approximation
\cite[\S2]{kastreva2025}.  Their reduction does not transfer: it encodes variable $X_j$ by the strings $0^{2j-1}$ and
$0^{2j}$, so the corpus has $\Theta(J^2)$ characters for $J$ variables, while the compressed length and the clause signal
are both $\Theta(J)$.  For pairs covered the optimum is $\Theta(J^2)$, so the induced gap ratio tends to $1$.
Theorem~\ref{thm:hard} uses an unbounded alphabet.  Over bounded alphabets, maximizing $f$ is NP-hard even for binary and
unary alphabets \cite[Thms~1, 5]{kastreva2025}, but it is not APX-hard over a fixed alphabet unless P$=$NP: Ion et al.\ give a
deterministic EPTAS for the same objective on every fixed alphabet \cite[Theorem~5.6]{ion2026}.
\end{remark}
\input{gt/hardness_app.tex}

\section{Fidelity to the released implementation}\label{app:impl}
We compared our model with the paper's Algorithms~1--3 \cite{lim2025} and with the authors' implementation
\texttt{pcatt} at commit \texttt{36b928c7} (16 April 2026), which contains an original solver
(\texttt{greedy\_builder.cpp}), the shipped solver (\texttt{pco\_tokenizer.cpp}) and an encoder
(\texttt{greedy\_encoder.cpp}).

\paragraph{Validity, gain and commit.}  The implementation labels each character with the committed occurrence that
contains it and tests validity by checking whether the two characters straddling a boundary carry the same label,
which is our condition.  It computes the gain as $\#\text{uncovered chars}+\#\text{distinct committed occurrences
inside}-1$, which equals the number of uncovered interior pairs by Lemma~\ref{lem:basic}(ii).  Alg.~3 of
\cite{lim2025} writes the index range as $\{i,\dots,i+|T|\}$; its Example~4 and the code make clear that the $|T|-1$
interior pair entries are meant.  A commit relabels the characters of each placed occurrence with a fresh label, and validity is re-tested against the
labels already written in the same pass.  This places exactly the same occurrences as the scan.  An occurrence that
overlaps the previously committed one in the pass has both characters of its left boundary pair inside that occurrence,
now with the same fresh label, so it is rejected, as the scan rejects it.  For a later non-overlapping occurrence, no
character of its right boundary pair or interior was written in the pass, and at most one character of its left
boundary pair was, with a fresh label equal to no older one; so the test gives the same answer as at time~$t$.

\paragraph{Ties and laziness.}  The original solver re-scores every candidate in every affected word, so it is the
exact greedy; on ties it takes the first maximum in hash-map order.  The shipped solver is lazy: it re-scores only
candidates whose words were touched since their last evaluation, and on ties it prefers the shorter token.  Since
$\Sc_t(x)$ depends only on the states of the words containing $x$, a stored score whose words were not touched is
still current; so every stored score is the true score at some earlier time, the accepted one is current, and Corollaries~\ref{cor:mono} and~\ref{cor:lazy} apply
to it as to Algorithm~1; since this rests on our reading of the code, we count it as an argument, not a proof.  Our bounds hold for any tie-breaking.  On a $k=3$ adversarial-tie instance in the supplement, a compiled run of the shipped solver attains the worst tie-break, ratio $\PcattThree$.  A
differential test of the compiled code against our simulator agreed on all $\PcattTOne$ score comparisons and all
$\PcattSteps$ greedy steps, and found no non-maximal choice of the lazy solver in $\PcattLazyN$ random instances
(Appendix~\ref{app:comp}).

\paragraph{Reading of Algorithm~1.}  Lines 16--18 of Algorithm~1 use $\tau$ and $I(\tau)$; we read this as a typo for
the accepted token $\hat\tau$, and the first popped token is committed too.  The positions $P(T)$ are visited left to
right within each word, as in Example~4 of \cite{lim2025} and in the implementation.  Read literally, the commit block
only ever writes $I(\tau)=0$ for the first token $\tau$, so it never changes the state, and Algorithm~1 returns the $k$
tokens of largest initial score.  The committed state then stays empty, so we measure such a run by the re-encoded
value $f(S_k)$; its worst-case ratio is then $1/k$: at
least $1/k$ by the argument of Proposition~\ref{prop:easy}, and a family with one long word and $k-1$ shorter words on
fresh letters approaches it.  Visiting positions in another order also breaks Lemma~\ref{lem:int}: on
\texttt{abababab} with $x=\texttt{abab}$, starting the scan at position $2$ scores $3$ instead of $6$.  Neither of these two alternative readings
matches the implementation.

\paragraph{Restricted candidates and the encoder.}  The implementation can cap token length and drop rare words; our
lower bounds then hold with $T$ and the corpus restricted accordingly, since every window of a token in $T$ is in $T$.
The exceptions are Theorem~\ref{thm:fixed}(a) and the results of Appendix~\ref{sec:alphabet} that use it, which need every
substring of length at most $M$ to be a candidate ($M\le3$ for Corollary~\ref{cor:byte} and Theorem~\ref{thm:half}, $M\approx\log_\sigma k$
for the fixed-alphabet consequence of Theorem~\ref{thm:fixed}(a)); so does Theorem~\ref{thm:capped-exhaustion}, whose
candidates are all substrings of lengths $2,\dots,M$.
Our lower bounds are for the committed value $G_k$ and hence also for the optimal re-encoding $f(S_k)\ge G_k$; the
upper-bound constructions of Appendix~\ref{sec:upper} bound the committed value $G_K$, and for J4 also the re-encoded
value $f(S_K)=G_K$ (Remark~\ref{rem:reencode}).  The released encoder labels
characters by rank only and can split a training word more finely than the committed state, e.g.\ with ranks
$(\texttt{ab},\texttt{abc},\texttt{ba})$ on \texttt{ababc}; our bounds do not cover it.

%% file: gt/hardness_app.tex
\textbf{Max $k$-VC} (edge-weighted, loops allowed): graph $H=(V,E)$, integer weights $w_e\ge1$, loops $\ell_u\ge0$;
$\cov(C)=\sum_{e\cap C\ne\emptyset}w_e+\sum_{u\in C}\ell_u$; maximize over $|C|\le k$.
Write $W_E=\sum_e w_e$, $W_\ell=\sum_u\ell_u$, $\deg_w(u)=\sum_{e\ni u}w_e+\ell_u$, $\Delta_w=\max_u\deg_w(u)$,
$w_{\max}=\max(\max_e w_e,\max_u\ell_u)$.

\subsection{With a candidate set: exactly Max \texorpdfstring{$k$}{k}-VC}
\begin{theorem}\label{hthm:restricted}
Given $(H,k)$, let the corpus have one word $Z\,u\,Z\,v\,Z$ of count $w_e$ per edge $e=\{u,v\}$ and one word
$Z\,u\,Z$ of count $\ell_u$ per loop, with $T=\{ZuZ:u\in V\}$ and $k_{\mathrm{tok}}=k$. Then $f(S)=2\cov(C_S)$ for every
$S\subseteq T$, where $C_S=\{u:ZuZ\in S\}$. Hence $\OPT(\textsc{Tok}_T)=2\,\OPT_{\mathrm{VC}}(H,k)$ and solutions
correspond one-to-one: every $(c,s)$-gap of Max $k$-VC transfers to a $(2c,2s)$-gap of \textsc{Tok}$_T$.
\end{theorem}
\begin{proof}
In $ZuZvZ$ the only candidates occurring are $ZuZ$ at $[0,2]$ and $ZvZ$ at $[2,4]$; they share position 2, so
at most one is used, covering 2 pairs; in $ZuZ$ the only candidate is the whole word (2 pairs).
\end{proof}
This is a weighted extension of Lim et al.'s NP-hardness construction (from Vertex Cover) read quantitatively; it is
approximation-preserving. Consequences: \textsc{Tok}$_T$ (coverage objective) is APX-hard (for $T$ the set of all input bigrams this also follows
from \cite[\S2]{kozma2024arxiv}); it is UG-hard to
approximate within $\approx0.929$, the tight UG-hardness of Max $k$-VC that Austrin and Stankovi\'c \cite{as19} derive from
monotone Max-2-Sat with a cardinality constraint (for the polynomially bounded weights their reduction produces); and it is NP-hard within $3355/3356+\varepsilon$ (Lemma~\ref{hlem:bk} with $V\le1344N$).
Since that bound is tight for Max $k$-VC (and Manurangsi's $0.92$-approximation \cite{manurangsi} already comes close),
$0.929$ is essentially the truth \emph{for these instances}; it says nothing about algorithms for \textsc{Tok}$_T$ in general.

\subsection{Unrestricted vocabularies}
\paragraph{Construction $I(H,k,\Lambda)$.} For $\Lambda \ge1$, every vertex $u$ gets $\Lambda$ fresh characters $u^1,\dots,u^\Lambda$;
one more character $Z$. Let $\Omega=\Lambda \Delta_w+w_{\max}$. Words (counts in brackets):
(E) $Z\,u^i\,Z\,v^j\,Z$ $[w_e]$ for every edge $e=\{u,v\}$ (one fixed orientation) and all $i,j\in[\Lambda]$;
(L) $Z\,u^i\,Z$ $[\Lambda \ell_u]$ for every $u,i$;
(F) $Z\,u^i$ $[\Omega]$ and $u^i\,Z$ $[\Omega]$ for every $u,i$.
Budget $k_{\mathrm{tok}}=2\Lambda n+\Lambda k$ ($n=|V|$). Call $Zx$, $xZ$ \emph{bigrams}, $T_x=ZxZ$ \emph{trigrams} ($x$ a copy character),
and every token containing two distinct copy characters an \emph{edge token}; each edge token occurs in exactly one
(E)-word (no two copy characters are adjacent, and $\{u^i,v^j\}$ determines the word), and no token other
than these occurs anywhere ($ZZ$ and $xy$ never occur).

\begin{lemma}[word values]\label{hlem:word}
Let $w=ZaZbZ$ be an (E)-word and $S$ arbitrary. (i) If $S$ contains its four bigrams and none of its edge tokens,
$\mathrm{cover}(w,S)=2+[T_a\in S\vee T_b\in S]$. (ii) Always
$\mathrm{cover}(w,S)\le 2+[T_a\in S\vee T_b\in S\vee \epsilon_w]+[\epsilon_w]$, $\epsilon_w$ = ``$S$ contains an edge token of $w$''.
For a loop word $ZaZ$: $\mathrm{cover}\le1+[T_a\in S]$, with equality if $Za,aZ\in S$.
\end{lemma}
\begin{proof}
$\mathrm{cover}=5-p$, $p$ the number of pieces. $p=1$ needs the token $w$ (an edge token). A 2-piece partition has
piece lengths $(1,4),(4,1)$ (edge tokens $aZbZ$, $ZaZb$), $(2,3)$ ($Za\cdot ZbZ$) or $(3,2)$ ($ZaZ\cdot bZ$); so cover
$=3$ needs $T_a$, $T_b$ or an edge token, and cover $\le 2$ otherwise. This gives (ii). For (i): $Za\cdot Zb\cdot Z$
gives 2, and $ZaZ\cdot bZ$ / $Za\cdot ZbZ$ give 3. Loop word: cover 2 needs the whole word $T_a$; 1 via $Za$.
\end{proof}

\begin{lemma}[blow-up, pipage]\label{hlem:pipage}
Let $H[\Lambda]$ be the weighted graph on copies with edge $\{u^i,v^j\}$ of weight $w_e$ and loop weight $\Lambda \ell_u$ at $u^i$.
For $A\subseteq V\times[\Lambda]$ put $x_u=|A\cap\{u^1..u^\Lambda \}|/\Lambda$. Then $\cov_{H[\Lambda]}(A)=\Lambda^2F(x)$ with
$F(x)=\sum_e w_e(1-(1-x_u)(1-x_v))+\sum_u\ell_ux_u$, and $\OPT_{H[\Lambda]}(\Lambda k)=\Lambda^2\OPT_H(k)$.
\end{lemma}
\begin{proof}
Edge $e$: $(\Lambda -\Lambda x_u)(\Lambda -\Lambda x_v)$ of its $\Lambda^2$ copies are uncovered; loops: $\Lambda  x_u$ copies, weight $\Lambda \ell_u$ each.
$F$ is monotone, and along $e_u-e_v$ it is linear if $uv\notin E$ and convex if $uv\in E$
(the $t^2$ coefficient is $+w_{uv}$). Pipage rounding (Ageev--Sviridenko \cite{ageev2004}) moves any $x\in[0,1]^V$ with $\sum x\le k$
to an integral point with $\sum x\le k$ without decreasing $F$, so $\max F=\OPT_H(k)$; the integral optimum is
realized by taking all copies of an optimal $C$.
\end{proof}

\begin{theorem}\label{hthm:main}
Let $B=2\Lambda n\Omega+\Lambda^2(2W_E+W_\ell)$. Then
\[ B+\Lambda^2\OPT_H(k)\ \le\ \OPT(\textsc{Tok}_\ast(I(H,k,\Lambda)))\ \le\ B+\Lambda^2\OPT_H(k)+w_{\max}\Lambda k. \]
\end{theorem}
\begin{proof}
\emph{Lower.} $S^\ast$ = all $2\Lambda n$ bigrams and $T_x$ for all copies $x$ of an optimal $C$ ($|S^\ast|=k_{\mathrm{tok}}$). By
Lemma~\ref{hlem:word}(i) and the loop clause, $f(S^\ast)=2\Lambda n\Omega+\sum_{(E)}w_e(2+[\text{covered}])+\sum_{(L)}\Lambda \ell_u(1+[\cdot])
=B+\cov_{H[\Lambda]}(\text{copies of }C)=B+\Lambda^2\OPT_H(k)$.

\emph{Upper.} Let $|S|\le k_{\mathrm{tok}}$, $b$ the number of bigrams in $S$, $d=2\Lambda n-b$, $A$ the copies with $T_x\in S$, and
$\mathcal E$ the (E)-words containing an edge token of $S$, $t=|\mathcal E|\le$ number of edge tokens in $S$. Then
$|A|+t\le k_{\mathrm{tok}}-b=\Lambda k+d$. Let $A'=A\cup\{$one endpoint copy of each word in $\mathcal E\}$, $|A'|\le \Lambda k+d$. By
Lemma~\ref{hlem:word}(ii), summing with counts,
\[ f(S)\le \Omega b+\Lambda^2(2W_E+W_\ell)+\cov_{H[\Lambda]}(A')+w_{\max}t . \]
Deleting $|A'|-\Lambda k\le d$ elements from $A'$ loses at most $\Lambda \Delta_w$ each (the weighted degree of a copy in $H[\Lambda]$),
so $\cov_{H[\Lambda]}(A')\le \Lambda^2\OPT_H(k)+d\,\Lambda \Delta_w$ by Lemma~\ref{hlem:pipage}. With $t\le \Lambda k+d$:
$f(S)\le B+\Lambda^2\OPT_H(k)+w_{\max}\Lambda k-d(\Omega-\Lambda \Delta_w-w_{\max})$, and $\Omega=\Lambda \Delta_w+w_{\max}$.
\end{proof}

\begin{corollary}[transfer]\label{hcor:transfer}
If it is NP-hard to distinguish $\OPT_H(k)\ge c$ from $\le s$ on a family with $D:=2n\Delta_w+2W_E+W_\ell$ and
$(n+k)\,w_{\max}$ polynomially bounded, then for every $\delta>0$ \textsc{Tok}$_\ast$ is NP-hard to approximate within
$\frac{D+s}{D+c}+\delta$ (take $\Lambda$ a large enough polynomial; $2\Lambda n\Omega=\Lambda^2\cdot2n\Delta_w+O(\Lambda)$).
\end{corollary}

\begin{lemma}[3-OCC-Max-2SAT $\to$ Max $k$-VC]\label{hlem:bk}
For a 2-CNF $\varphi$ on $V$ variables in which every variable occurs in exactly three clauses, let $H_\varphi$ have
a vertex per literal, an edge $\{x,\bar x\}$ of weight 1 per variable, an edge $\{l_1,l_2\}$ per clause $l_1\vee l_2$
($l_1\ne l_2$; parallel edges merge by adding weights) and a loop at $l$ per clause $l\vee l$; $k=V$. Then
$\OPT_{H_\varphi}(V)=V+\OPT_{\mathrm{sat}}(\varphi)$ and $w_{\max}\le\Delta_w\le4$ (repeated clauses merge, so $w_{\max}$ can reach 4; it only enters $O(\Lambda)$ terms).
\end{lemma}
\begin{proof}
$\ge$: the literals true under an assignment. $\le$: if $C$ ($|C|\le V$) contains both literals of some $x$ and none of
some $y$, delete the literal of $x$ lying in fewer clauses not containing the other literal (at most one such clause,
since $x$ is in three clauses; the variable edge stays covered) and add a literal of $y$ (gain $\ge1$ from its
variable edge). Repeat, then complete; the result is an assignment with $\cov=V+\#$satisfied. Degrees: $1$ plus at
most three clauses (parallel edges, including $x\vee\bar x$ merged into the variable edge, only add weights).
\end{proof}

\begin{theorem}\label{hthm:const}
For every $\delta>0$, \textsc{Tok}$_\ast$ (unrestricted vocabulary, pairs-covered objective, unbounded alphabet) is
NP-hard to approximate within $\frac{31579}{31580}+\delta\ (=1-1/31580+\delta)$. In particular it has no PTAS. It is also APX-hard: the upper-bound proof is constructive and yields $C$ with $\OPT_H-\cov(C)\le(\OPT_{\mathrm{tok}}-f(S))/\Lambda^2+w_{\max}k/\Lambda$; choosing $\Lambda >2w_{\max}k$ and using
integrality of $\cov$, the additive term is absorbed, and $\OPT_{\mathrm{tok}}=O(\Lambda^2N)$; this is an approximation-preserving (AP-) reduction
(its parameters depend on $\Lambda$, so it is not an L-reduction in the strict sense).
\end{theorem}
\begin{proof}
Berman and Karpinski \cite[\S7]{bk} show that for every $\varepsilon>0$ it is NP-hard to decide whether an instance of
3-OCC-MAX 2SAT with $2016N$ clauses has an assignment satisfying at least $(2012-\varepsilon)N$ clauses or at most
$(2011+\varepsilon)N$.  They define $k$-OCC as \emph{at most} $k$ occurrences, but their instances come from the gadget of
\cite[\S7]{bk}, which creates four variables per edge of a cubic graph, each occurring in exactly three of the clauses
(two in the edge gadget and one in the ring of the incident node).  So every variable occurs exactly three times, and
$3V=2\cdot2016N$ gives $V=1344N$. With Lemma~\ref{hlem:bk},
$c=V+(2012-\varepsilon)N$, $s=V+(2011+\varepsilon)N$, and $D\le 2(2V)4+2(V+2016N)=18V+4032N$. The ratio
$\frac{D+s}{D+c}$ increases in $D$ and $V$, so it is at most
$\frac{19\cdot1344+4032+2011}{19\cdot1344+4032+2012}+O(\varepsilon)=\frac{31579}{31580}+O(\varepsilon)$; apply
Corollary~\ref{hcor:transfer}.
\end{proof}

%% file: app/comp.tex
\section{Verification provenance and computations}\label{sec:provenance}\label{app:comp}
\emph{Proved on paper.}  Theorem~\ref{thm:main} and everything it uses (the shift lemma, the history lemma, the
structural theorem with its peeling and partial-target inductions, the zigzag of Theorem~\ref{thm:E0}, and the envelope
computation) are written proofs; the constant is a closed form.  The same holds for the junction theorems
(Theorems~\ref{thm:j2}, \ref{thm:j4}, \ref{thm:opt} and~\ref{thm:unique}, whose numeric side conditions were checked in exact
arithmetic), for the results of Appendix~\ref{sec:alphabet} except the numbers of Corollary~\ref{cor:byte}, and for
Theorem~\ref{thm:hard}.  The fixed-alphabet results of Section~\ref{sec:fixed} (Lemma~\ref{lem:dom}, Theorems~\ref{thm:fixed},
\ref{thm:binary}, \ref{thm:string}, \ref{thm:repair} and~\ref{thm:capped-exhaustion}, Corollaries~\ref{cor:shortest}
and~\ref{cor:eptas}) are written proofs with no computational step.

\emph{Computed.}  Proposition~\ref{prop:cxL20} rests on an explicit rational mixture and distribution, checked in exact
arithmetic against all patterns of $\mathcal Q(\mathcal W)$, which are found by filtering all $2^{21}$ patterns.  The table and
the threshold of Corollary~\ref{cor:byte} are an exact integer recursion and an interval-arithmetic evaluation of a closed
form.  The instances of Table~\ref{tab:up} and Proposition~\ref{prop:smallk} are exact replays of \GT{} runs.  Each of these
computations was carried out by two separately written programs.

\emph{Tested.}  The written proofs were also tested by computation, which is not part of the proofs.  The history lemma was
compared with a direct simulation of \GT{} on millions of occurrences in random and exhaustively enumerated small
corpora, with no mismatch; deliberately wrong variants of the middle process (no skipping rule, reversed class order) were
detected.  The witnesses of Theorems~\ref{thm:H2} and~\ref{thm:PH} were built explicitly by their recursions and
evaluated in exact rational arithmetic against the complete history family, for more than $10^5$ random, clustered and
exhaustively enumerated ordered window sets ($L\le26$ for $\beta=1$, $L\le20$ for $\beta\ge2$), with no failure and minimum slack exactly $0$; the same $\beta=1$ witness evaluated against
the history-free patterns of the $L=20$ counterexample fails, as Proposition~\ref{prop:cxL20} predicts.  The per-step inequalities of
Appendix~\ref{sec:alphabet} were checked at every step of exhaustive tie-break searches on small corpora.
Table~\ref{tab:comp} lists every computation, and the supplementary archive contains standalone checkers.

\subsection{Computations and artifacts}
Table~\ref{tab:comp} lists every computation behind a statement of the paper, and every computational test of a written
proof.  All rest on exact simulators of \GT{}, which implement the leftmost scan, the commit rule and every tie-break,
with exact evaluation of $f$ and of $\OPT_k$ (by enumeration or MILP), and on exact rational or integer arithmetic.  Rows
marked $^\ast$ were re-checked with separately written code.  Rows marked ``test'' check a written proof and are not
part of it.

{\small\renewcommand{\arraystretch}{1.15}
\begin{longtable}{p{0.23\textwidth}p{0.44\textwidth}p{0.22\textwidth}}
\caption{Computations and computational tests.}\label{tab:comp}\\\toprule
Claim & Check & Scale\\\midrule\endfirsthead
\toprule Claim & Check & Scale\\\midrule\endhead
Theorem~\ref{thm:main}$^\ast$ (constant) & closed form of $y(1)$ evaluated in interval arithmetic; independent RK4 integration of
$y'=\varphi(y)$; $\gamma_2,\dots,\gamma_8$ in exact rationals & ---\\
Lemma~\ref{lem:2P}$^\ast$ (test) & pattern of every occurrence compared with $[0,p)\cup(q,L]\cup M(p,q)$ along random and
exhaustive \GT{} runs; wrong variants (no skipping rule, reversed or merged classes, newest class dropped) detected &
$2.1\cdot10^6$ random and $6.6\cdot10^7$ exhaustive occurrence checks (all words of length $\le7$ over two letters, $\le5$
over three); no mismatch\\
Theorem~\ref{thm:E0}$^\ast$ (test) & loads, value and mass of the zigzag in exact arithmetic; $\Phi_Z\ge V$ over all of
$\mathcal Q(\emptyset)$; the short-token distribution & $\beta\le6$, $L\le150$\\
Theorems~\ref{thm:H2}, \ref{thm:PH}$^\ast$ (test) & witnesses built by the recursions of the proofs and evaluated exactly
over the whole history family, asserting $D\ge0$ and the mass bound; Lemmas~\ref{lem:three} and~\ref{lem:restrict} checked
directly on patterns; negative controls (wrong peeled window; history-free patterns; witnesses scaled by $\frac{19}{20}$)
fail & $\beta=1$: $1.7\cdot10^6$ witness checks on $1.7\cdot10^5$ ordered window sets, $L\le26$, exhaustive for $\le3$ windows at
$L=8,9$; $\beta\ge2$: ${>}2\cdot10^4$ sets, $L\le20$; no failure, minimum slack $0$\\
Lemmas~\ref{lem:shift}, \ref{lem:int}, \ref{lem:chosen} (test) & pairwise check of the shift lemma; scores against the
maximum-weight disjoint valid family & $\VSINST$ random states; all binary words of length $\le11$, ternary $\le7$, three
commits ($\EXSTATES$ states); no violation\\
Corollary~\ref{cor:mono} (test) & exhaustive search for a score increase & $\LZBIGchecks$ checks on short words; none found\\
Corollary~\ref{cor:lazy} (test) & lazy loop vs.\ exact greedy over all tie-breaks & $\LZinstances$ random instances; lazy value
always in the exact range\\
Theorem~\ref{thm:short}$^\ast$ & exact $\gamma^{\rm s}_k$; goodness of $D_2,D_3$; brute force $\GTv_k\ge\gamma^{\rm s}_k\cdot\OPT^{(\le3)}_k$
over all tie-breaks & $4{,}800$ corpora with $k\le5$, $22{,}338$ with $k\le4$; no violation\\
Proposition~\ref{prop:cxL20}$^\ast$ & exact mixture $\mu$ ($32$ patterns) and distribution $D_{\mathcal W}$ ($22$ windows)
checked against $\mathcal Q(\mathcal W)$, obtained by filtering all $2^{21}$ patterns; $\alpha^{\rm H}(20;\emptyset;1)=\frac12$ by
Lemma~\ref{lem:mono} and an exact $66$-pattern mixture; standalone checker in the supplement & $L=20$, $\beta=1$\\
Theorem~\ref{thm:fixed}(a), Lemmas~\ref{lem:dom}, \ref{lem:short}, Proposition~\ref{prop:S}$^\ast$ (test) & every inequality of the
proofs at every state of exhaustive tie-break searches; controls with weakened constants fire & $16{,}000$ corpora over
$2$--$4$ letters ($7.2\cdot10^4$ states) and $1.6\cdot10^4$ random corpora ($2\cdot10^5$ steps); no violation\\
Corollary~\ref{cor:byte}$^\ast$ & recursion in integers scaled by $10^{18}$, every branch rounded down; threshold by
interval arithmetic & $k=2^{14},\dots,2^{22}$\\
Proposition~\ref{prop:U} (test) & exact runs of explicit members, five tie rules, against the bounds (a)--(c) & $U(s,x,r)$
with $s\le4$; $42$ runs\\
Theorem~\ref{thm:opt}$^\ast$ & exact LP-duality certificates of $\OPT_K=f(Y)$; exact MILP & J2 $n\in\{\OcertJtwo\}$, J4
$n\in\{\OcertJfour\}$\\
Theorems~\ref{thm:j2}, \ref{thm:j4}$^\ast$ & exact replay of every pick (maximum over all candidates, tie set recorded); $f(Y)$
by DP; families rebuilt independently & J2: $n=10,12,20,30$; J4: $n=10,12,14,16,20$\\
Proposition~\ref{prop:recode}$^\ast$ & injectivity of the recoding on all substrings; exact replay of the recoded runs & J4
$n=10,12,14,16,20$, J2 $n=30$\\
Lemma~\ref{lem:bonus}, Theorem~\ref{thm:unique}$^\ast$ & bonus words added to J2/J4 witnesses; uniqueness of the maximizer
asserted at every step & $1{,}210$ random runs; no failure\\
Table~\ref{tab:up}, Prop.~\ref{prop:smallk}$^\ast$ & exact replay; full tie tree for $k\le4$; $\OPT_k$ by MILP and brute force &
up to $88{,}000$ words\\
Lemmas~\ref{lem:copies}, \ref{lem:copies-strict} & exact runs on copies, all tie-breaks, exact OPT & up to $k=\CsKmax$\\
Proposition~\ref{prop:covfam} & exact runs of explicit members & $\CovN$ members, $k=\CovKmin$--$\CovKmax$\\
Table~\ref{tab:exh} & exhaustive enumeration, OPT over all $k$-subsets & see table\\
Theorem~\ref{thm:hard} (test) & CP-SAT over all substrings; control instances with forcing weight $1$ & $\HardInst$ instances,
$\HardCtrl$ controls\\
Appendix~\ref{app:impl} & released C++ (\texttt{pcatt} at \texttt{36b928c7}) vs.\ our simulator & $\PcattTOne$ scores,
$\PcattSteps$ steps, $\PcattLazyN$ lazy runs; all agree\\
\bottomrule\end{longtable}}

\begin{table}[!htb]\centering
\input{gt/exh_table.tex}
\caption{Exact minima of $\GTv_k/\OPT_k$ over bounded classes (binary alphabet, distinct words, counts with $\gcd=1$, one
representative per $a\leftrightarrow b$ swap).  ``Worst ties'': adversarial tie-breaking; ``no ties'': instances where every
greedy choice is unique.  These small classes are far from extremal; compare Table~\ref{tab:up}.}\label{tab:exh}
\end{table}

\paragraph{Reproducibility.}  All code, result files and witnesses are in the supplementary archive\ifdefined\NAMEDBUILD{} of the
Zenodo record \href{https://doi.org/10.5281/zenodo.24451478}{\texttt{doi:10.5281/zenodo.24451478}}\fi.  Its folder \texttt{main\_bound\_checkers/} contains standalone exact
programs for the constant of Theorem~\ref{thm:main}, for Theorem~\ref{thm:E0}, for the history lemma, for the witnesses of
the structural theorem, and for Corollary~\ref{cor:byte}; the folder \texttt{counterexample\_L20/} checks
Proposition~\ref{prop:cxL20}.  Checksums and commands are in the supplement's README.

%% file: gt/exh_table.tex
\begin{tabular}{rrrrrcc}\toprule $k$ & words & max len & max count & instances & worst ties & no ties\\\midrule
2 & 2 & 7 & 4 & 148,830 & $3/4$ & $15/19$\\
2 & 3 & 5 & 3 & 327,775 & $13/18$ & $7/9$\\
2 & 3 & 6 & 3 & 2,852,575 & $13/18$ & $23/30$\\
2 & 4 & 4 & 3 & 585,548 & $2/3$ & $15/19$\\
2 & 5 & 4 & 2 & 1,015,343 & $2/3$ & $13/16$\\
3 & 2 & 7 & 4 & 148,830 & $7/8$ & $1$\\
3 & 3 & 5 & 2 & 91,777 & $3/4$ & $7/8$\\
3 & 4 & 4 & 2 & 111,180 & $13/18$ & $7/8$\\
3 & 4 & 4 & 3 & 585,548 & $13/18$ & $19/23$\\
\bottomrule\end{tabular}

%% file: main.bbl
\begin{thebibliography}{10}

\bibitem{ageev2004}
Alexander~A. Ageev and Maxim~I. Sviridenko.
\newblock Pipage rounding: A new method of constructing algorithms with proven
  performance guarantee.
\newblock {\em Journal of Combinatorial Optimization}, 8(3):307--328, 2004.

\bibitem{as19}
Per Austrin and Aleksa Stankovi{\'c}.
\newblock Global cardinality constraints make approximating some {Max-2-CSPs}
  harder.
\newblock In {\em Approximation, Randomization, and Combinatorial Optimization
  (APPROX 2019)}, 2019.
\newblock arXiv:1907.04165.

\bibitem{bpetight}
Kundan Baliga, Aiden Xie, and Aditya Byju.
\newblock Greedy {BPE} with replacement counts saves at least
  $\frac12(1-(1-\frac2k)^k)$ of the optimum, which is tight as $k\to\infty$.
\newblock Preprint, 2026.

\bibitem{complexity}
Kundan Baliga, Aiden Xie, and Aditya Byju.
\newblock Pair-merge tokenisation is hard to approximate on binary strings and
  strongly {NP}-hard on unary datasets.
\newblock arXiv:2611.13568v1, 2026.
\newblock URL: \url{https://arxiv.org/abs/2611.13568}.

\bibitem{bastan2026}
Atieh Bastan, Xiaoling He, A.~Pavan, and N.~V. Vinodchandran.
\newblock Tokenization: Hardness and bicriteria approximation algorithms.
\newblock arXiv:2611.15074v1, 2026.
\newblock URL: \url{https://arxiv.org/abs/2611.15074}.

\bibitem{bk}
Piotr Berman and Marek Karpinski.
\newblock On some tighter inapproximability results, further improvements.
\newblock Electronic Colloquium on Computational Complexity, Report TR98-065,
  1998.

\bibitem{bian2017}
Andrew~An Bian, Joachim~M. Buhmann, Andreas Krause, and Sebastian Tschiatschek.
\newblock Guarantees for greedy maximization of non-submodular functions with
  applications.
\newblock In {\em International Conference on Machine Learning (ICML)}, 2017.

\bibitem{compressed}
Miko{\l}aj Boja{\'n}czyk and Markus Lohrey.
\newblock Transducers on compressed strings.
\newblock In {\em International Colloquium on Automata, Languages, and
  Programming (ICALP)}, 2026.
\newblock \href {https://doi.org/10.4230/LIPIcs.ICALP.2026.170}
  {\path{doi:10.4230/LIPIcs.ICALP.2026.170}}.

\bibitem{fstat}
Marco Cognetta and Naoaki Okazaki.
\newblock Tokenization as finite-state transduction.
\newblock arXiv:2410.15696, 2024.
\newblock URL: \url{https://arxiv.org/abs/2410.15696}.

\bibitem{dask2011}
Abhimanyu Das and David Kempe.
\newblock Submodular meets spectral: Greedy algorithms for subset selection,
  sparse approximation and dictionary selection.
\newblock In {\em International Conference on Machine Learning (ICML)}, 2011.

\bibitem{ganczorz}
Micha{\l} Ga{\'n}czorz.
\newblock Entropy bounds for grammar compression.
\newblock arXiv:1804.08547v2, 2020.
\newblock URL: \url{https://arxiv.org/abs/1804.08547}.

\bibitem{hjl}
Danny Hucke, Artur Je{\.z}, and Markus Lohrey.
\newblock Approximation ratio of {RePair}.
\newblock arXiv:1703.06061, 2017.
\newblock URL: \url{https://arxiv.org/abs/1703.06061}.

\bibitem{ion2026}
Andrei~Robert Ion, Rares Mateiu, Adrian Miclaus, and Alexandru Popa.
\newblock How hard is tokenisation? {A}lgorithms and complexity beyond
  {NP}-hardness.
\newblock arXiv:2611.12348v1, 2026.
\newblock URL: \url{https://arxiv.org/abs/2611.12348}.

\bibitem{jain2003}
Kamal Jain, Mohammad Mahdian, Evangelos Markakis, Amin Saberi, and Vijay~V.
  Vazirani.
\newblock Greedy facility location algorithms analyzed using dual fitting with
  factor-revealing {LP}.
\newblock {\em Journal of the ACM}, 50(6):795--824, 2003.

\bibitem{kastreva2025}
Violeta Kastreva, Philip Whittington, Dennis Komm, and Tiago Pimentel.
\newblock Tokenisation over bounded alphabets is hard.
\newblock In {\em International Conference on Learning Representations (ICLR)},
  2026.
\newblock arXiv:2511.15709v2.

\bibitem{kozma2024arxiv}
L{\'a}szl{\'o} Kozma and Johannes Voderholzer.
\newblock Theoretical analysis of byte-pair encoding.
\newblock Extended version, arXiv:2411.08671v2, 2026.
\newblock URL: \url{https://arxiv.org/abs/2411.08671v2}.

\bibitem{kozma2026esa}
L{\'a}szl{\'o} Kozma and Johannes Voderholzer.
\newblock Theoretical analysis of byte-pair encoding.
\newblock In {\em 34th Annual European Symposium on Algorithms (ESA 2026)},
  volume 388 of {\em LIPIcs}, pages 80:1--80:17, 2026.
\newblock \href {https://doi.org/10.4230/LIPIcs.ESA.2026.80}
  {\path{doi:10.4230/LIPIcs.ESA.2026.80}}.

\bibitem{lim2025}
Jia~Peng Lim, Shawn Tan, Davin Choo, and Hady~W. Lauw.
\newblock A partition cover approach to tokenization.
\newblock In {\em Advances in Neural Information Processing Systems (NeurIPS)},
  2025.
\newblock arXiv:2501.06246v3.

\bibitem{manurangsi}
Pasin Manurangsi.
\newblock A note on {Max} $k$-vertex cover: Faster {FPT-AS}, smaller
  approximate kernel and improved approximation.
\newblock In {\em Symposium on Simplicity in Algorithms (SOSA 2019)}, 2019.
\newblock arXiv:1810.03792.

\bibitem{nr}
Gonzalo Navarro and Lu{\'\i}s M.~S. Russo.
\newblock Re-pair achieves high-order entropy.
\newblock In {\em Data Compression Conference (DCC)}, 2008.
\newblock \href {https://doi.org/10.1109/DCC.2008.79}
  {\path{doi:10.1109/DCC.2008.79}}.

\bibitem{sennrich2016}
Rico Sennrich, Barry Haddow, and Alexandra Birch.
\newblock Neural machine translation of rare words with subword units.
\newblock In {\em Proceedings of the 54th Annual Meeting of the Association for
  Computational Linguistics}, pages 1715--1725, 2016.
\newblock \href {https://doi.org/10.18653/v1/P16-1162}
  {\path{doi:10.18653/v1/P16-1162}}.

\bibitem{storer1982}
James~A. Storer and Thomas~G. Szymanski.
\newblock Data compression via textual substitution.
\newblock {\em Journal of the ACM}, 29(4):928--951, 1982.

\bibitem{whittington2025}
Philip Whittington, Gregor Bachmann, and Tiago Pimentel.
\newblock Tokenisation is {NP}-complete.
\newblock In {\em Proceedings of the 63rd Annual Meeting of the Association for
  Computational Linguistics (Long Papers)}, pages 28133--28153, 2025.
\newblock arXiv:2412.15210.

\bibitem{zouhar2023formal}
Vil{\'e}m Zouhar, Clara Meister, Juan Gastaldi, Li~Du, Tim Vieira, Mrinmaya
  Sachan, and Ryan Cotterell.
\newblock A formal perspective on byte-pair encoding.
\newblock In {\em Findings of the Association for Computational Linguistics:
  ACL 2023}, pages 598--614, 2023.

\end{thebibliography}
